\documentclass[sigconf,edbt]{acmart-edbt2023_nofooter}

\usepackage{ifthen}
\newboolean{fullversion}
\setboolean{fullversion}{true}

\def\BibTeX{{\rm B\kern-.05em{\sc i\kern-.025em b}\kern-.08em
    T\kern-.1667em\lower.7ex\hbox{E}\kern-.125emX}}

\usepackage{booktabs} 

\ifthenelse{\boolean{fullversion}}{
  \setcopyright{none}
  \fancyfoot{}
  \thispagestyle{empty}
}{
\setcopyright{rightsretained}

\acmDOI{}

\acmISBN{978-3-89318-088-2}

\acmConference[EDBT 2023]{26th International Conference on Extending Database Technology (EDBT)}{28th March-31st March, 2023}{Ioannina, Greece}
\acmYear{2023}
}

\usepackage{amsmath}
\usepackage{amsthm}
\usepackage{xcolor}
\usepackage{adjustbox}
\usepackage{calc}
\usepackage{tikz}
\usetikzlibrary{decorations.pathreplacing,angles,quotes,positioning,shapes.misc,decorations.text}
\usepackage{subcaption}
\usepackage{xspace}
\usepackage{url}

\usepackage[dutch,english]{babel}

\usepackage{wasysym}

\usepackage[normalem]{ulem} 

\usepackage[ruled,vlined,noend]{algorithm2e}

\usepackage[labelfont=bf]{caption}

\usepackage{verbatimbox}

\usepackage{ifthen}

\theoremstyle{plain}
\newtheorem{cond}[theorem]{Condition}
\theoremstyle{definition}

\usepackage{listings}

\newcommand{\phantomtodo}[1]{}

\renewcommand{\epsilon}{\varepsilon}

\newcommand{\noncounterflow}{\ensuremath{\textit{non-counterflow}}}
\newcommand{\counterflow}{\ensuremath{\textit{counterflow}}}
\newcommand{\adjacentcfp}{adjacent-counterflow pair\xspace}
\newcommand{\orderedcfp}{ordered-counterflow pair\xspace}
\newcommand{\adjacentcf}{adjacent-counterflow\xspace}

\newcommand{\typeone}{type-I\xspace}
\newcommand{\typetwo}{type-II\xspace}

\newcommand{\schRel}{\textsf{Rels}}
\newcommand{\schFK}{\textsf{FKeys}}

\newcommand{\Attr}[1]{\ensuremath{\text{Attr}(#1)}}
\newcommand{\rel}[1]{\text{rel}(#1)}

\newcommand{\tuplesres}[1]{I(#1)} 
\newcommand{\tversions}[1]{V({#1})}

\newcommand{\x}{\mathtt{t}}
\newcommand{\y}{\mathtt{u}} 

\newcommand{\myl}{\mathtt{l}}

\newcommand{\dom}[1]{\text{\it dom}(#1)}
\newcommand{\range}[1]{\text{\it range}(#1)}

\newcommand{\myR}{\ensuremath{\mathtt{R}}}
\newcommand{\myW}{\ensuremath{\mathtt{W}}}

\newcommand{\myIN}{\ensuremath{\mathtt{I}}}
\newcommand{\myDE}{\ensuremath{\mathtt{D}}}
\newcommand{\myPR}{\ensuremath{\mathtt{PR}}}
\newcommand{\R}[2][i]{\myR_{#1}\mathtt{[#2]}}
\newcommand{\W}[2][i]{\myW_{#1}\mathtt{[#2]}}

\newcommand{\IN}[2][i]{\myIN_{#1}\mathtt{[#2]}}
\newcommand{\DE}[2][i]{\myDE_{#1}\mathtt{[#2]}}
\newcommand{\PR}[2][i]{\myPR_{#1}{[#2]}}
\newcommand{\CT}[1][i]{\mathtt{C}_{#1}}

\newcommand{\ReadSet}[1]{\ensuremath{\text{ReadSet}(#1)}}
\newcommand{\WriteSet}[1]{\ensuremath{\text{WriteSet}(#1)}}
\newcommand{\PReadSet}[1]{\ensuremath{\text{PReadSet}(#1)}}
\newcommand{\RGSet}[1]{\ensuremath{\text{Attr}(#1)}} 
\newcommand{\WGSet}[1]{\ensuremath{\text{Attr}(#1)}} 
\newcommand{\PGSet}[1]{\ensuremath{\text{Attr}(#1)}} 
\newcommand{\BGSet}[1]{\ensuremath{\text{Attr}(#1)}} 

\newcommand{\trans}[1][i]{T_{#1}}

\newcommand{\transset}{{\mathcal{T}}}

\newcommand{\chunks}[1]{\textit{Chunks}(#1)}

\newcommand{\schedule}{s}
\newcommand{\schop}[1][\schedule]{O_{#1}}
\newcommand{\schord}[1][\schedule]{\leq_{#1}}
\newcommand{\schords}[1][\schedule]{<_{#1}}
\newcommand{\schinit}[1][\schedule]{\textit{init}_{#1}}
\newcommand{\schvord}[1][\schedule]{\ll_{#1}}
\newcommand{\schvf}[1][\schedule]{v_{#1}}

\newcommand{\schrvf}[1][\schedule]{v^r_{#1}}
\newcommand{\schwvf}[1][\schedule]{v^w_{#1}}
\newcommand{\schpvf}[1][\schedule]{\textit{Vset}_{#1}}

\newcommand{\dependson}[3][\schedule]{#2 \rightarrow_{#1} #3}

\newcommand{\prog}[1][]{P_{#1}}
\newcommand{\progset}{\mathcal{P}}
\newcommand{\schedules}[2]{\textit{schedules}(#1, #2)}

\newcommand{\type}[1]{\text{type}(#1)}

\newcommand{\selectset}[1]{\textit{select-set}(#1)}
\newcommand{\keycond}[1]{\textit{key-condition}(#1)}
\newcommand{\predcond}[1]{\textit{predicate-condition}(#1)}
\newcommand{\expr}[1]{\textit{expr}(#1)}
\newcommand{\subprogram}{\textit{subprogram}}

\newcommand{\obsset}[1]{\ReadSet{#1}}
\newcommand{\modset}[1]{\WriteSet{#1}}
\newcommand{\predset}[1]{\PReadSet{#1}}

\newcommand{\qins}{\text{ins}}
\newcommand{\qkdel}{\text{key del}}
\newcommand{\qpdel}{\text{pred del}}
\newcommand{\qkupd}{\text{key upd}}
\newcommand{\qpupd}{\text{pred upd}}
\newcommand{\qksel}{\text{key sel}}
\newcommand{\qpsel}{\text{pred sel}}
\newcommand{\qkmod}{\qkupd} 
\newcommand{\qpmod}{\qpupd} 

\newcommand{\unfold}[1]{\textit{Unfold}_{\leq 2}(#1)}

\newcommand{\readcom}{{\normalfont\textsc{read committed}}\xspace}
\newcommand{\readmvcom}{{\normalfont\textsc{multiversion read committed}}\xspace}

\newcommand{\snapshot}{{\normalfont\textsc{snapshot isolation}}\xspace}

\newcommand{\serializable}{{\normalfont\textsc{serializable}}\xspace}

\newcommand{\parsnapshot}{{\normalfont\textsc{parallel snapshot isolation}}\xspace}

\newcommand{\mvrc}{\textsc{mvrc}\xspace}
\newcommand{\MVRC}{\mvrc}

\newcommand{\btp}{\text{BTP}\xspace}
\newcommand{\ltp}{\text{LTP}\xspace}

\newcommand{\algonameconstructsg}{\sc{constructSuG}}

\newcommand{\ie}{\textit{i.e.}}
\newcommand{\eg}{\textit{e.g.}}
\newcommand{\ssize}[1]{|#1|}

\newcommand{\cg}[1]{SeG(#1)}
\newcommand{\sg}[1]{SuG(#1)}

\newcommand{\cyclesym}{\Gamma}

\newcounter{conditioncounter}

\newcommand{\Account}{\ensuremath{\text{Account}}\xspace}
\newcommand{\Savings}{\ensuremath{\text{Savings}}\xspace}
\newcommand{\Checking}{\ensuremath{\text{Checking}}\xspace}

\begin{document}

\title{Detecting Robustness against MVRC \\ for Transaction Programs with Predicate Reads}

\author{Brecht Vandevoort}
\affiliation{%
  \institution{UHasselt, Data Science Institute, ACSL}
  \country{Belgium}
}

\author{Bas Ketsman}
\affiliation{%
  \institution{Vrije Universiteit Brussel}
  \country{Belgium}
}

\author{Christoph Koch}
\affiliation{%
  \institution{\'Ecole Polytechnique F\'ed\'erale de Lausanne}
  \country{Switzerland}
}

\author{Frank Neven}
\orcid{0000-0002-7143-1903}
\affiliation{\institution{UHasselt, Data Science Institute, ACSL}
  \country{Belgium}
}

\begin{abstract}
  {The transactional robustness problem revolves around deciding whether, for a given workload, a lower isolation level than Serializable is sufficient to guarantee serializability. The paper presents a new characterization 
    for robustness against isolation level (multi-version) Read Committed.} It supports
  transaction programs with control structures (loops and conditionals) and
  inserts, deletes, and predicate reads -- scenarios that trigger the phantom
  problem, which is known to be hard to analyze in this context.
  The characterization is graph-theoretic and not unlike previous decision
  mechanisms known from the concurrency control literature that database researchers and practicians are comfortable with.
  We show experimentally that our characterization pushes the frontier in allowing to recognize more and more complex workloads as robust than before.
\end{abstract}

\maketitle


\section{Introduction}
\label{sec:introduction}

{The gold standard for desirable transactional semantics is serializability, and much research and technology development has gone into creating systems that provide the greatest possible transaction throughput. Nevertheless, in practice, a hierarchy of alternative isolation levels of different strengths is available, allowing users to trade off semantic guarantees for better performance. One example is the isolation level (multi-version) Read Committed (\mvrc), which does not guarantee serializability but which can be implemented more efficiently than isolation level Serializable.
  The central question that we address in this paper is: \emph{When is it safe to run a transactional workload under \mvrc?} }
%

Recently, a number of researchers have studied the so-called transactional
robustness problem~\cite{Alomari:2008:CSP:1546682.1547288,DBLP:conf/aiccsa/AlomariF15,DBLP:conf/cav/BeillahiBE19,DBLP:conf/concur/BeillahiBE19,DBLP:conf/concur/0002G16,DBLP:conf/concur/Cerone0G15,DBLP:journals/tods/FeketeLOOS05,cerone_et_al:LIPIcs:2017:7794,DBLP:conf/pods/Fekete05,
  VLDBpaper,ICDTpaper}, which revolves around deciding whether, for a given workload, a lower isolation level than Serializable is sufficient to guarantee serializability. Specifically, a set of transactions is called robust against a given isolation level if every possible interleaving of the transactions under consideration that is allowed under the specified isolation level is serializable. That there is a real chance
that nontrivially robust workloads do exist is probably best demonstrated by the fact that the
well-known benchmark TPC-C is robust for Snapshot Isolation~\cite{DBLP:journals/tods/FeketeLOOS05}.

Robustness is a static property of workloads involving an offline analysis. A workload (the set of transaction programs at the application level) is analyzed by its developers during development time, and the insight into its robustness for a given low isolation level is later used to consistently deploy it with a database server using a specific isolation level weaker than
serializable. Robustness is a hard problem and undecidability is reached quite quickly~\cite{ICDTpaper}.
%
For exact characterizations of robustness, the possibility of phantom problem anomalies makes the problem very difficult, and, typically, research on the robustness problem has excluded insertions, deletions, and predicate reads~\cite{DBLP:conf/pods/Fekete05,ICDTpaper,VLDBpaper,DBLP:conf/pods/Ketsman0NV20}, in addition to assuming that transaction programs are linear sequences of reads and writes without any control structures.


{
To allow for the inclusion of predicate reads, sound
robustness tests based on sufficient conditions have been developed~\cite{DBLP:conf/aiccsa/AlomariF15,DBLP:journals/tods/FeketeLOOS05,DBLP:conf/concur/0002G16,DBLP:conf/concur/Cerone0G15,Cerone:2018:ASI:3184466.3152396}. Such conditions are based on the following observation. When a \emph{schedule} is not serializable, then the serialization graph constructed from that schedule contains a cycle satisfying a property specific to the isolation level under consideration: {\em dangerous structure}\/~\cite{DBLP:journals/tods/FeketeLOOS05} for \snapshot and the presence of a counterflow edge for \MVRC~\cite{DBLP:conf/aiccsa/AlomariF15}.
This approach is extended to
a workload of \emph{transaction programs} via a
so-called static dependency graph summarizing all possible serialization graphs for all possible executions allowed under the isolation level at hand. In this static dependency graph, each program is represented by a node, and there is a conflict edge from one program to another if there can be a schedule that gives rise to that conflict. The absence of a cycle satisfying the condition specific to that isolation level then guarantees robustness, while the presence of a cycle does not necessarily imply non-robustness. Indeed, every counterexample cycle in a serialization graph is witnessed by a cycle in this static dependency graph, but the converse is not necessarily true.
A major obstacle preventing direct application to practical workloads is that the construction of the static dependency graph is a manual step that should be performed by a database specialist. This is a difficult problem as the decision to place an edge requires reasoning over all possible schedules.
In this paper, we build further upon the just mentioned line of work by (i) identifying a more specific condition that holds for all cycles found in the serialization graph of a schedule allowed under \mvrc, thereby allowing to identify more workloads as robust against \mvrc, and (ii), by providing a more formal approach to construct these static dependency graphs, thereby facilitating automatic construction for a given set of transaction programs.
}

{In this paper, we study the robustness problem for \mvrc and 
  obtain a sound
  robustness detection algorithm that improves over the state-of-the-art in that it (i) can detect larger sets of transaction programs to be robust; (ii) incorporates operations like insert, delete and predicate reads that, to the best of our knowledge, have not been considered before
  thereby, allowing to verify robustness for a wider range of workloads, including for example TPC-C; and, (iii) can readily be implemented and applied in practice as the static dependency graph (called summary graph in this work) can be automatically constructed based on a formalization of transaction programs, called \btp.} The precise formalisation facilitates the applicability to any kind of transaction programs consisting of operations for which the following information can be derived (when applicable): type of operation, set of observed and modified attributes, set of attributes used in a predicate read, and implied foreign key constraints.
In other words, our techniques require only this information, and do not need to keep and analyze intermediate representations of the transaction program code.





\vspace{1mm}
\noindent
{\it Outline and contributions.}
{To make the paper more readable, we introduce the main ideas behind our formalisation and the algorithm by means of a running example in Section~\ref{sec:runningex} before introducing the necessary definitions in Section~\ref{sec:defs}.}
%
In Section~\ref{sec:ser:graph},
we show that when a schedule allowed under \mvrc is not serializable, then it must contain a cycle satisfying a certain condition (Theorem~\ref{theo:cg:cycles}). This improves over the graph-based condition presented in \cite{DBLP:conf/aiccsa/AlomariF15}.
In Section~\ref{sec:robustness}, we introduce the formalism of basic transaction programs (\btp{}s) incorporating inserts, deletes, predicate reads and control structure.
In Section~\ref{sec:detecting:robustness}, we provide algorithms for constructing the summary graph (Algorithm~\ref{alg:graph}) and testing robustness (Algorithm~\ref{alg:robust:check}) based on the sufficient condition obtained in Section~\ref{sec:ser:graph}.
We show through experiments in Section~\ref{sec:experiments} on two well known transaction benchmarks, TPC-C and Smallbank, that our approach detects strictly more sets of programs as robust compared to earlier work~\cite{DBLP:conf/aiccsa/AlomariF15}.
We furthermore introduce a new synthetic benchmark where the number of programs is parameterized. Based on this benchmark, we show that our algorithm scales to larger sets of programs as well and can test for robustness in a matter of seconds.
We discuss related work in Section~\ref{sec:relwork} and conclude in Section~\ref{sec:conclusions}.
\ifthenelse{\boolean{fullversion}}{}{
Missing proofs can be found in \cite{fullversionVDLB2022}.
}

\section{Running Example}
\label{sec:runningex}


To illustrate our approach, we introduce a running example based on an auction service. The database schema consists of three relations: Buyer(\underline{id}, calls), Bids(\underline{buyerId}, bid), and Log(\underline{id}, buyerId, bid), where the primary key for each relation is underlined and buyerId in Bids and Log is a foreign key referencing Buyer(id). The relation Buyer lists all potential buyers, Bids keeps track of the current bid for each potential buyer, and Log keeps a register of all bids. Each buyer can interact with the auction service through API calls. For logging purposes, the attribute Buyer(calls) counts the total number of calls made by the buyer.
The API interacts with the database via two transaction programs: FindBids($B$, $T$) and PlaceBid($B$, $V$)
whose SQL code is given in Figure~\ref{fig:sql_auction_progs}.
FindBids returns all current bids above threshold $T$, whereas PlaceBids increases the bid of buyer $B$ to value $V$ (if $V$ is {higher} than the current bid, otherwise the current bid remains unchanged) and inserts this newly placed bid as a new tuple in Log.
Both programs increment the number of calls for $B$.

\begin{figure}

  \begin{minipage}[t]{.50\linewidth-1em}
    \small
    \begin{verbatim}
FindBids(:B, :T):
  UPDATE Buyer --q1
  SET calls = calls + 1
  WHERE id = :B;

  SELECT bid --q2
  FROM Bids
  WHERE bid >= :T;

  COMMIT;
\end{verbatim}
    \begin{center}
      \begin{tabular}{|l|}
        \hline
        \multicolumn{1}{|c|}{Auction schema} \\
        \hline
        Buyer(\underline{id},calls)          \\
        Bids(\underline{buyerId}, bid)       \\
        Log(\underline{id},buyerId,bid)      \\
        \hline
        \multicolumn{1}{|c|}{Foreign keys}   \\
        \hline
        $f_1$: Bids(BuyerId) $\to$ Buyer(id) \\
        $f_2$: Log(BuyerId) $\to$ Buyer(id)  \\
        \hline
      \end{tabular}
    \end{center}

    \begin{center}
      \begin{tabular}{|l|l|}
        \hline
                 & \btp                      \\
        \hline
        FindBids & $q_1; q_2$                \\
        PlaceBid &
        $q_3; q_4; (q_5 \mid \epsilon); q_6$ \\
        \hline
      \end{tabular}
    \end{center}

  \end{minipage}%
  \hfill%
  \begin{minipage}[t]{.50\linewidth-1em}
    \small
    \begin{verbatim}
PlaceBid(:B, :V): 
  UPDATE Buyer --q3
  SET calls = calls + 1
  WHERE id = :B;

  SELECT bid into :C --q4
  FROM Bids
  WHERE buyerId = :B;

  IF :C < :V: --q5
    UPDATE Bids
    SET bid = :V
    WHERE buyerId = :B;
  ENDIF;
    
  :logId = uniqueLogId();
    
  INSERT INTO Log --q6
  VALUES(:logId, :B, :V);

  COMMIT;
\end{verbatim}

  \end{minipage}

  \caption{Auction schema, SQL code and \btp formalization for FindBids($B$, $T$) and PlaceBid($B$, $V$)}
  \label{fig:sql_auction_progs}
\end{figure}

\begin{figure}
  {\small
    \begin{tabular}[t]{| l | l | l | l | l | l |}
      \hline
      $q$   & $\type{q}$ & $\rel{q}$ & $\predset{q}$ & $\obsset{q}$ & $\modset{q}$   \\
      \hline
      \hline
      \multicolumn{6}{|c|}{\bf FindBids}                                             \\
      \hline
      $q_1$ & $\qkmod$   & Buyer     & $\bot$        & \{calls\}    & \{calls\}      \\
      \hline
      $q_2$ & $\qpsel$   & Bids      & \{bid\}       & \{bid\}      & $\bot$         \\
      \hline


      \multicolumn{6}{|c|}{\bf PlaceBid}                                             \\
      \hline
      $q_3$ & $\qkmod$   & Buyer     & $\bot$        & \{calls\}    & \{calls\}      \\
      \hline
      $q_4$ & $\qksel$   & Bids      & $\bot$        & \{bid\}      & $\bot$         \\
      \hline
      $q_5$ & $\qkmod$   & Bids      & $\bot$        & \{\}         & \{bid\}        \\
      \hline
      $q_6$ & $\qins$    & Log       & $\bot$        & $\bot$       & \{id, buyerId, \\
            &            &           &               &              & bid\}          \\
      \hline
    \end{tabular}
  }
  \caption{Query details for BTPs FindBids and PlaceBid.
    \label{fig:btp_queries_placebid} \label{fig:btp_queries_findbids}}
\end{figure}

\paragraph*{Basic Transaction Programs}
We introduce the formalism of \emph{basic transaction programs} (BTP)
to overestimate the set of schedules that can arise when executing
transaction programs as given in Figure~\ref{fig:sql_auction_progs}. A \btp{} is a sequence of statements that only retains the information necessary to detect robustness against \mvrc: the type of statement (insert, key-based selection/update/delete, or predicate-based selection/update/delete), the relation that is referred to, and the attributes that are read from, written to, and that are used in predicates. {In particular, \btp{}s ignore the concrete predicate selection condition.}

Formally, a BTP is a sequence of statements $q_1;\ldots; q_k$.
For example, FindBids
is modeled by 
$q_1; q_2$, where $q_1$ and $q_2$ are two statements reflecting the corresponding SQL statements in Figure~\ref{fig:sql_auction_progs}.
Each statement $q_i$ is supplemented with additional information as detailed in Figure~\ref{fig:btp_queries_findbids}.
There, $\type{q_i}$ refers to the type of statement: an insert, a key-based or predicate-based selection, update or delete;
$\rel{q_i}$ is the relation under consideration;
$\ReadSet{q_i}$ are the attributes
  {read}
by $q_i$;
$\WriteSet{q_i}$ those
  {written}
by $q_i$; and,
$\PReadSet{q_i}$ the attributes used for predicates in the WHERE part of the query. 
We use $\bot$ to indicate that a specific function is not applicable to a statement.
For example, $q_1$ in FindBids is a key-based update over relation Buyer, since the corresponding SQL query selects exactly one tuple based on the primary key attribute Buyer(id). This statement reads and then overwrites the value for attribute Buyer(calls), and therefore $\ReadSet{q_1} = \WriteSet{q_1} = \{\text{calls}\}$. Since this statement is not predicate-based, we have $\PReadSet{q_1} = \bot$.
Statement $q_2$ is a predicate-based selection over relation Bids. The predicate \texttt{id = :B} in the corresponding SQL statement only uses the attribute Bids(bid), and therefore $\PReadSet{q_2} = \{\text{bid}\}$.
Therefore, $\ReadSet{q_2} = \{\text{bid}\}$.

\btp{}s incorporate conditional branching and loops as well. Indeed, 
PlaceBid is modeled by
$q_3; q_4; (q_5 \mid \epsilon); q_6$
supplemented
with additional information as depicted in Figure~\ref{fig:btp_queries_placebid}.
Here,
$(q_5 \mid \epsilon)$ denotes the branching corresponding to the IF-statement in the SQL program: either $q_5$ is executed (if the condition in the SQL program evaluates to true), or nothing is executed (if the condition evaluates to false). We note that an
ELSE-clauses can be modeled by replacing $\epsilon$ by a corresponding statement. Analogously, \btp{}s allow $\text{loop}(P)$ to express iteration, where $P$ is an arbitrary sequence of statements. Intuitively, $\text{loop}(P)$ specifies that $P$ can be repeated for an arbitrary yet finite number of iterations. We refer to Section~\ref{sec:robustness} for a formal definition of \btp{}s.

\begin{figure*}[t]
  \begin{minipage}[t]{.65\linewidth}
    \vspace{-2.4cm}
    {\footnotesize
      \addtolength{\arraycolsep}{-2mm}
      $
        \begin{array}{ l l l l l l l}
          \trans[1]:  \underbrace{\R[1]{\x_1} \W[1]{\x_1}}_{q_3} \underbrace{\R[1]{\y_1}}_{q_4} \underbrace{\IN[1]{\myl_1}}_{q_6} \CT[1]                                                                                                                                                                        \\
          \trans[2]: & \underbrace{\R[2]{\x_1} \W[2]{\x_1}}_{q_3} \underbrace{\R[2]{\y_1}}_{q_4} &                                            & \underbrace{\W[2]{\y_1}}_{q_5} &                                                                            & \underbrace{\IN[2]{\myl_2}}_{q_6} \CT[2]          \\
          \trans[3]: &                                                                           & \underbrace{\R[3]{\x_2} \W[3]{\x_2}}_{q_1} &                                & \underbrace{\PR[3]{\text{Bids}} \R[3]{\y_1} \R[3]{\y_2} \R[3]{\y_3}}_{q_2} &                                          & \CT[3]
        \end{array}
      $
      \addtolength{\arraycolsep}{+2mm}
    }
    \caption{Example schedule $\schedule$
      where $\trans[1]$ and $\trans[2]$ are {instantiations} of PlaceBid and $\trans[3]$ is an {instantiation} of FindBids.}
    \label{fig:ex:sched}
  \end{minipage}
  \hfill
  \begin{minipage}[t]{.30\linewidth}
    \resizebox{0.99\textwidth}{!}{
      \begin{tikzpicture}[sgnode/.style={draw}, sgedge/.style={draw,->},sgcedge/.style={sgedge,dashed},qnode/.style={font=\footnotesize}]
        \node[sgnode] (FindBids1) at (0,2) {FindBids};
        \node[sgnode] (PlaceBidIf1) at (-2.45,0) {$\text{PlaceBid}_1$};
        \node[sgnode] (PlaceBidElse1) at (2.45,0) {$\text{PlaceBid}_2$};
        \path[sgedge] (FindBids1) to[loop above]
        node[qnode,xshift=0,yshift=0] {$q_1 \rightarrow q_1$}
        (FindBids1);
        \path[sgedge] (FindBids1) to[bend left=45]
        node[qnode,xshift=6,yshift=2,rotate=320] {$q_1 \rightarrow q_3$}
        (PlaceBidElse1);
        \path[sgedge] (PlaceBidElse1) to[bend left=-30]
        node[qnode,xshift=-7,yshift=-2,rotate=320] {$q_3 \rightarrow q_1$}
        (FindBids1);
        \path[sgcedge] (FindBids1) to[bend left=25]
        node[qnode,xshift=8,yshift=0,rotate=40] {$q_2 \rightarrow q_5$}
        (PlaceBidIf1);
        \path[sgedge] (FindBids1) to[bend left=0]
        node[qnode,xshift=8,yshift=0,rotate=40] {$q_1 \rightarrow q_3$}
        (PlaceBidIf1);
        \path[sgedge] (FindBids1) to[bend left=-20]
        node[qnode,xshift=8,yshift=0,rotate=40] {$q_2 \rightarrow q_5$}
        (PlaceBidIf1);
        \path[sgedge] (PlaceBidIf1) to[bend left=35]
        node[qnode,xshift=0,yshift=5,rotate=40] {$q_3 \rightarrow q_1$}
        (FindBids1);
        \path[sgedge] (PlaceBidIf1) to[bend left=52]
        node[qnode,xshift=0,yshift=5,rotate=40] {$q_5 \rightarrow q_2$}
        (FindBids1);
        \path[sgedge] (PlaceBidElse1) to[loop right,in=10,out=350,looseness=4]
        node[qnode,xshift=-9,yshift=-7] {$q_3 \rightarrow q_3$}
        (PlaceBidElse1);
        \path[sgedge] (PlaceBidElse1) to[bend left=30]
        node[qnode,xshift=0,yshift=-4] {$q_4 \rightarrow q_5$}
        (PlaceBidIf1);
        \path[sgedge] (PlaceBidElse1) to[bend left=45]
        node[qnode,xshift=0,yshift=-4] {$q_3 \rightarrow q_3$}
        (PlaceBidIf1);
        \path[sgedge] (PlaceBidIf1) to[bend left=-5]
        node[qnode,xshift=0,yshift=4] {$q_5 \rightarrow q_4$}
        (PlaceBidElse1);
        \path[sgedge] (PlaceBidIf1) to[bend left=-15]
        node[qnode,xshift=0,yshift=4] {$q_3 \rightarrow q_3$}
        (PlaceBidElse1);
        \path[sgedge] (PlaceBidIf1) to[loop left,in=170,out=155,looseness=7]
        node[qnode,xshift=12.5,yshift=-7.5] {$q_5 \rightarrow q_4$}
        (PlaceBidIf1);
        \path[sgedge] (PlaceBidIf1) to[loop above,in=140,out=110,looseness=5]
        node[qnode,xshift=-3,yshift=-2] {$q_3 \rightarrow q_3$}
        (PlaceBidIf1);
        \path[sgedge] (PlaceBidIf1) to[loop left,in=205,out=190,looseness=7]
        node[qnode,xshift=15,yshift=-5] {$q_5 \rightarrow q_5$}
        (PlaceBidIf1);
        \path[sgedge] (PlaceBidIf1) to[loop below,in=250,out=220,looseness=5]
        node[qnode,xshift=5,yshift=2] {$q_4 \rightarrow q_5$}
        (PlaceBidIf1);

      \end{tikzpicture}
    } 

    \caption{{Summary graph containing a \typeone but no \typetwo cycles.}}
    \label{fig:ex:sg}
  \end{minipage}
\end{figure*}

A set of transaction programs $\progset$ induces an infinite set of possible schedules where each transaction in the schedule is an {instantiation} of a program in $\progset$ as informally explained next by means of an example.
We refer to Section~\ref{sec:robustness} for a formal treatment.
Consider the schedule $\schedule$ over transactions $\trans[1]$, $\trans[2]$ and $\trans[3]$ presented in Figure~\ref{fig:ex:sched}.
Here, $\trans[1]$ and $\trans[2]$ are {instantiations} of 
PlaceBid and $\trans[3]$ is an {instantiation} of 
FindBids (when considered as a \btp). Furthermore,
$\x_1$ and $\x_2$ are tuples of relation Buyer, $\y_1$, $\y_2$ and $\y_3$ are tuples of Bids, and $\myl_1$ and $\myl_2$ are tuples of  Log.
The operation $\R[1]{\x_1}$ (respectively $\W[1]{\x_1}$) indicates that transaction $\trans[1]$ reads (respectively writes to) tuple $\x_1$, and operation $\IN[1]{\myl_1}$ indicates that $\trans[1]$ inserts a new tuple $\myl_1$ into the database. The operation $\PR[3]{\text{Bids}}$ in $\trans[3]$ is a predicate read that evaluates a predicate over all tuples in relation Bids.

Figure~\ref{fig:ex:sched} further illustrates how each statement in a BTP leads to one or more operations over tuples.
For example, the key-based update $q_3$ in PlaceBid results in two operations $\R[1]{\x_1}$ and $\W[1]{\x_1}$. Notice in particular that these two operations are over the same tuple $\x_1$ of relation Buyer = $\rel{q_3}$, where the first operation reads the value for attribute Buyer(calls) and the second operation overwrites the value for this attribute, as indicated by $\ReadSet{q_3}$ and $\WriteSet{q_3}$.
The predicate-based selection statement $q_2$ of FindBids results in a larger number of operations in $\trans[3]$. First, the predicate read $\PR[3]{\text{Bids}}$ evaluates a predicate
over all tuples in Bids = $\rel{q_2}$, where only attribute Bids(bid) is used in the predicate, indicated by $\PReadSet{q_2}$.
This predicate intuitively corresponds to the WHERE clause of the corresponding SQL statement, but in our formalism, we will only specify the attributes needed in the predicate rather than the predicate itself.
Then, $\trans[3]$ reads 
three tuples of relation Bids. For each such tuple, only the value of attribute Bids(Bid) is read, as specified by $\ReadSet{q_2}$.
Also notice how $\trans[1]$ is an instantiation of PlaceBid where the if-condition evaluates to false, whereas for $\trans[2]$ it evaluates to true, witnessed by the presence of $q_5$ in $\trans[2]$ and its absence in $\trans[1]$.

\vspace{-2mm}
\paragraph*{Foreign Keys}
Schedules should respect foreign keys.
Two {instantiations} of PlaceBid that access the same tuple $\x_1$ of relation Bids also need to access the same Buyer $\y_1$
as Bids(buyerId) is a foreign key referencing Buyer(Id).
Such information can be used to rule out inadmissible schedules (that could otherwise inadvertently cause a set of transaction programs to not be robust). 
For example, the schedule $\schedule'$ obtained from $\schedule$ by substituting $\x_1$ with $\x_2$ in $\trans[1]$ violates the foreign key constraint and is therefore not admissible.
We refer to Section~\ref{sec:robustness} for a more formal treatment of how we handle foreign keys in BTPs.

\vspace{-2mm}
\paragraph*{\mvrc, Dependencies and Conflict Serializability}
When a database is operating under isolation level Multiversion Read Committed (\mvrc), each read operation reads the most recently committed version of a tuple, and write operations cannot overwrite uncommitted changes. For example, under the assumption that $\schedule$ in Figure~\ref{fig:ex:sched} is allowed under \mvrc, $\R[2]{\x_1}$ will observe the version of $\x_1$ written by $\W[1]{\x_1}$, as $\trans[1]$ committed before $\R[2]{\x_1}$. Read operation $\R[3]{\y_1}$ on the other hand will not see the changes made by $\W[2]{\y_1}$, as the commit of $\trans[2]$ occurs after $\R[3]{\y_1}$.

We say that two operations occurring in two different transactions are conflicting if they are over the same tuple, access a common attribute of this tuple, and at least one of these two operations overwrites the value for this common attribute. These conflicts introduce dependencies between operations.
For example, $\W[1]{\x_1}$ in $\trans[1]$ and $\R[2]{\x_1}$ in $\trans[2]$ are conflicting, as the former modifies the value for attribute Buyer(calls) and the latter reads this value. We therefore say that there is a wr-dependency from $\W[1]{\x_1}$ to $\R[2]{\x_1}$, denoted by $\dependson{\W[1]{\x_1}}{\R[2]{\x_1}}$.
Similarly, since we assume that $\schedule$ is allowed under \mvrc, $\R[3]{\y_1}$ observes a version of $\y_1$ before the changes made by $\W[2]{\y_1}$. We therefore say that there is an rw-antidependency from $\R[3]{\y_1}$ to $\W[2]{\y_1}$, denoted by $\dependson{\R[3]{\y_1}}{\W[2]{\y_1}}$. The serialization graph $\cg{s}$ contains transactions as nodes and edges correspond to dependencies.
It is well-known that a schedule is conflict serializable if there is no cycle
in 
$\cg{s}$. A more formal definition of dependencies, conflict serializability and \mvrc can be found in Section~\ref{sec:defs}.

A dependency from a transaction $\trans[i]$ to a transaction $\trans[j]$ is counterflow if $\trans[j]$ commits before $\trans[i]$ (that is, the direction of the dependency is opposite to the commit order). In our running example, the dependency $\dependson{\R[3]{\y_1}}{\W[2]{\y_1}}$ is a counterflow dependency, as $\trans[3]$ commits after $\trans[2]$. Alomari and Fekete~\cite{DBLP:conf/aiccsa/AlomariF15}
showed that if a schedule is allowed under \mvrc, then every cycle in the serialization graph contains at least one counterflow dependency.
  {We refer to cycles containing at least one counterflow dependency as a \typeone cycle.}
In Theorem~\ref{theo:cg:cycles}, we refine this condition
and show that every such cycle
must either contain an \adjacentcfp or an \orderedcfp, as well as a non-counterflow dependency, {and refer to the latter as a \typetwo cycle
    (formal definitions are given in Section~\ref{sec:ser:graph}).
    {As every \typetwo cycle is a \typeone cycle but not vice-versa, this refinement will allow us to identify larger sets of programs to be robust against \mvrc. In Section~\ref{sec:experiments}, we will show that our approach indeed leads to practical improvements for all considered benchmarks.}}

\vspace{-2mm}
\paragraph*{Linear Transaction Programs}
We refer to BTPs without branching and loops as linear transaction programs (LTP). 
For each BTP an equivalent set of LTPs can be derived by unfolding all branching statements and loops.
FindBids is also an LTP and PlaceBid can be unfolded into two LTPs $\text{PlaceBid}_1:= q_3; q_4; q_5; q_6$ and $\text{PlaceBid}_2:= q_3; q_4; q_6$.
Loop unfolding gives rise to an infinite number of LTPs.
However, we will show that for detecting robustness against \mvrc it suffices to limit loop unfoldings to at most two iterations.

\vspace{-3mm}
\paragraph*{Detecting Robustness against \mvrc}
A set $\progset$ of LTPs is robust against \mvrc if every allowed schedule is serializable. We therefore lift
the just mentioned condition from serialization graphs to
summary graphs. The summary graph $\sg{\progset}$ summarizes all serialization graphs for all possible schedules allowed under \mvrc over transactions instantiated from programs in $\progset$. Here, nodes in $\sg{\progset}$ are programs in $\progset$ and if a schedule allowed under \mvrc exists with a dependency $\dependson[]{b_i}{a_j}$, then an edge is added from $P_i$ to $P_j$ where $b_i$ is an operation in transaction $\trans[i]$ instantiated from a program $P_i \in \progset$ and $a_j$ is an operation in transaction $\trans[j]$ instantiated from $P_j \in \progset$.
That edge is annotated with statements $P_i$ and $P_j$ 
and is dashed when the dependency is counterflow.
The summary graph for the three LTPs FindBids, $\text{PlaceBid}_1$ and $\text{PlaceBid}_2$ is visualized in Figure~\ref{fig:ex:sg}.
If we consider for example the dependency $\dependson{\W[1]{\x_1}}{\R[2]{\x_1}}$, we see that $\sg{\progset}$ has a corresponding edge from $\text{PlaceBid}_2$ to $\text{PlaceBid}_1$, labeled with $q_3$ and $q_3$.
Analogously, the counterflow dependency $\dependson{\R[3]{\y_1}}{\W[2]{\y_1}}$ is witnessed by the counterflow edge from FindBids to $\text{PlaceBid}_1$ in $\sg{\progset}$. We present a formal algorithm constructing the graph $\sg{\progset}$ for a given set of LTPs in Section~\ref{sec:sum:graph}.

Let $\schedule$ be an arbitrary schedule allowed under \mvrc where transactions are {instantiations} of $\progset$. As each dependency 
in the 
serialization graph $\cg{\schedule}$ is witnessed by an edge in the summary graph $\sg{\progset}$, it immediately follows that each cycle in $\cg{\schedule}$ is witnessed by a cycle in $\sg{\progset}$. 
So, when $\sg{\progset}$ does not contain a {\typetwo cycle},
we can safely conclude that $\progset$ is robust against \mvrc. Indeed, the absence of such cycles indicates (by Theorem~\ref{theo:cg:cycles}) that no schedule allowed under \mvrc exists with a cycle in its serialization graph, implying that every such schedule is serializable.
The presence of a {\typetwo cycle}
does not necessarily imply non-robustness as there might not be a single schedule in which the corresponding cycle is realized. However, in that case, the conservative approach is to attest non-robustness to avoid false positives.
Algorithm~\ref{alg:robust:check} follows this conservative approach and determines $\progset$ to be robust iff $\sg{\progset}$ does not contain a {\typetwo} cycle.

We show in Section~\ref{sec:detecting:robustness}
the summary graph in Figure~\ref{fig:ex:sg} does not contain a \typetwo cycle.
The set $\{\text{FindBids},\text{PlaceBid}\}$ is therefore identified by Algorithm~\ref{alg:robust:check} as robust against \mvrc. The SQL programs presented in Figure~\ref{fig:sql_auction_progs} can thus be safely executed under isolation level \mvrc, without risking non-serializable behavior.
This 
improves over earlier work, as the summary graph does contain a \typeone cycle (e.g., between FindBids and $\text{PlaceBid}_1$), and, hence, the method of \cite{DBLP:conf/aiccsa/AlomariF15} can not identify $\{\text{FindBids},\text{PlaceBid}\}$ as robust.

\section{Definitions}
\label{sec:defs}

Our formalization of transactions and conflict serializability is closely related to the formalization presented by Adya et al.~\cite{DBLP:conf/icde/AdyaLO00}.
We extend upon the definitions presented in~\cite{VLDBpaper} and include three additional types of operations: predicate reads, inserts and deletes.

\subsection{Databases}

A \emph{relational schema} is a pair $(\schRel, \schFK)$, where $\schRel$ is a set of relation names and $\schFK$ is a set of foreign keys.
Then, $\Attr{R}$ denotes the finite set of %
attribute names.
We fix an infinite set $\tuplesres{R}$ of abstract objects called tuples, for
each $R\in\schRel$.
We assume that $\tuplesres{R} \cap \tuplesres{S} = \emptyset$ for all  $R,S\in\schRel$ with $R\neq S$.
When $t\in\tuplesres{R}$, we say that $\x$ is of \emph{type} $R$ and denote the latter by $\rel{\x}=R$.
We often refer to tuples $\x$ without mentioning their type, in which case the definition implies
there is a unique relation $R \in \schRel$ with $\x\in\tuplesres{R}$.

We associate to $\x$ an infinite set $\tversions{\x}$ that
conceptually represents the different versions that are created when $\x$ is changed over time.
We require that $\tversions{\x} \cap \tversions{\y} = \emptyset$ for all {tuples $\x \neq \y$}.
Each set $\tversions{\x}$ contains two special versions that we refer to as the \emph{unborn} and \emph{dead} version. We refer to all other versions
as \emph{visible} versions. Intuitively, the unborn version represents the state of $\x$ before it is inserted in the database, the dead version
represents the state
after the tuple is deleted, and the visible versions %
are the versions of $\x$ that can be read by transactions.
For a foreign key $f \in \schFK$, $\dom{f} \in \schRel$ and $\range{f} \in \schRel$ denote the associated domain and range of $f$, and %
$f$ itself is a mapping associating each tuple $\x \in \tuplesres{\dom{f}}$ to a tuple in $f(\x) \in \tuplesres{\range{f}}$.

\subsection{Operations over Tuples and Relations}

For a tuple $\x$, we distinguish four operations $\R[]{\x}$, $\W[]{\x}$,
$\IN[]{\x}$ and $\DE[]{\x}$, %
denoting that %
$\x$ is read, written, %
inserted or deleted, respectively, and say that the operation is on the tuple $\x$.
We also assume a special \emph{commit} operation denoted by $\CT[]$.
We will use the following terminology:
a \emph{read operation} is an $\R[]{\x}$, %
and a \emph{write operation} is a $\W[]{\x}$,
an $\IN[]{\x}$ or a $\DE[]{\x}$. Furthermore,
an \myR-operation is an $\R[]{\x}$, a \myW-operation is a $\W[]{\x}$,
an \myIN-operation is an $\IN[]{\x}$, and a \myDE-operation is a $\DE[]{\x}$.
To every operation $o$ on a tuple of type $R$, we associate a set of attributes {$\BGSet{o}\subseteq\Attr{R}$ to denote the attributes that $o$ reads from or writes to.}
Furthermore, when $o$ is an $\myIN$-operation or a $\myDE$-operation then $\WGSet{o} = \Attr{R}$.

For a relation $R \in \schRel$, a predicate read $\PR[]{R}$ is an operation that evaluates a predicate over each tuple of type $R$, and %
$\PGSet{\PR[]{R}}\subseteq\Attr{R}$ contains the set of attributes over which the predicate is evaluated.

\subsection{Transactions and Schedules}
\label{sec:chunks}

For %
$i,j\in \mathbb{N}$ with $i\leq j$, denote by $[i,j]$ the set $\{i,\ldots,j\}$.

A \emph{transaction} $\trans[]$ %
is a sequence of read and write operations on tuples, as well as predicate read operations on relations in $\schRel$, followed by a special commit operation denoted by $\CT[]$.
Formally, we model a transaction as a linear order $(\trans[],\leq_{\trans[]})$, where $\trans[]$ is the set of (read, write, predicate read and commit) operations occurring in the transaction and $\leq_{\trans[]}$ encodes the ordering of the operations. As usual, we use $<_{\trans[]}$ to denote the strict ordering.
Throughout the paper, we interchangeably consider transactions both as linear orders as well as sequences.

\sloppy

Let $a$ and $b$ be two operations in a transaction $\trans[]$ with $a \leq_{\trans[]} b$. An \emph{atomic chunk} $(a, b)$ represents a sequence of operations that cannot be interleaved by other concurrent transactions. Formally, an atomic chunk
is a pair $(a, b)$ that denotes the restriction of $\trans[]$ to all operations $o$ with $a\leq_{\trans[]} o \leq_{\trans[]} b$.
{In this paper, we only consider chunks encapsulating
specific sequences of operations:
\begin{itemize}
    \item \emph{key-based update}: $\R[]{\x}\W[]{\x}$ with $\rel{\x} = R$;
    \item \emph{predicate-based selection}: $\PR[]{R} \R[]{\x_1} \ldots
              \R[]{\x_n}$ for an arbitrary number of tuples $\x_i$ with $\rel{\x_i}=R$;
    \item \emph{predicate-based update}: $\PR[]{R}
              {\R[]{\x_1}\W[]{\x_1} \ldots \R[]{\x_n}\W[]{\x_n}}$ for an arbitrary
          number of tuples $\x_i$ with $\rel{\x_i}=R$; and
    \item \emph{predicate-based deletion}: $\PR[]{R} \DE[]{\x_1} \ldots
              \DE[]{\x_n}$ for an arbitrary number of tuples $\x_i$ with $\rel{\x_i}=R$.
\end{itemize}
We refer to Section~\ref{sec:assumptions} for a discussion on the assumptions we make on a DBMS (including chunks). We denote by $\chunks{\trans[]}$ the set of atomic chunks associated to $\trans[]$.
For instance, the transactions in Figure~\ref{fig:ex:sched} have the following chunks: $\chunks{\trans[1]}=\{(\R[1]{\x_1}, \W[1]{\x_1})\}$,
$\chunks{\trans[2]}=\{(\R[2]{\x_1}, \W[2]{\x_1})\}$, and $\chunks{\trans[3]}=\{(\R[3]{\x_2},$ $\W[3]{\x_2})$, $(\PR[3]{\text{Bids}} %
    ,\R[3]{\y_3})\}.$

\fussy

When considering a set $\transset$ of transactions, we assume that every transaction in the set has a unique id $i$ and write $\trans$ to make this id explicit. Similarly, to distinguish the operations from different transactions, we add this id as an index to the operation.
That is, we write $\W{\x}$, $\R{\x}$, %
$\IN{\x}$ and $\DE{\x}$ to denote respectively a write operation, read operation,
insert or delete on tuple $\x$ occurring in transaction $\trans$; similarly, $\PR{R}$ denotes a predicate read on relation $R$ in transaction $\trans[i]$ and $\CT[i]$ denotes the commit operation in transaction $\trans[i]$.
This convention is consistent with the literature (see, \eg\
\cite{DBLP:conf/sigmod/BerensonBGMOO95,DBLP:conf/pods/Fekete05}).
To avoid ambiguity of notation, we assume that a transaction performs at most one read operation and at most one write operation per {tuple}.
The latter is a common assumption (see, \eg~\cite{DBLP:conf/pods/Fekete05}). All our results carry over to the more general setting in which multiple writes and reads per tuple are allowed.

A \emph{(multiversion) schedule} $\schedule$ over a set $\transset$ of transactions is a tuple $(\schop, \schord, \schinit, \schwvf, \schrvf, \schpvf, \schvord)$ where
\emph{(i)} $\schop$ is the set containing all operations of transactions in $\transset$;
\emph{(ii)} $\schord$ encodes the ordering of these operations;
\emph{(iii)} $\schinit$ is the \emph{initial version set} associating each tuple $\x$ to a version $\schinit(\x) \in \tversions{\x}$ {which is either the unborn or any visible version of $\x$};
\emph{(iv)} $\schwvf$ is a \emph{write version function} mapping each write operation over a tuple $\x$ in $\schop$ to the version in $\tversions{\x}$ that this operation created;
\emph{(v)} $\schrvf$ is a \emph{read version function} mapping each read operation over a tuple $\x$ in $\schop$ to the version in $\tversions{\x}$ that this operation observed;
\emph{(vi)}
$\schpvf$ is a function mapping each predicate read operation $a\in\schop$ to a \emph{version set} containing the version of each tuple that is observed by $a$, or, more formally, for each tuple $\x \in \tuplesres{R}$ a version in $\tversions{\x}$ where $a$ is over a relation $R$;
\emph{(vi)} $\ll_\schedule$ is a \emph{version order} providing for each tuple $\x$ a total order over all the versions in $\tversions{\x}$ with the unborn and dead version of $\x$ being respectively the first and last version according to $\ll_\schedule$ for $\x$.  %

We furthermore require that
\begin{itemize}
    \item the order of operations in $s$ is consistent with the order of operations in every transaction $\trans[]\in\transset$. That is, $a <_{\trans[]} b$ implies $a \schords b$ for every $\trans[] \in \transset$ and every $a,b \in \trans[]$;
    \item {atomic chunks are not interleaved by operations of other transactions.} That is, for every $\trans[i] \in \transset$ and for each atomic chunk $(a_i, b_i) \in \chunks{\trans[i]}$, there is no operation $c$ with $a_i \schords c \schords b_i$ and $c \not\in \trans[i]$;
    \item {each write operation creates a version that is newer (according to $\ll_s$) than the initial version and that is different
          from versions created by other write operations. Furthermore, $\myDE$-operations always create the dead version for a tuple. Formally, for each write operation $a \in \schop$ over a tuple $\x$, we have $\schinit(\x) \ll_s \schwvf(a)$ and there is no other write operation $b \in \schop$ over $\x$ with $\schwvf(a) = \schwvf(b)$. If $a$ is a $\myDE$-operation, then $\schwvf(a)$ is the dead version};
    \item read and predicate read operations always observe visible versions of tuples that are already installed. That is,
          for each read and predicate read operation $a \in \schop$, the read version $v$ of a tuple $t$ (being either $\schrvf(a)$ or as defined by $\schpvf(a)$) is visible
          and either equals $\schinit(\x)$ or there is a write operation $b \in \schop$ over $\x$ with $b \schords a$ and $v = \schwvf(b)$.
    \item {an operation creates the first visible version of a tuple if and only if it is an $\myIN$-operation.}
          Formally, for each write operation $a \in \schop$ over a tuple $\x$, {$a$ is an $\myIN$-operation if and only if} there is no other write operation $b\in \schop$ over $\x$ with $\schwvf(b) \schvord \schwvf(a)$ and $\schinit(\x)$ is the unborn version.
\end{itemize}
Notice that it follows immediately from these requirements that there can be at most one $\myIN$-operation and at most one $\myDE$-operation in $\schop$ over each tuple.
A schedule $\schedule$ is a \emph{single version schedule} if
versions are installed in the order that they are written and
every (predicate) read operation always observes the most recent version of all relevant tuples.
Formally,
\emph{(i)} for each pair of write operations $a$ and $b$ on the same tuple, $\schwvf(a) \ll_\schedule \schwvf(b)$ iff $a <_\schedule b$; \emph{(ii)} for every read operation $a$ there is no write operation $c$ on the same tuple as $a$
with $c \schords a$ and $\schrvf(a) \schvord \schwvf(c)$; and,
\emph{(iii)} for every predicate read operation $a$ over a relation $R$ and tuple $\x$ of type $R$ there is no write operation $c$ on $\x$
with $c \schords a$ and $\x_i \schvord \schwvf(c)$, with $\x_i$ the version of $\x$ in $\schpvf(a)$.

A {serial} schedule over a set of transactions $\transset$ is a single version schedule %
in which operations from transactions are not interleaved with operations from other transactions. That is, for every $a,b,c \in {O_\schedule}$ with $a <_{\schedule}
    b<_{\schedule} c$ and $a,c \in \trans[]$ implies $b \in \trans[]$ for every
$\trans[] \in \transset$.

The absence of aborts in our definition %
is consistent with the common assumption~\cite{DBLP:conf/concur/0002G16,DBLP:conf/pods/Fekete05} that an underlying recovery mechanism will roll back transactions that interfere with aborted transactions.

\subsection{Conflict Serializability}
\label{sec:conser}

Let $a_j$ and $b_i$ be two operations from different transactions $\trans[j]$ and $\trans[i]$ in a set of transactions $\transset$.  We say that $a_j$ \emph{depends on} $b_i$ (or that there is a dependency from $b_i$ to $a_j$) in a schedule $\schedule$ over $\transset$, denoted $b_i \rightarrow_\schedule a_j$ if one of the following holds:

\begin{itemize}
    \item \emph{(ww-dependency)} {$b_i$ and $a_j$ are write operations on the same tuple with $\WGSet{b_i} \cap \WGSet{a_j} \neq \emptyset$ and $\schwvf(b_i) \schvord \schwvf(a_j)$}; 
    \item \emph{(wr-dependency)} {$b_i$ is a write operation and $a_j$ is a read operation on the same tuple with
              $\WGSet{b_i} \cap \RGSet{a_j} \neq \emptyset$ and either $\schwvf(b_i) = \schrvf(a_j)$ or $\schwvf(b_i) \schvord \schrvf(a_j)$}; 
    \item \emph{(rw-antidependency)} {$b_i$ is a read operation and $a_j$ is a write operation on the same tuple with $\RGSet{b_i} \cap \WGSet{a_j} \neq \emptyset$ and $\schrvf(b_i) \schvord \schwvf(a_j)$}; 
    \item \emph{(predicate wr-dependency)} {$b_i$ is a write operation on a
              tuple of type $R$, $a_j$ is a predicate read on relation $R$, $b_i$ is
              over a tuple $\x$ and $\schwvf(b_i) = \x_i$ or $\schwvf(b_i) \schvord
                  \x_i$ with $\x_i$ the version of $\x$ in $\schpvf(a_j)$}, and if
          $b_i$ is not an $\myIN$ or $\myDE$ operation, then
          $\WGSet{b_i} \cap \PGSet{a_j} \neq \emptyset$; or,
    \item \emph{(predicate rw-antidependency)}  $b_i$ is a predicate read on a relation $R$, $a_j$ is a write operation on a tuple of type $R$, $a_j$ is over a tuple $\x$ and $\x_i \schvord \schwvf(a_j)$ with $\x_i$ the version of $\x$ in $\schpvf(b_i)$, and if
          $a_j$ is not an $\myIN$ or $\myDE$ operation, then
          $\PGSet{b_i} \cap \WGSet{a_j} \neq \emptyset$.
\end{itemize}

Intuitively, a ww-dependency from $b_i$ to $a_j$ implies that $a_j$ writes a version of a tuple {that is installed} after the version written by $b_i$.
A (predicate) wr-dependency from $b_i$ to $a_j$ implies that $b_i$ either writes the version observed by $a_j$, or it writes a version that is {installed} before the version observed by $a_j$.
A (predicate) rw-antidependency from $b_i$ to $a_j$ implies that $b_i$ observes a version {installed} before the version written by $a_j$.

Notice that dependencies essentially lift the well-known notion of conflicting operations (\ie, two operations from different transactions over a same tuple with at least one being a write operation) to multi-version schedules. Indeed, ignoring predicate reads, if $a_j$ depends on $b_i$ then $a_j$ and $b_i$ are conflicting; for a single-version schedule $s$, an operation $a_j$ depends on $b_i$ if and only if $a_j$ and $b_i$ are conflicting with $b_i <_s a_j$.

Two schedules $\schedule$ and $\schedule'$ are \emph{conflict equivalent} if they are over the same set $\transset$ of transactions and for every pair of 
operations $a_j$ and $b_i$ from different transactions, $b_i \rightarrow_\schedule a_j$ iff $b_i \rightarrow_{\schedule'} a_j$.

These dependencies intuitively imply a specific order on pairs of transactions in conflict equivalent serial schedules.
That is, when an operation $a_j \in \trans[j]$ depends on an operation $b_i \in \trans[i]$ in a schedule $\schedule$, then in every serial schedule $\schedule'$ conflict equivalent to $\schedule$, transaction $\trans[i]$ should occur before transaction $\trans[j]$.

\begin{definition}
    A schedule $\schedule$ is \emph{conflict serializable} if it is conflict equivalent to a serial schedule.
\end{definition}
A {\emph{serialization graph}} $\cg{\schedule}$ for schedule $\schedule$ over a set of transactions $\transset$ is the graph whose nodes are the transactions in $\transset$ and where there is an edge from $T_i$ to $T_j$ if {$T_j$ has an operation $a_j$ that depends on an operator $b_i$ in $T_i$, thus with $b_i \rightarrow_\schedule a_j$.}
Since we are usually not only interested in the existence of
    {dependencies between operations}, but also in the operations themselves, we assume the
existence of a labeling function $\lambda$ mapping each edge to a set of pairs
of operations. Formally, $(b_i,a_j)\in \lambda(T_i,T_j)$ iff there is an
operation $a_j\in T_j$ that depends on an operation $b_i\in T_i$.
For ease of notation, we choose to
represent $\cg{\schedule}$ as a set of quadruples $(T_i,b_i,a_j,T_j)$ denoting
all possible pairs of these transactions $T_i$ and $T_j$ with all possible
choices of operations {with $\dependson{b_i}{a_j}$}. Henceforth, we refer to these quadruples simply as edges. Notice that edges cannot contain commit operations.

A \emph{cycle} $\cyclesym$ in $\cg{\schedule}$ is a non-empty sequence of edges $$(T_1,b_1,a_2,T_2),(T_2,b_2,a_3,T_3),\ldots, (T_n,b_n,a_1,T_1)$$ in $\cg{\schedule}$, in which every transaction is mentioned exactly twice. Note that cycles are by definition simple. Here, transaction $T_1$ starts and concludes the cycle. For a transaction $T_i$ in $\cyclesym$, we denote by $\cyclesym[T_i]$ the cycle obtained from $\cyclesym$ by letting $T_i$ start and conclude the cycle while otherwise respecting the order of transactions in $\cyclesym$. That is, $\cyclesym[T_i]$ is the sequence
\begin{multline*}    
    (T_i,b_i,a_{i+1},T_{i+1}),\cdots,(T_n,b_n,a_1,T_1),\\
    (T_1,b_1,a_2,T_2),
    \cdots,  (T_{i-1},b_{i-1},a_i,T_i).
\end{multline*}

\begin{theorem}[implied by \cite{DBLP:conf/icde/AdyaLO00}]\label{theo:not-conflict-serializable}
    A schedule $\schedule$ is conflict serializable iff $\cg{\schedule}$ is acyclic.
\end{theorem}

\subsection{Multiversion Read Committed}
\label{sec:def:mvrc}

Let $\schedule$ be a schedule for a set $\transset$ of transactions.
Then, $\schedule$ \emph{exhibits a dirty write}
iff there are two 
    {write} operations $a_j$ and $b_i$ in $\schedule$ on the same tuple $\x$,
$a_j \in \trans[j]$, $b_i \in \trans[i]$ and $\trans[j] \neq \trans[i]$
such that
\[b_i <_\schedule a_j <_\schedule \CT[i].\]
That is, transaction $T_j$ writes to 
a tuple that has
been modified earlier by $T_i$, but $T_i$ has not yet issued a commit.

For a schedule $\schedule$, the version order $\ll_\schedule$ {is consistent with} the commit order in $\schedule$ if for every pair of write operations $a_j \in \trans[j]$ and $b_i \in \trans[i]$, we have $\schwvf(b_i) \schvord \schwvf(a_j)$ iff $\CT[i] <_\schedule \CT[j]$.
We say that a schedule $\schedule$ is \emph{read-last-committed (RLC)} if the following conditions hold:
\begin{itemize}
    \item $\ll_\schedule$ is consistent with the commit order;
    \item for every read operation $a_j$ in $\schedule$ on some tuple $\x$:
          \begin{itemize}
              \item $\schrvf(a_j) = \schinit(\x)$ or $\CT[i] \schords a_j$ with $\schrvf(a_j) = \schwvf(b_i)$ for some write operation $b_i \in \trans[i]$, and
              \item there is no write 
                    operation $c_k \in \trans[k]$ on $\x$ with $\CT[k] <_\schedule a_j$ and $\schrvf(a_j) \schvord \schwvf(c_k)$; and
          \end{itemize}
    \item for every predicate read operation $a_j$ in $\schedule$ on relation $R$ and tuple $\x$ of type $R$, with $\x_j$ the version of $\x$ in $\schpvf(a_j)$:
          \begin{itemize}
              \item $\x_j = \schinit(\x)$ or $\CT[i] \schords a_j$ with $\x_j = \schwvf(b_i)$ for some write operation $b_i \in \trans[i]$; and
              \item there is no write operation $c_k \in \trans[k]$ on $\x$ with $\CT[k] \schords a_j$ and $\x_j \schvord \schwvf(c_k)$.
          \end{itemize}
\end{itemize}
That is, each (predicate) read operation $a_j$ observes for each relevant tuple the version that was committed most recently (according to the order of commits) before $a_j$.

\begin{definition} \label{def:mvisolationlevels}
    A schedule is \emph{allowed under isolation level} \readmvcom (\mvrc) if
    it is read-last-committed and does not exhibit dirty writes.
\end{definition}


\section{Serialization Graphs under MVRC}
\label{sec:ser:graph}

Towards a sufficient condition for robustness against \mvrc (
c.f. Section~\ref{sec:detecting:robustness}), we 
present a condition that holds
for all cycles in a {serialization} graph $\cg{\schedule}$ when $\schedule$ is allowed under \mvrc.

Let $a_j$ and $b_i$ be two operations occurring in a schedule $\schedule$ with $a_j \in \trans[j]$ and $b_i \in \trans[i]$ such that $\dependson[\schedule]{b_i}{a_j}$. We say that this dependency is a \emph{counterflow dependency} if $\CT[j] \schords \CT[i]$~\cite{DBLP:conf/aiccsa/AlomariF15}. That is, the direction of the dependency is opposite to the commit order. The following Lemma is a generalization of a result in~\cite{DBLP:conf/aiccsa/AlomariF15} to include dependencies based on predicate reads:
\begin{lemma}\label{lem:counterflow}
    {In a schedule allowed under \mvrc, only (predicate) rw-antidependencies can be counterflow.}
\end{lemma}

The following theorem presents a property of cycles that must occur in $\cg{\schedule}$ when a schedule $\schedule$ allowed under $\mvrc$ is not serializable. The robustness detection method of Section~\ref{sec:detecting:robustness} then tests for the absence of such cycles to establish robustness for transaction programs.
The theorem is a refinement of
~\cite{DBLP:conf/aiccsa/AlomariF15}, where it was proven that {a cycle} must contain at least one counterflow dependency. Our refined property allows to detect larger sets of transaction programs to be robust as we show in Section~\ref{sec:experiments}.

\begin{theorem}\label{theo:cg:cycles}
    Let $\cyclesym$ be a cycle in $\cg{\schedule}$ for some schedule $\schedule$ allowed under \mvrc. Then $\cyclesym$ contains at least one non-counterflow dependency and at least one of the following two conditions hold:
    \begin{enumerate}
        \item \label{c1} there are two adjacent counterflow dependencies in $\cyclesym$; or
        \item \label{c2} there are two adjacent dependencies $\dependson[\schedule]{b_{i-1}}{a_i}$ and $\dependson[\schedule]{b_i}{a_{i+1}}$ in $\cyclesym$, where $\dependson[\schedule]{b_i}{a_{i+1}}$ is a counterflow dependency and {either} $b_i <_{\trans[i]} a_i$ in the corresponding transaction $\trans[i]$ {or $b_{i-1}$ is an $\myR$- or $\myPR$-operation}.
    \end{enumerate}
\end{theorem}

{To see why Theorem~\ref{theo:cg:cycles} holds, note that not every dependency in $\cyclesym$ can be counterflow, as otherwise the implied order on the commits in $\cyclesym$ leads to a transaction committing before itself. The remaining conditions are based on an analogous analysis.}

We refer to a pair of dependencies satisfying
condition (\ref{c1}) (resp., condition (\ref{c2})) as an \emph{\adjacentcfp}  (\emph{\orderedcfp}).

{
    \begin{definition}
        A cycle $\cyclesym$ in $\cg{\schedule}$ for some schedule $\schedule$ is a \emph{\typetwo} cycle if it has at least one non-counterflow dependency as well as either an \adjacentcf pair or an \orderedcfp,
        and $\cyclesym$ is a \emph{\typeone} cycle if it has at least one counterflow dependency.
    \end{definition}
}
Every \typetwo cycle is a \typeone cycle but not vice-versa, and the absence of a {\typeone cycle} implies the absence of a {\typetwo cycle}.
{Theorem~\ref{theo:cg:cycles} now implies that if a schedule $\schedule$ is allowed under \mvrc, then every cycle in $\cg{\schedule}$ is a \typetwo cycle (and therefore a \typeone cycle as well). Conflict serializability of $\schedule$ therefore coincides with the absence of \typetwo cycles in $\cg{\schedule}$.}


\section{Robustness for Transaction Programs}
\label{sec:robustness}



\subsection{Basic Transaction Programs}

A basic transaction program (\btp) adheres to the following syntax:\footnote{Appendix~A\ifthenelse{\boolean{fullversion}}{}{ in \cite{fullversionVDLB2022}} provides an overview of the SQL transactions that inspired the definition of \btp.
}
\vspace{-.5em}
$$P \enskip\leftarrow \enskip \text{loop}(P) \enskip\mid\enskip (P \mid P) \enskip\mid\enskip {(P \mid \epsilon)} \enskip\mid\enskip P;P \enskip\mid\enskip q$$
\vspace{-.5em}
where $q$ is a statement with the following associated functions:
\begin{itemize}
    \item $\rel{q}$: the relation name the statement is over;

    \item $\predset{q}$: the subset of attributes
          from $\Attr{\rel{q}}$ used in selection predicates in $q$, or symbol
          $\bot$ (for undefined);
    \item $\obsset{q}$: the subset of attributes from $\Attr{\rel{q}}$ that are
          observed by $q$, or symbol $\bot$;
    \item $\modset{q}$: the subset of attributes from $\Attr{\rel{q}}$
          that are modified by $q$, or symbol $\bot$; and

    \item $\type{q} \in \{\qins,
              \qkdel, \qpdel, \qksel, \qpsel,
              \qkmod$, $\qpmod\}$ the type of statement.
\end{itemize}

Statements $q$ can be of one of the following types: insertion, deletion, selection or update. Apart from insertion, each statement depends on a retrieval of tuples at the start of the statement. That retrieval can be a key-based look-up (always returning exactly one tuple) or can be a predicate-based look-up (returning an arbitrary number of tuples). We refer to those types of statements, respectively, as key-based and predicate-based updates, deletions, and selections.
Figure~\ref{fig:btp:types:constraints} details how
$\type{q}$ constrains $\predset{q}$, $\obsset{q}$, and $\modset{q}$.
For instance, when $\type{q}=\qins$, then $\modset{q}$ are all attributes and
$\obsset{q}$ and $\predset{q}$ are undefined. The notation $S: \emptyset \subseteq S$ (resp., $S: \emptyset \subsetneq S$) indicates that the set $S$ under consideration can be empty (resp., can not be empty).

\begin{figure}

    \begin{center}
        \begin{tabular}{lccc}
            $\type{q}$ & $\modset{q}$                 & $\obsset{q}$               & $\predset{q}$              \\
            \hline
            $\qins$    & $\Attr{\rel{q}}$             & $\bot$                     & $\bot$                     \\
            $\qkdel$   & $\Attr{\rel{q}}$             & $\bot$                     & $\bot$                     \\
            $\qpdel$   & $\Attr{\rel{q}}$             & $\bot$                     & $S: \emptyset \subseteq S$ \\
            $\qksel$   & $\bot$                       & $S: \emptyset \subseteq S$ & $\bot$                     \\
            $\qpsel$   & $\bot$                       & $S: \emptyset \subseteq S$ & $S: \emptyset \subseteq S$ \\
            $\qkmod$   & $S: \emptyset \subsetneqq S$ & $S: \emptyset \subseteq S$ & $\bot$                     \\
            $\qpmod$   & $S: \emptyset \subsetneqq S$ & $S: \emptyset \subseteq S$ & $S: \emptyset \subseteq S$
        \end{tabular}
    \end{center}
    \caption{\label{fig:btp:types:constraints} Constraints relative to $\type{q}$.}

\end{figure}

A BTP $\prog$ can furthermore be annotated by a set of foreign key constraints. Each such constraint is an expression of the form $q_j = f(q_i)$, where $q_i$ and $q_j$ are statements occurring in $\prog$ and $f$ is a foreign key in $\schFK$. In addition, we require that $\rel{q_i} = \dom{f}$, $\rel{q_j} = \range{f}$, and $q_j$ must be a key-based statement.

    {In our running example, the foreign key constraints $q_3=f_1(q_4)$, {$q_3=f_1(q_5)$ and $q_3=f_2(q_6)$ are} added to the \btp given in Figure~\ref{fig:sql_auction_progs} where $f_1$ is the foreign key Bids(buyerId)$\to$ Buyer(id) {and $f_2$ is the foreign key Log(buyerId)$\to$ Buyer(id)}. Notice, that there is no foreign key constraint $q_1=f_1(q_2)$ as $q_2$ does not refer to buyerId.}

\subsection{{Instantiations} and schedules}
\label{sec:instances:schedules}

Robustness for a set $\progset$ of $\btp$s is defined in the next subsection w.r.t.\ the set of all possible schedules over $\progset$ that result from transactions that are {instantiations} of $\btp$s in $\progset$. We first define {instantiations of statements and \btp{}s.} 

Intuitively, an instantiation of a \btp{} $P$ is a transaction consisting of a sequence of chunks, which are instantiations of the statements that it consists of.
For a formal treatment, we observe that all operations encapsulated in a chunk $c$ are over the same relation, say $\rel{c}$. Similarly, since all operations in a chunk are of the same type (i.e., $\myR$, $\myW$, $\myDE$,
$\myPR$), they agree on the set $\BGSet{\cdot}$, and we can thus unambiguously define $\ReadSet{c}$ to denote $\RGSet{\R[]{\x_i}}$ (in case of selection and update) or $\bot$ (otherwise); $\WriteSet{c}$ to denote $\WGSet{\W[]{\x_i}}$ (in case of an insert, deletion or update) or $\bot$ (otherwise); and $\PReadSet{c}$ denoting $\PGSet{\PR[]{R}}$ (in case there is a predicate read) or $\bot$ (otherwise).

An \emph{instantiation} of a \btp $P$ is a transaction that can be obtained by applying the following rules:
\begin{itemize}
    \item $\text{loop}(P)$: unfold with an arbitrary {finite} number of instantiations of $P$.

    \item $P_1 \mid P_2$: replace with either an
          instantiation of $P_1$ or $P_2$;

    \item {$P_1 \mid \epsilon$: replace with either an
          instantiation of $P_1$ or the empty sequence;} %
          {
    \item $q$, with $\type{q} \in \{\qins\}$: replace by operation $a = \IN[]{\x}$ for some tuple $\x$ with $\rel{\x} = R$ and $\WGSet{a} = \modset{q}$;
    \item $q$, with $\type{q} \in \{\qksel\}$: replace by operation $a = \R[]{\x}$ for some tuple $\x$ with $\rel{\x} = R$ and $\RGSet{a} = \obsset{q}$;
    \item $q$, with $\type{q} \in \{\qkdel\}$: replace by operation $a = \DE[]{\x}$ for some tuple $\x$ with $\rel{\x} = R$ and $\WGSet{a} = \modset{q}$};
    \item $q${, otherwise}: replace by an arbitrary chunk $c$ (as defined in Section~\ref{sec:chunks}, and with arbitrary tuple instantiations) of type $\type{q}$ with
          $\rel{c} = \rel{q}$, $\predset{c} = \predset{q}$, $\obsset{c} =
              \obsset{q}$, and $\modset{c} = \modset{q}$.
\end{itemize}

If $\prog$ is annotated with a foreign key constraint $q_j = f(q_i)$, then we furthermore require for every $\myR$-, $\myW$-, 
$\myIN$- and $\myDE$-operation over a tuple $\x_i$ instantiated from $q_i$ and for every $\myR$-, $\myW$-, 
$\myIN$- and $\myDE$-operation over a tuple $\x_j$ instantiated from $q_j$ that $\x_j = f(\x_i)$ (i.e., every {instantiation} of $\prog$ must respect the foreign key constraints of $\prog$).
In our running example, $T_1$ and $T_2$ are instantiations of PlaceBid where $f_1(\y_1)=\x_1$, and $T_3$ is an instantiation of FindBids.
Indeed, e.g., for $T_1$, $q_3$ is replaced by $\R[1]{\x_1} \W[1]{\x_1}$, $q_4$ by $\R[1]{\y_1}$, $q_5$ by $\varepsilon$, and $q_6$ by $\IN[1]{\myl_1}$.
A set of transactions $\mathcal{T}$
is an {instantiation} of $\progset$ if for every $T\in \mathcal{T}$ there is a $P\in\progset$ such that $T$ is an {instantiation} of $P$.
Now,
$\schedules{\progset}{\mvrc}$ consists of all schedules $s$ allowed under
$\mvrc$ for all {finite} sets of transactions that are {instantiations} of $\progset$.

\subsection{Robustness}

We are now ready to define robustness on the level of \btp{}s:
\begin{definition}[Robustness]\label{def:robustness}
    A set of \btp{}s 
    $\progset$ is \emph{robust against \mvrc} if every schedule in $\schedules{\progset}{\mvrc}$ is conflict serializable.
\end{definition}


We need to address how robustness for \btp{}s relates to robustness for the SQL programs they model. To this end, we first establish in the following proposition, that robustness over a set of schedules implies robustness over each subset:

\begin{proposition}\label{prop:robust}
    Let $\schedules{\progset}{\mvrc}\subseteq \schedules{\progset'}{\mvrc}$ for $\progset$, $\progset'$ sets of \btp{}s. 
    If $\progset'$ is robust against \mvrc, then $\progset$ is robust against \mvrc as well.
\end{proposition}

The running example in Section~\ref{sec:runningex} already provides an idea on how to translate a set of SQL-programs $\progset_{\text{SQL}}$ into the  corresponding set $\progset$ of \btp{}s (Appendix~A \ifthenelse{\boolean{fullversion}}{}{of \cite{fullversionVDLB2022}} provides a general construction). From this construction, it follows that, as \btp{}s abstract away from the concrete conditions used for instance in WHERE-clauses, that $\schedules{\progset_{\text{SQL}}}{\mvrc}\subseteq
    \schedules{\progset}{\mvrc}$. Therefore, when $\progset$ is robust against \mvrc, so is $\progset_{\text{SQL}}$ and the results in this paper can be directly applied to the considered SQL fragment. 

\vspace{-0.9em}
\subsection{Assumptions on the DBMS}
\label{sec:assumptions}
Our definitions as well as our formalism of program {instantiations} impose requirements on how
the database management system operates. In this section, we discuss these requirements in more detail and argue why they are reasonable.

For a schedule $\schedule$ to be allowed under \mvrc, we deliberately require that every (predicate) read operation in $\schedule$ observes the most recently committed version of all relevant tuples, rather than an arbitrary committed version. Although this assumption rules out distributed settings where such a requirement cannot be guaranteed, this more strict definition of \mvrc is often necessary to detect larger fragments that are robust against \mvrc (without it, we could deliberately choose to observe older versions to facilitate constructing a non-serializable counterexample). For non-distributed systems, 
this is a reasonable assumption as returning an outdated version when the most recently committed version is available anyway would make little to no sense. 

When instantiating transactions from programs, each predicate-based statement is replaced by a number of operations in one atomic chunk, thereby requiring this set of operations to not be interleaved by operations from other transactions. Without this assumption, a predicate-based selection statement over a relation $R$, for example, could see an inconsistent view of $R$. Indeed, the read operations instantiated from this statement could be interleaved by a transaction $\trans[j]$ updating tuples of $R$, thereby resulting in a statement where the updates of $\trans[j]$ are only partially observed. We emphasize that our assumption does not rule out concurrent execution of statements from different programs, as long as the concurrent execution leads to a schedule 
equivalent to a schedule where the atomic chunks are respected. In Postgres\ifthenelse{\boolean{fullversion}}{}{\footnote{{Although the Postgres implementation deviates slightly from this presumption, our findings are still accurate.
            We refer to~\cite{fullversionVDLB2022} for more details.}}}
and Oracle, for example, each SQL statement is evaluated over a snapshot taken just before the statement started and can therefore not be influenced by concurrent updates from other transactions that committed while the statement is being evaluated.
\ifthenelse{\boolean{fullversion}}{
    {For the sake of completeness, it should be noted that the actual implementation of Postgres does not follow this assumption to the letter. In particular, Postgres evaluates the predicate twice: first to select tuples, and, if the tuple is changed by another transaction in the meantime, a second time right before changing the tuple to evaluate whether the tuple still satisfies the predicate. We emphasize that this does not break our results presented in Section~\ref{sec:detecting:robustness}, but merely requires a small addition to our instantiation from predicate-based update statements to atomic chunks of operations (i.e., predicate updates are instantiated with two chunks, where the first chunk is only a predicate read and the second chunk is the conventional predicate read followed by write operations over tuples). This change does not alter the types of dependencies that can arise between two statements (cf. Algorithm~\ref{alg:graph} and Table~\ref{alg:conditions}), and consequently, the eventual summary graph (cf. Section~\ref{sec:sum:graph}) would remain exactly the same.}
}{}

For key-based statements, we assume each tuple is uniquely identified by a (primary) key that cannot be altered by update statements, and each key-based statement accesses exactly one tuple (i.e., if no tuple with the specified key exists, the transaction must abort). All benchmarks considered in Section~\ref{sec:experiments} satisfy these assumptions. Our BTP formalism remains applicable if these assumptions are not guaranteed, but in this case each such statement $q$ should be modeled as a predicate-based statement, where $\PReadSet{q}$ contains the key attributes.
Note that this over-approximation allows instantiations of $q$ to access more than one tuple, which cannot occur in practice, but one could easily extend BTPs with an additional type of statement accessing at most one tuple. Our robustness results presented in Section~\ref{sec:detecting:robustness} remain applicable under such an extension, merely requiring additional checks in Algorithm~\ref{alg:graph}.
{Our formalism can also be easily extended to multi-relation statements (e.g.\ joins).
}
\vspace{-.2em}

\section{Detecting Robustness} 
\label{sec:detecting:robustness}



\vspace{-.2em}
\subsection{Linear Transaction Programs}
Towards an algorithm to detect robustness against \mvrc for arbitrary sets of \btp{}s, we first introduce linear transaction programs (\ltp{}s): a restriction of \btp{}s where loops and branching are not allowed. More formally, an \ltp adheres to the following syntax:
\[
    P \quad\leftarrow \quad P;P \quad\mid\quad q
\]
where $q$ represents a statement as before.

Obviously, for every set of \btp{}s $\progset$, we can construct a (possibly infinite) set of \ltp{}s $\progset'$ 
such that $\schedules{\progset}{\mvrc} = \schedules{\progset'}{\mvrc}$ by considering all possible unfoldings of loops and conditional statements. However, w.r.t.\ robustness testing, we show in Proposition~\ref{pro:equi} that it suffices to restrict attention to loop unfoldings of size at most two as defined next.

For a BTP $\prog$, let $\unfold{\prog}$ denote the set of \ltp{}s obtained by repeated application of the following rules:
\begin{itemize}
    \item $\text{loop}(P_1)$: replace with zero, one or two repetitions of $P_1$;

    \item $P_1 \mid P_2$: replace with either $P_1$ or $P_2$;
    \item $P_1 \mid \epsilon$: replace with either $P_1$ or the empty sequence.
\end{itemize}
By slight abuse of notation, we use $\unfold{\progset}$ for a set of \btp{}s $\progset$ to denote the set of \ltp{}s obtained by applying $\unfold{\prog}$ to each $\prog \in \progset$. More formally:
\[
    \unfold{\progset} = \bigcup_{\prog \in \progset} \unfold{\prog}.
\]
Since each $\text{loop}(\prog[1])$ is replaced by at most two repetitions of $\prog[1]$, it immediately follows that $\unfold{\progset}$ is a finite set. In practice, unfolding does not increase the size too much, e.g., for TPC-C the number of
transaction programs increases from 5 to 13.
By construction, it follows that $\schedules{\unfold{\progset}}{\mvrc} \subseteq \schedules{\progset}{\mvrc}$.

\begin{proposition}\label{pro:equi}
    Far a set $\progset$ of \btp{}s, the following are equivalent:
    \begin{enumerate}
        \item $\progset$ is robust against \mvrc; \item $\unfold{\progset}$ is robust against \mvrc.
    \end{enumerate}
\end{proposition}

{To see why two iterations of each loop suffice, note that we are looking for a cycle. Since in each transaction only two operations are important for this cycle (one for the incoming edge, one for the outgoing edge), all other iterations not involving one of these two operations can be removed.}


We introduce a \emph{summary graph} $\sg{\progset}$ summarizing all possible serialization graphs for schedules in $\schedules{\progset}{\mvrc}$. This summary graph is closely related to the dependency graph used by Alomari and Fekete~\cite{DBLP:conf/aiccsa/AlomariF15} but differs in two aspects. We add additional
information to edges necessary to detect \typetwo cycles, and, whereas \cite{DBLP:conf/aiccsa/AlomariF15} relies on a domain specialist that can predict possible conflicts to construct the graph, we provide a formal construction based on the formalism of \ltp{}s 
(Algorithm~\ref{alg:graph}).

Formally, $\sg{\progset}$ is a graph where each program in $\progset$ is represented by a node, and potential dependencies between two {instantiations} of programs in $\progset$ are represented by edges. Since we are not only interested in the existence of these dependencies, but also in the type of dependency (counterflow or not) and the two statements that give rise to this dependency, we assume an edge labeling function $\lambda$. The function $\lambda$ maps each edge in $\sg{\progset}$ from a program $P_i$ to a program $P_j$ to a set of tuples $(c, q_i, q_j)$ where $q_i \in P_i$, $q_j \in P_j$, and $c \in \{\counterflow, \noncounterflow\}$.
We will often represent these edges as a quintuple $(P_i, q_i, c, q_j, P_j)$.

The summary graph $\sg{\progset}$ should be constructed in such a way that the following condition holds:

\begin{cond}\label{cond:sg:prop}
    Let $\dependson[\schedule]{b_i}{a_j}$ be a dependency occurring between transaction $\trans[i]$ and $\trans[j]$ in a schedule $\schedule \in \schedules{\progset}{\mvrc}$. Let $P_i$ and $P_j$ be the programs in $\progset$ from which $\trans[i]$ and $\trans[j]$ were instantiated, and let $q_i$ and $q_j$ be the two statements in respectively $P_i$ and $P_j$ from which operations $b_i$ and $a_j$ were instantiated. Then, $\sg{\progset}$ must have an edge $(P_i, q_i, c, q_j, P_j)$, where $c$ is $\textit{counterflow}$ iff $\dependson[\schedule]{b_i}{a_j}$ is a counterflow dependency.
\end{cond}

\begin{table*}[tp]
    \begin{subtable}[t]{0.48\textwidth}
        {
            \setlength{\tabcolsep}{3pt}
            \footnotesize
            \begin{center}\begin{tabular}{l|c|c|c|c|c|c|c}
                    $q_i$ \textbackslash{} $q_j$ & $\qins$ & $\qksel$ & $\qpsel$ & $\qkmod$ & $\qpmod$ & $\qkdel$ & $\qpdel$ \\
                    \hline
                    $\qins$                      & false   & $\bot$   & true     & $\bot$   & true     & $\bot$   & true     \\
                    \hline
                    $\qksel$                     & false   & false    & false    & $\bot$   & $\bot$   & $\bot$   & $\bot$   \\
                    \hline
                    $\qpsel$                     & true    & false    & false    & $\bot$   & $\bot$   & true     & true     \\
                    \hline
                    $\qkmod$                     & false   & $\bot$   & $\bot$   & $\bot$   & $\bot$   & $\bot$   & $\bot$   \\
                    \hline
                    $\qpmod$                     & true    & $\bot$   & $\bot$   & $\bot$   & $\bot$   & true     & true     \\
                    \hline
                    $\qkdel$                     & false   & false    & true     & false    & true     & false    & true     \\
                    \hline
                    $\qpdel$                     & true    & false    & true     & $\bot$   & true     & true     & true
                \end{tabular}\end{center}}
        \caption{\label{table:dep}$\textsc{ncDepTable}$}
    \end{subtable}
    \hfill
    \begin{subtable}[t]{0.48\textwidth}
        {
            \footnotesize\centering
            \setlength{\tabcolsep}{3pt}
            \begin{center}\begin{tabular}{l|c|c|c|c|c|c|c}
                    $q_i$ \textbackslash{} $q_j$ & $\qins$ & $\qksel$ & $\qpsel$ & $\qkmod$ & $\qpmod$ & $\qkdel$ & $\qpdel$ \\
                    \hline
                    $\qins$                      & false   & false    & false    & false    & false    & false    & false    \\
                    \hline
                    $\qksel$                     & false   & false    & false    & $\bot$   & $\bot$   & $\bot$   & $\bot$   \\
                    \hline
                    $\qpsel$                     & true    & false    & false    & $\bot$   & $\bot$   & true     & true     \\
                    \hline
                    $\qkmod$                     & false   & false    & false    & false    & false    & false    & false    \\
                    \hline
                    $\qpmod$                     & true    & false    & false    & $\bot$   & $\bot$   & true     & true     \\
                    \hline
                    $\qkdel$                     & false   & false    & false    & false    & false    & false    & false    \\
                    \hline
                    $\qpdel$                     & true    & false    & false    & $\bot$   & $\bot$   & true     & true
                \end{tabular}\end{center}}
        \caption{\label{table:cdep}$\textsc{cDepTable}$}
    \end{subtable}

    \caption{\label{alg:conditions} Condition tables used in
        Algorithm~\ref{alg:graph}.}
\end{table*}

\subsection{Constructing the Summary Graph}
\label{sec:sum:graph}

\begin{algorithm}[t]
    \SetAlgoLined
    \SetKwProg{Fn}{Function}{ }{end}
    {\Fn{\textsc{ncDepConds}($q_i$, $q_j$) : Boolean}{
            \Return{
                $\WriteSet{q_i} \cap \WriteSet{q_j} \ne \emptyset$ 
                or
                $\WriteSet{q_i}\cap\ReadSet{q_j}\ne \emptyset$ 
                or
                $\WriteSet{q_i}\cap\PReadSet{q_j}\ne \emptyset$ 
                or
                $\ReadSet{q_i}\cap\WriteSet{q_j}\ne \emptyset$ 
                or
                $\PReadSet{q_i}\cap\WriteSet{q_j}\ne \emptyset$
            }\;
        }}
    \BlankLine

    \SetKwProg{Fn}{Function}{ }{end}
    \Fn{\textsc{cDepConds}($q_i$, $q_j$) : Boolean}{
        \If{$\PReadSet{q_i}\cap\WriteSet{q_j}\ne \emptyset$}{
            \Return{\upshape{\textbf{true}}}\;
        }
        \If{$\ReadSet{q_i}\cap\WriteSet{q_j}\ne \emptyset$}{
            \For{foreign key constraints $q_k = f(q_i)$ for $\prog[i]$ and $q_\ell = f(q_j)$ for $\prog[j]$}{
                \If{$\type{q_k}, \type{q_\ell} \in \{\qkupd, \qkdel, \qins\}$ and $q_k <_{\prog[i]} q_i$ and $q_\ell <_{\prog[j]} q_j$}{
                    \Return{\upshape{\textbf{false}}}\;
                }
            }
            \Return{\upshape{\textbf{true}}}\;
        }
        \Return{\upshape{\textbf{false}}}\;
    }
    \BlankLine
    \Fn{\algonameconstructsg($\progset$) : $\sg{\progset}$}{
        $S\mathrel{:=} \emptyset$\;
        \For{$P_i \in \progset$, $P_j \in \progset$, $q_i \in P_i$, and $q_j \in P_j$
            with $\rel{q_i} = \rel{q_j}$}{
            \If{$\textsc{ncDepTable}[q_i, q_j] =
                    \text{\bf\upshape{true}}$ or ($\textsc{ncDepTable}[q_i, q_j] =
                    \bot$ and $\textsc{ncDepConds}(q_i, q_j)$)}{
                add $(P_i, q_i, \noncounterflow, q_j, P_j)$ to $S$\;

            }
            \If{$\textsc{cDepTable}[q_i, q_j] =
                    \text{\bf\upshape{true}}$ or ($\textsc{cDepTable}[q_i, q_j] =
                    \bot$ and $\textsc{cDepConds}(q_i, q_j)$)}{
                add $(P_i, q_i, \counterflow, q_j, P_j)$ to $S$\;
            }
        }
        \Return{S}\;
    }
    \caption{\label{alg:graph} Construction of $\sg{\progset}$ for a set $\progset$ of \ltp{}s.}
\end{algorithm}

The algorithm to construct the summary graph $\sg{\progset}$ for a given set of \ltp{}s $\progset$ is given in Algorithm~\ref{alg:graph}.
We discuss how the edges in the graph $\sg{\progset}$ are constructed. 
To this end, let $q_i$ and $q_j$ be two (not necessarily different) statements in respectively programs $P_i$ and $P_j$ with $\rel{q_i} = \rel{q_j}$. The basic idea underlying the construction of $\sg{\progset}$ is to add an edge $(P_i, q_i,
    c, q_j, P_j)$ with $c \in \{\textit{non-counterflow}, \textit{counterflow}\}$
if $P_i$ and $P_j$ could have instantiations that admit a $c$
dependency for operations in the transaction fragments instantiated by $q_i$ and $q_j$,
respectively.

For  $c=\textit{non-counterflow}$ the conditions
are relatively straightforward and mostly analogous to the definition of
dependency, since every type of dependency listed in Section~\ref{sec:conser} can be
(and sometimes must be) non-counterflow.
More precisely, Table~(\ref{table:dep}) details when the types of $q_i$
and $q_j$ imply that a {\it non-counterflow} dependency can be admitted
(entry is {\it true}), may not not be admitted (entry is {\it false}), or
when additional checks need to be performed regarding
the intersections of involved read, write and predicate read
attributes (entry is $\bot$). Algorithm~\ref{alg:graph}, function
    {\sc ncDepConds}$(q_i, q_j)$ gives the precise condition of these additional checks.

For $c={\it counterflow}$ the approach is similar. Table~(\ref{table:cdep})
shows if a {\it counterflow} dependency can be admitted based on the types of $q_i$ and $q_j$.
In case of $\bot$, it is tested if the intersection between the (predicate) read attributes of $q_i$ and
write attributes of $q_j$ is non-empty, which is analogous to the condition of
a (predicate) rw-antidependency (c.f., Section~\ref{sec:conser}) which are the
only dependencies that can be counterflow. 
In this case also a
check on the foreign keys of the programs is performed, see \textsc{cDepConds} in Algorithm~\ref{alg:graph}.

We remark that, since the edges added to  $\sg{\progset}$ are based on
conditions that are independent of a particular schedule, two statements can at the same time allow a counterflow as well as non-counterflow dependency. The following proposition shows that the construction is sound:

\begin{proposition}\label{prop:algocorrectness}
    For a set of \ltp{}s $\progset$, the summary graph $\sg{\progset}$
    constructed by Algorithm~\ref{alg:graph} satisfies Condition~\ref{cond:sg:prop}.
\end{proposition}

\subsection{Detecting Robustness for Linear Transaction Programs}

We start by lifting Theorem~\ref{theo:cg:cycles} to \ltp{}s:
\begin{theorem}\label{theo:sufcond:robustness}
    A set of \ltp{}s $\progset$ is robust against \mvrc{} if there is no cycle
    $\cyclesym$ in $\sg{\progset}$ containing at least one non-counterflow
    edge for which at least one of the following two conditions holds:
    \begin{itemize}
        \item there are two adjacent counterflow edges in $\cyclesym$; or
        \item there are two adjacent edges $(P_{i-1}, q_{i-1}, \textit{non-counterflow},\allowbreak q_i, P_i)$ and $(P_i, q'_i, \textit{counterflow}, q_{i+1}, P_{i+1})$ in $\cyclesym$, where either $q'_i <_{P_i} q_i$ in the corresponding program $P_i$, or $\type{q_{i-1}} \in \{\qksel,\allowbreak \qpsel, \qpmod, \qpdel\}$.
    \end{itemize}
\end{theorem}
{The proof relies on Proposition~\ref{prop:algocorrectness} to show how these
    properties about dependencies between operations as in Theorem~\ref{theo:cg:cycles} can be lifted to
    properties over edges in $\sg{\progset}$. In particular, Condition~\ref{cond:sg:prop} implies that for every schedule $\schedule$ allowed under \mvrc, every cycle in $\cg{\schedule}$ is witnessed by a cycle in $\sg{\progset}$.}
It should be noted that the cycle $\cyclesym$ in the theorem above is allowed to visit the same nodes/edges multiple times.
Note that such a cycle $\cyclesym$ corresponds to a \typetwo cycle described in Theorem~\ref{theo:cg:cycles} lifted to summary graphs. For convenience, we will therefore refer to these cycles in $\sg{\progset}$ as \typetwo cycles as well.
Figure~\ref{fig:ex:sg} does not contain a \typetwo cycle whereas it clearly contains a \typeone cycle.

Based on Theorem~\ref{theo:sufcond:robustness}, Algorithm~\ref{alg:robust:check} then tests for the absence of \typetwo cycles as a proxy for robustness against \mvrc. Notice that Algorithm~\ref{alg:robust:check} is sound but incomplete: it can return false negatives but never a false positive, as formally shown in Propostion~\ref{prop:soundness}. We demonstrate in Section~\ref{sec:experiments} that it can detect strictly larger sets of \btp{}s to be robust than the state-of-the-art.
Even though the complexity is {$\mathcal{O}(n^6)$ with $n$ the total number of statements in $\unfold{\progset}$, we show that a proof-of-concept implementation runs in a matter of seconds.}

\begin{proposition}\label{prop:soundness}
    For a set $\progset$ of \btp{}s,
    if the algorithm returns true, then $\progset$ is robust against \mvrc.
\end{proposition}


\begin{algorithm}[t]
    \SetAlgoLined%
    \SetKwInOut{Input}{input}
    \SetKwInOut{Output}{output}
    \Input{ a set $\progset$ of \btp{}s}
    \Output{ true if $\sg{\progset}$ does not contain a \typetwo cycle,
        false otherwise
    }

    $G \leftarrow \text{\upshape{\algonameconstructsg}}(\unfold{\progset})$\;
    \For{$(P_1, q_1, \text{non-counterflow}, q_2, P_2) \in G$}{
        \For{$(P_3, q_3, c, q_4, P_4) \in G$}{
            \If{$P_3$ is reachable from $P_2$ in $G$}{
                \For{$(P_4, q_4', {\counterflow}, q_5, P_5) \in G$}{
                    \If{$P_1$ is reachable from $P_5$ in $G$ and
                        ($c = \text{counterflow}$ or $q_4' <_{P_4} q_4$ or
                        $\type{q_3}\in \{\qksel,\allowbreak \qpsel, \qpmod,
                            \qpdel\}$)}{
                        \Return{\upshape{\textbf{false}}}\;
                    }
                }
            }
        }
    }
    \Return{\upshape{\textbf{true}}}\;
    \caption{\label{alg:robust:check}
        Testing robustness.
    }
\end{algorithm}


\section{Experimental Validation}
\label{sec:experiments}



\subsection{Benchmarks}\label{sec:exp:benchmarks}
We implemented Algorithm~\ref{alg:robust:check} in Python
and tested it on three benchmarks  
whose characteristics are given in Table~\ref{table:benchmark:char}.  
Appendix~E\ifthenelse{\boolean{fullversion}}{}{ of \cite{fullversionVDLB2022}} contains a detailed description of their schema, the SQL transaction programs as well as their translation into \btp{}s and foreign key constraints.
{Since our experimental validation is based on static program analysis, benchmark configuration parameters influencing database size (e.g.\ number of warehouses for TPC-C) and how often different transactions occur are irrelevant to our experiments. If robustness is detected, serializability is guaranteed for all such possible configurations.}

{\it SmallBank~\cite{Alomari:2008:CSP:1546682.1547288}.} The schema 
consists of three relations, where each relation has two attributes. SmallBank models a banking application where customers can interact with their savings and checking accounts through five different transaction programs: Balance, Amalgamate, DepositChecking, TransactSavings and WriteCheck. These programs do not contain insert or delete statements, and there is no branching or iteration. Furthermore, tuples are always accessed through their primary key, implying that there are no predicate reads. In this more limited setting, 
the machinery developed in~\cite{VLDBpaper} can completely \emph{decide} robustness against \mvrc (that is, never results in  false negatives). 
A comparison with the results of~\cite{VLDBpaper} can thus provide insight on the completeness of Algorithm~\ref{alg:robust:check}.

{\it TPC-C~\cite{TPCC}.} 
This benchmark models a multi-warehouse wholesale operation. The database schema consists of nine different relations, where each relation has between 3 and 21 attributes. Five transaction programs (NewOrder, Delivery, Payment, OrderStatus and StockLevel) model different actions, such as creating and delivering orders, handling customer payments, as well as read-only programs collecting information about orders and stock levels.

{\it Auction.} 
The Auction benchmark is presented in Section~\ref{sec:runningex}. In Section~\ref{sec:exp:scalability} we describe an alternative version of this benchmark where the total number of transaction programs can be scaled.

\subsection{Detecting Robustness against \mvrc}

{\bf Different settings.}
In this paper, as in \cite{VLDBpaper}, we deviate from the literature by considering dependencies between operations on the granularity of individual attributes, as it allows to detect larger sets of transaction programs to be robust. To assess this advantage, we also compare with the setting where dependencies are defined on the level of complete tuples, that is, operations over the same tuple are no longer required to access a common attribute for a dependency to occur.
We stress that when our algorithm determines a set of transaction programs to be robust, that set will still be robust on systems that assure \mvrc with tuple-level database objects, for the simple reason that every conflict on the granularity of attributes implies a conflict on the granularity of tuples. As a result, every schedule that can be created by these systems is allowed under our definition of \mvrc.
We consider 
four different settings: `tpl dep', `attr dep', `tpl dep + FK' and `attr dep + FK'. The first two settings ignore foreign key constraints, and the settings `tpl dep' and `tpl dep + FK' consider dependencies on the granularity of tuples rather than that of attributes.  


{\bf Maximal robust subsets.}
We test robustness 
for each possible subset of programs for all three benchmarks to detect maximal robust subsets. 
Figure~\ref{fig:table:robust} summarizes the subsets detected as robust against \mvrc by Algorithm~\ref{alg:robust:check} for each benchmark and setting. 
Here, transactions are represented by their abbreviations (e.g., NO stands for NewOrder).
Visualizations of these summary graphs can be found in Appendix~E\ifthenelse{\boolean{fullversion}}{}{ of \cite{fullversionVDLB2022}}.

For both SmallBank and TPC-C, we identify a subset consisting of three (out of five) programs as robust against \mvrc for setting `attr dep + FK', and for the Auction benchmark, we are even able to detect the complete benchmark as robust against \mvrc. 
When comparing the different settings, we can make the following observations. Attribute-granularity is required for TPC-C to detect a maximal possible robust subset of size 3 (row `attr dep + FK'). On the other hand, attribute-granularity does not provide additional benefit over tuple-granularity for SmallBank and Auction.  
This is not unexpected, as relations in both benchmarks have only a limited number of attributes each whereas TPC-C contains many more attributes per relation. Furthermore, foreign key constraints are necessary to derive the largest robust subsets for TPC-C and Auction (compare the rows `attr dep' with `attr dep + FK'). This underlies the utility of foreign key constraints and the effectiveness of our approach, especially when taking into account that deciding robustness against \mvrc w.r.t.\ foreign key constraints is undecidable~\cite{ICDTpaper}.


{\bf Comparison with \cite{DBLP:conf/aiccsa/AlomariF15}.}
Alomari and Fekete~\cite{DBLP:conf/aiccsa/AlomariF15} detect robustness through the absence of cycles involving at least one counterflow edge, which we refer to as \typeone cycles. A direct comparison would be unfair as that work does not include predicate reads or atomic updates, and does not consider attribute-granularity. Furthermore, no formal method is provided to construct a summary graph. 
Towards an unbiased comparison, we report in Figure~\ref{fig:table:robust:counterflow-cycle} the maximal robust subsets that can be detected via the absence of \typeone cycles in the corresponding summary graphs (as constructed through Algorithm~\ref{alg:graph}) for the different settings.
When comparing 
to Figure~\ref{fig:table:robust}, we see that our technique detects more and larger subsets as robust for all benchmarks. Subsets not detected by \cite{DBLP:conf/aiccsa/AlomariF15} are displayed in bold in 
Figure~\ref{fig:table:robust}. Notice in particular
that {Algorithm~\ref{alg:robust:check}}
correctly identifies the Auction benchmark as a whole as robust against \mvrc, whereas~\cite{DBLP:conf/aiccsa/AlomariF15} only detects singleton sets as robust against \mvrc.

{\bf False negatives.} Algorithm~\ref{alg:robust:check} is based on a sufficient condition and can result in false negatives. Earlier work \cite{VLDBpaper} provided a complete characterization for deciding robustness against \mvrc for benchmarks satisfying certain restrictions: tuples can only be accessed through key-based lookup (ruling out predicate-based dependencies) and the value of keys is not allowed to be changed.
As discussed earlier, SmallBank can be captured by this restricted formalism and \cite{VLDBpaper} therefore lists the actual robust subsets. Comparing with 
Figure~\ref{fig:table:robust}, we can report that Algorithm~\ref{alg:robust:check} finds \emph{all} maximal robust subsets and does not report any false negatives.
That is, for each subset of SmallBank not detected as robust by Algorithm~\ref{alg:robust:check}, a counterexample schedule exists that is allowed under \mvrc but not conflict serializable.
{
Sometimes, specific details such as predicate conditions can lead to robustness not detected by our algorithm.
For the TPC-C benchmark for example, we identified \{Delivery\} as a false negative. The reason is that, for each district, Delivery first identifies the oldest open order through a predicate read, followed by deleting this tuple from relation NewOrder and handling the order. Because of this, no two instances of Delivery over the same warehouse can be concurrent: if they are, they would select the same oldest open order, and the second one to delete it would have to abort, since the tuple no longer exists.
}

\begin{table}[t]
    \begin{center}
        \footnotesize
            \begin{tabular}{l||c|c|c|c}
                & SmallBank & TPC-C & Auction & Auction($n$)\\
                \hline
                \hline
                relations & 3 & 9 &3 &3\\
                attributes per relation& 2& 3--21 & 2&2\\
                transaction programs &5 & 5& 2 & $2n$\\
                nodes / unfolded tr pr & 5& 13& 3& $3n$\\ 
                edges (counterflow) & 56 (12) & 396 (83) & 17 (1) & {$8n+9n^2$ ($n$)}\\
            \end{tabular}
    \end{center}
    \caption{\label{table:benchmark:char} Benchmark characteristics.
            }
    \end{table}

\begin{figure}
    \begin{center}
        \begin{footnotesize}
            \begin{tabular}{l||l|l|l}
                Alg~\ref{alg:robust:check} & SmallBank & TPC-C & Auction\\
                \hline
                \hline
                tpl dep & \{Am, DC, TS\}, \textbf{\{Bal, DC\}}, & \{OS, SL\}, \{NO\} & \{FB\}\\
                & \textbf{\{Bal, TS\}} & & \\
                \hline
                attr dep & \{Am, DC, TS\}, \textbf{\{Bal, DC\}}, & \{OS, SL\}, \{NO\} & \{FB\}\\
                & \textbf{\{Bal, TS\}} & & \\
                \hline
                tpl dep + FK & \{Am, DC, TS\}, \textbf{\{Bal, DC\}}, & \{OS, SL\}, \{NO\} & \textbf{\{FB, PB\}}\\
                & \textbf{\{Bal, TS\}} & & \\
                \hline
                attr dep + FK & \{Am, DC, TS\}, \textbf{\{Bal, DC\}}, & \textbf{\{OS, Pay, SL\}}, & \textbf{\{FB, PB\}}\\
               & \textbf{\{Bal, TS\}} & \{NO, Pay\} & \\
            \end{tabular}
        \end{footnotesize}
    \end{center}
    \caption{{Robust subsets based on absence of \typetwo cycles 
    (Algorithm~\ref{alg:robust:check}). Subsets for which the summary graph contains a \typeone cycle, 
    that are thus not detected by \cite{DBLP:conf/aiccsa/AlomariF15}, 
    are in bold.}
    \label{fig:table:robust}}
\end{figure}

\begin{figure}
    \begin{center}
        \begin{footnotesize}
            \begin{tabular}{l||l|l|l}
                Method of \cite{DBLP:conf/aiccsa/AlomariF15} & SmallBank & TPC-C & Auction\\
                \hline
                \hline
                tpl dep & \{Am, DC, TS\}, \{Bal\} & \{OS, SL\}, \{NO\} & \{FB\}\\
                \hline
                attr dep & \{Am, DC, TS\}, \{Bal\} & \{OS, SL\}, \{NO\} & \{FB\}\\
                \hline
                tpl dep + FK & \{Am, DC, TS\}, \{Bal\} & \{OS, SL\}, \{NO\} & \{PB\}, \{FB\}\\
                \hline
                attr dep + FK & \{Am, DC, TS\}, \{Bal\} & \{NO, Pay\}, \{Pay, SL\}, & \{PB\}, \{FB\}\\
                & & \{OS, SL\} & \\
            \end{tabular}
        \end{footnotesize}
    \end{center}
    \caption{Robust subsets wrt absence of 
    {\typeone cycles} (\cite{DBLP:conf/aiccsa/AlomariF15}).
    \label{fig:table:robust:counterflow-cycle}}
\end{figure}

\subsection{Scalability}
\label{sec:exp:scalability}

We reiterate that robustness is static property and involves an offline analysis where a set of transaction programs can be tested at design time. There is no need to perform online robustness testing during transaction processing. Execution times in the order of milliseconds are therefore not required. 
Previous work~\cite{VLDBpaper,DBLP:conf/aiccsa/AlomariF15} has already established the performance benefit of executing transactions 
under the lower isolation level \mvrc over executing them under a higher isolation level such as \snapshot or \serializable, so we do not repeat such experiments here.

Table~\ref{table:benchmark:char}
describes for each benchmark the size of the summary graph in terms of the number of nodes as well as the number of (counterflow) edges. Since programs with loops and branches are unfolded, the number of nodes can be larger than the number of programs at the application level. For each of the benchmarks, our implementation runs in a fraction of a second. To better illustrate the feasibility of our approach for larger sets of programs (and, consequently, larger summary graphs), we next present a modification of the Auction benchmark, referring to it as Auction($n$), where the total number of programs depends on a scaling parameter $n$, {which should be contrasted with the benchmarks presented in Section~\ref{sec:exp:benchmarks} where the number of programs is fixed (5 for SmallBank and TPC-C, and 2 for Auction).}

Auction($n$) extends upon Auction, as presented in Section~\ref{sec:runningex}, by modelling the auction of $n$ different items, where the bids for each item $i$ are stored in a separate relation $\text{Bids}^i$(\underline{buyerId}, bid), rather than having only one relation Bids.\footnote{Alternatively, we can still assume that all bids are stored in one relation Bids and each $\text{Bids}^i$ acts as a view over this relation, disjoint with all other views.} For each item $i$, Auction($n$) has two different programs: $\text{FindBids}^i$ and $\text{PlaceBid}^i$. The meaning of these programs as well as the program details remain as discussed in Section~\ref{sec:runningex}, with the only difference that they are now over item $i$ and corresponding relation $\text{Bids}^i$. The statement details of the corresponding BTP programs are as presented in Figure~\ref{fig:btp_queries_findbids}, with the only exception that $\rel{q_2}$, $\rel{q_4}$ and $\rel{q_5}$ are now the corresponding $\text{Bids}^i$. Notice that Auction(1) corresponds to the Auction benchmark as introduced in Section~\ref{sec:runningex}. 
By construction, the number of BTPs in Auction($n$) is $2\cdot n$, and since each $\text{PlaceBid}^i$ is unfolded in two LTPs, the derived set of LTPs has size $3 \cdot n$. 

Algorithm~\ref{alg:robust:check} detects Auction($n$) as robust against \mvrc for each $n$. We emphasize that the summary graph of Auction($n$) does not just consist of $n$ connected components, where each such component is equivalent to the graph given in Figure~\ref{fig:ex:sg}. Indeed, since each statement still writes to the relation Buyer, the summary graph will have a non-counterflow edge between each pair of programs, even if they are over different items. The skeleton of the summary graph for Auction$(n)$ is given \ifthenelse{\boolean{fullversion}}{in Appendix~E.}{in~\cite{fullversionVDLB2022}.}

Figure~\ref{fig:exp:scaled} shows the execution time of our implementation as well as the resulting number of edges in the summary graph for Auction($n$) for different values of $n$. For each value of $n$, the experiment was repeated 10 times, and the graph shows both the average value as well as the 95\% confidence interval. These results demonstrate that our approach can be applied to larger sets of programs. We reiterate that execution times of seconds (or even larger) are acceptable, as robustness detection is a form of static program analysis that has no influence on the actual transaction throughput once programs are being executed under \mvrc. Furthermore, we stress that the parameter $n$ refers to the number of transaction programs in the benchmark (which is unlikely to be a three figure number in practice), and does not refer to the concurrent online execution of transactions which of course can be several orders of magnitude larger.
{Our experiments do not cover scalability according to the complexity of transaction programs, as such experiments would require a benchmark where the complexity (number of nested loops and
branching) of the programs can be scaled. We are not aware of such a benchmark. To be even more precise, it is the size of the resulting summary graph that influences the required time to analyze the workload (cf. Table~\ref{table:benchmark:char}). While more nested loops and branching leads to more unfolded nodes in the graph, the same increasing in the size of the summary graph can be achieved by simply adding programs as well (which is what we did in our Auction($n$) benchmark).}

\begin{figure}
    \centering
    \includegraphics[width=.85\linewidth]{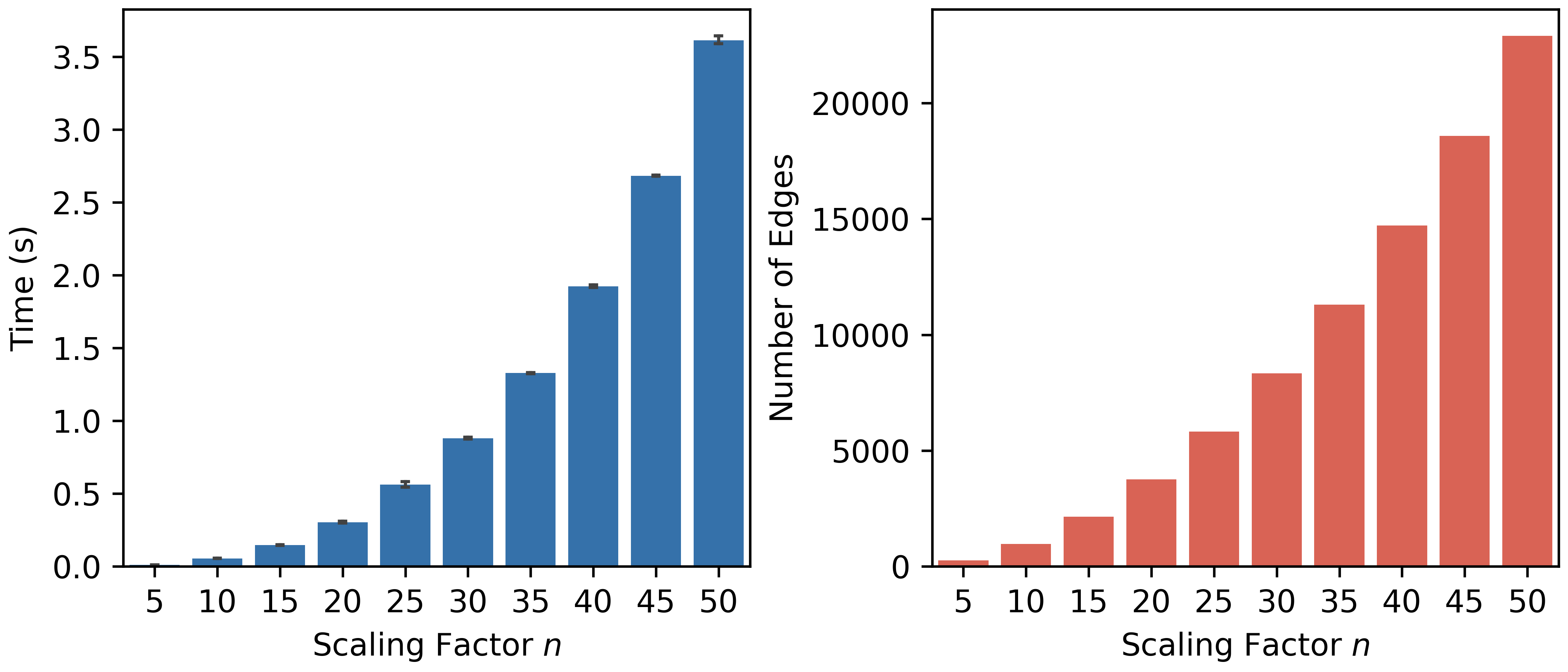}
    \caption{\emph{(left)} Time required to verify robustness against \mvrc for Auction($n$) for different scaling factors. \emph{(right)} Number of edges in the corresponding summary graphs.}
    \label{fig:exp:scaled}
\end{figure}


\vspace{-.5em}
\section{Related Work}
\label{sec:relwork}

\ifthenelse{\boolean{fullversion}}{
\subsection{Static robustness checking on the application level}
}{
    }
As mentioned in the introduction, previous work on static robustness testing~\cite{DBLP:journals/tods/FeketeLOOS05,DBLP:conf/aiccsa/AlomariF15} for transaction programs is based on testing for the absence of cycles in a static dependency graph containing some dangerous structure. This paper builds further upon the above ideas but is different in two key aspects: (1) Through the formalism of \btp{}s, our approach can be readily implemented and does not require a database expert for the construction of the summary graph. The only manual step that is required is to model SQL code in terms of \btp{}s and foreign key constraints. (2) For the first time inserts, deletes as well as predicate reads are incorporated providing a significant step towards the utilization of robustness testing in practice.

{
Our earlier work~\cite{VLDBpaper} provides a complete algorithm for deciding robustness against \mvrc but is restricted to the setting where tuples can only be accessed through 
key-based lookup and key attributes are not allowed to change. That approach can not be extended to include inserts, deletes, or predicate reads. In fact, we show in \cite{ICDTpaper} that the extension to foreign key constraints already renders the problem undecidable. Undecidability is circumvented in this paper by devising a sound but incomplete algorithm. 
}
The work in \cite{DBLP:conf/pods/Ketsman0NV20} considers robustness on the level of transactions rather than transaction programs and is based on locking rather than versioning as a concurrency control mechanism.

{Gan et al.~\cite{DBLP:journals/pvldb/GanRRB020} present IsoDiff, a tool to detect and resolve potential anomalies caused by executing transactions under \readcom or \snapshot instead of \serializable.
Similar to our approach, IsoDiff is based on detecting cycles with a specific structure. For \readcom, IsoDiff searches for type-I cycles, but includes additional timing constraints and correlation constraints to reduce the number of false positives.
Contrasting our work, IsoDiff derives potential transactions from a database SQL trace, while we derive potential transactions through our formalism of BTPs. A potential pitfall of analyzing a trace is that it may overlook transactions that are rarely executed, thereby incorrectly considering an application to be robust.
The correlation constraints IsoDiff derives from these traces correspond to the foreign key constraints expressed over BTPs.
A more subtle difference is that the timing constraints proposed as part of IsoDiff assume that a dependency $\dependson{b_i}{a_j}$ always implies that operation $b_i$ occurs before $a_j$ in $\schedule$, thereby implicitly assuming a single version implementation of \readcom, rather than \mvrc as discussed in this paper. In particular, \mvrc allows for situations where $b_i$ occurs after $a_j$ in $\schedule$, if $\dependson{b_i}{a_j}$ is a rw-antidependency.}


\ifthenelse{\boolean{fullversion}}{
Cerone et al.~\cite{DBLP:conf/concur/Cerone0G15} provide a framework for uniformly specifying different isolation levels in a declarative way. 
A key assumption is \emph{atomic visibility} requiring that either all or none of the updates of each transaction are visible to other transactions.
Based on this framework, Bernardi and Gotsman~\cite{DBLP:conf/concur/0002G16} provide sufficient conditions for robustness against these isolation levels. 
Similar to the work of Fekete et al.~\cite{DBLP:journals/tods/FeketeLOOS05}, they first identify specific properties admitted by cycles in the dependency graphs of schedules that are allowed by the isolation level but not serializable. While analyzing robustness for a given set of program instances, they assume that each program instance is overestimated by three sets of tuples: those that might be read or written to by the program instance, and those that must be written to by the program instance. based on these sets, a static dependency graph is constructed. Analogous to~\cite{DBLP:journals/tods/FeketeLOOS05}, the absence of cycles with the property related to an isolation level in this graph guarantees that the set of program instances is robust against that isolation level. When analyzing robustness for a set of programs instead of specific program instances, a summary dependency graph is constructed, where each program is represented by a node. This graph is similar to static dependy graphs, but has additional information on the edges related to how the programs should be instantiated to create a specific conflict. This additional information reduces the number of workloads that are falsely identified to be non-robust.
Continuing on this line of work, Cerone and Gotsman~\cite{Cerone:2018:ASI:3184466.3152396} later studied the problem of robustness against \parsnapshot towards \snapshot (i.e., whether for a given workload every schedule allowed under \parsnapshot is allowed under \snapshot).
\emph{This declarative framework cannot be used to study robustness against \MVRC, as \MVRC does not admit \emph{atomic visibility}.}
}
{
 Other work studies robustness within a framework for uniformly specifying different isolation levels in a declarative way~\cite{DBLP:conf/concur/Cerone0G15,DBLP:conf/concur/0002G16,Cerone:2018:ASI:3184466.3152396}. A key assumption here is \emph{atomic visibility} requiring that either all or none of the updates of each transaction are visible to other transactions.
 {This work aims at higher isolation levels and cannot be used for \MVRC, as \MVRC does not admit \emph{atomic visibility}.}
 }

{
}}

\ifthenelse{\boolean{fullversion}}{

}{
Transaction chopping splits transactions into smaller pieces to obtain performance benefits and is correct if, for every serializable execution of the chopping, there 
is an equivalent serializable execution of the original transactions~\cite{DBLP:journals/tods/ShashaLSV95}.  Cerone et al. \cite{Cerone:2018:ASI:3184466.3152396,DBLP:conf/wdag/CeroneGY15}  studied chopping under various isolation levels. {Transaction chopping has no direct relationship with robustness testing against \MVRC.}
}

\ifthenelse{\boolean{fullversion}}{
\subsection{Other approaches}
}{
}

\ifthenelse{\boolean{fullversion}}{
{Instead of weakening the isolation level, other approaches to increasing transaction throughput without sacrificing ACID guarantees have been studied as well.}
Transactions can for example be split in smaller pieces to obtain performance benefits. However, this approach poses a new challenge, as not every serializable execution of these chopped transactions is necessarily equivalent to some serializable execution over the original transactions. A chopping of a set of transactions is correct if for every serializable execution of the chopping there exists an equivalent serializable execution of the original transactions. Shasha et al.~\cite{DBLP:journals/tods/ShashaLSV95} provide a graph based characterization for this correctness problem.
{This problem has been studied for different isolation levels such as \snapshot~\cite{Cerone:2018:ASI:3184466.3152396} and \parsnapshot~\cite{DBLP:conf/wdag/CeroneGY15} as well. However, in this case a correct chopping does not guarantee serializability. Instead, it verifies whether every execution of the chopped transactions allowed under an isolation level is equivalent to some execution of the original transactions allowed under this isolation level.}
\emph{Transaction chopping has no direct relationship with robustness testing against \MVRC.}

{Another approach is to modify existing algorithms that guarantee serializability. One notable example is a modification of S2PL where a transaction might release some locks before it acquired all locks. Wolfson~\cite{DBLP:journals/jal/Wolfson86,DBLP:journals/jal/Wolfson87} uses a sufficient condition to determine for a given workload at which point each lock acquired by a transaction might be released without risking anomalies.}

{When semantic knowledge of the transaction programs is available, it can be used to weaken the serializability requirement. Farrag and \"Ozsu~\cite{DBLP:journals/tods/FarragO89} use semantic knowledge of allowed interleavings between transactions to construct a new concurrency control algorithm that guarantees relatively consistent schedules. These relatively consistent schedules always preserve consistency, but do not necessarily guarantee serializability. Lu et al.~\cite{DBLP:journals/tkde/LuBL04} provide sufficient conditions under which every execution over a set of transactions under a given lock-based isolation level is semantically correct. A schedule is semantically correct if it has the same semantic effect as a serial schedule. As such, semantic correctness does not necessarily guarantee traditional serializability.
}
}{}

{
Many approaches to increase transaction throughput without sacrificing serializability have been proposed: improved or novel pessimistic (cf., e.g., ~\cite{DBLP:journals/pvldb/YanC16,DBLP:journals/pvldb/TianHMS18,DBLP:conf/sigmod/RenFA16,DBLP:journals/pvldb/RenTA12,DBLP:journals/pvldb/JohnsonPA09})
or optimistic (cf., e.g., ~\cite{DBLP:conf/sigmod/SharmaSD18,DBLP:conf/sigmod/YuPSD16,DBLP:journals/pvldb/GuoCWQZ19,DBLP:journals/pvldb/HuangQKLS20,DBLP:journals/pvldb/LarsonBDFPZ11,DBLP:conf/sigmod/DiaconuFILMSVZ13,DBLP:conf/sigmod/BernsteinDDP15,DBLP:conf/cidr/BernsteinRD11,DBLP:journals/pvldb/SadoghiCBNR14,DBLP:conf/cloud/DingKDG15,DBLP:conf/sigmod/0001MK15,DBLP:conf/sigmod/KimWJP16,DBLP:conf/sigmod/LimKA17,DBLP:conf/sigmod/JonesAM10,DBLP:journals/pvldb/YuanWLDXBZ16})
algorithms, 
as well as approaches based on coordination avoidance (cf., e.g., ~\cite{DBLP:journals/pvldb/FaleiroAH17,DBLP:conf/sigmod/PrasaadCS20,DBLP:journals/pvldb/LuYCM20,DBLP:conf/sigmod/ShengTZP19,DBLP:journals/pvldb/FaleiroA15,DBLP:journals/pvldb/RenLA19,DBLP:conf/sigmod/ThomsonDWRSA12, DBLP:journals/tkde/YaoA0LOWZ16,DBLP:conf/middleware/QadahS18}). 
Robustness differs from these approaches in that it can be applied to standard DBMS's without any modifications to the database internals. Instead, the robustness property is leveraged to guarantee serializability even though the database system provides a lower isolation level. 

}

{Orthogonal to robustness detection, tools such as Elle~\cite{DBLP:journals/pvldb/AlvaroK20} aim at detecting anomalies that should not occur under a given isolation level. These tools can be used to detect whether a database system implements the declared isolation levels correctly, whereas robustness assumes that the isolation level is implemented correctly to decide whether every possible execution of a given workload is serializable.
}

\ifthenelse{\boolean{fullversion}}{
\subsection{Formalization}
\label{sec:def:adya}
}
{
}
Our formalization of transactions and conflict serializability is closely related to the formalization presented by Adya et al.~\cite{DBLP:conf/icde/AdyaLO00}, but with some important differences, which we discuss next. 
We assume a total rather than a partial order over the operations in a schedule, and the different types of write operations are made more explicit by introducing inserts and deletes. In particular, we require that only an insert operation can create the first visible version after the unborn version, and only a delete operation can create the dead version in a schedule. Our definitions consider an atomic update operation as well, which is essentially a read operation followed by a write operation on the same object, and which cannot be interleaved by other operations in a schedule. Atomic chunks take this assumption one step further by allowing arbitrary sequences of operations in a transaction to act as one atomic operation.
We furthermore assume that all operations are over concrete (database) tuples rather than abstract objects, and keep track of the specific attribute values that each operation observes or modifies.
As illustrated in~\cite{VLDBpaper}, explicitly taking into account these atomic update operations as well as the attributes that are accessed can greatly increase the effectiveness of robustness detection.
\ifthenelse{\boolean{fullversion}}{
Due to these changes relative to the formalization presented by Adya et al.~\cite{DBLP:conf/icde/AdyaLO00}, there are some notational differences as well. One particular difference is that we will not use the subscript notations $\W[]{\x_i}$ and $\R[]{\x_i}$ to indicate the version $\x_i$ of a tuple $\x$ that is respectively written or observed.
Instead, we will define two functions $\schrvf$ and $\schwvf$ mapping each operation over a tuple $\x$ to the version of $\x$ it respectively observed or created.}{}

\vspace{-.6em}
\section{Conclusions}
\label{sec:conclusions}
\vspace{-.3em}

The present paper makes a significant step towards robustness testing in practice: through a formal approach based on \btp{}s, we provide an algorithm for robustness testing that (1) can be readily implemented; and (2) improves over the state-of-the-art in that it incorporates a larger set of operations (inserts, deletes, predicate reads) and can detect larger sets of transaction programs to be robust against \mvrc. In the future, we plan to cover more expressive transaction programs.

\vspace{-.6em}

\begin{acks}
  \vspace{-.3em}
  This work is funded by FWO-grant G019921N.
\end{acks}


\bibliographystyle{ACM-Reference-Format}
\bibliography{references}

\ifthenelse{\boolean{fullversion}}{
  \clearpage
  \onecolumn
  \appendix

  \section*{APPENDIX}


\section{Format of SQL transactions}
\label{app:sql}

Here, we provide an overview of the SQL transactions that inspired the definition of basic transaction programs and their translation into \btp:

\begin{itemize}
    \item \textbf{key-based selection}\\
    $q:=$\begin{verbatim}
        SELECT <select-set(q)>
          FROM R
         WHERE <key-condition(q)>
    \end{verbatim}
    Where $R$ is a relation in $\schRel$, $\selectset{q} \subseteq \Attr{R}$, and $\keycond{q}$ is a condition intended to find a tuple by its primary key attributes of $R$.\\

    Then,
    \begin{itemize}
        \item $\type{q}=\qksel$;
        \item $\rel{q}=R$;
        \item $\predset{q}=\emptyset$ (as the selection is not predicate-based);
        \item $\obsset{q}=\selectset{q}$; and,
        \item $\modset{q}=\emptyset$.
    \end{itemize}
    \medskip

    \item \textbf{predicate-based selection}\\
    $q:=$\begin{verbatim}
        SELECT <select-set(q)>
          FROM R
         WHERE <predicate-condition(q)>
    \end{verbatim}
    Where $R$ is a relation in $\schRel$, $\selectset{q} \subseteq \Attr{R}$, and $\predcond{q}$ is a condition over (a subset of) the attributes in $\Attr{R}$.\\

    Then,
    \begin{itemize}
        \item $\type{q}=\qpsel$;
        \item $\rel{q}=R$;
        \item $\predset{q}$ equals the attributes mentioned in $\predcond{q}$;
        \item $\obsset{q}=\selectset{q}$; and,
        \item $\modset{q}=\emptyset$.
    \end{itemize}
    \medskip

    \item \textbf{key-based update}\\
    $q:=$
    \begin{verbatim}
        UPDATE R
           SET A1 = <expr(q,1)>, ..., An = <expr(q,n)>
         WHERE <key-condition(q)>
     RETURNING <select-set(q)>
    \end{verbatim}
    Where $R$ is a relation in $\schRel$, $\{A_1, \ldots, A_n\} \subseteq \Attr{R}$, $\selectset{q} \subseteq \Attr{R}$, $\keycond{q}$ is a condition intended to find a tuple by its primary key attributes of $R$, and each $\expr{q,i}$ is an expression over (a subset of) attributes in $\Attr{R}$.\\

    Then,
    \begin{itemize}
        \item $\type{q}=\qkmod$;
        \item $\rel{q}=R$;
        \item $\predset{q}=\emptyset$ (as the selection is not predicate-based);
        \item $\obsset{q}$ corresponds to the attributes occurring in $\selectset{q}$ as well as each $\expr{q,j}$;                
        \item $\modset{q}=\{A_1, \ldots, A_n\}$.
    \end{itemize}
    \medskip

    \item \textbf{predicate-based update}\\
    $q:=$\begin{verbatim}
        UPDATE R
           SET A1 = <expr(q,1)>, ..., An = <expr(q,n)>
         WHERE <predicate-condition(q)>
     RETURNING <select-set(q)>
    \end{verbatim}
    Where $R$ is a relation in $\schRel$, $\{A_1, \ldots, A_n\} \subseteq \Attr{R}$, $\selectset{q} \subseteq \Attr{R}$, $\predcond{q}$ is a condition over (a subset of) the attributes in $\Attr{R}$, and each $\expr{q,i}$ is an expression over (a subset of) attributes in $\Attr{R}$.\\
    
    Then,
    \begin{itemize}
        \item $\type{q}=\qpmod$;
        \item $\rel{q}=R$;
        \item $\predset{q}$ equals the attributes mentioned in $\predcond{q}$;        
        \item $\obsset{q}$ corresponds to the attributes occurring in $\selectset{q}$ as well as each $\expr{q,j}$                
        \item $\modset{q}=\{A_1, \ldots, A_n\}$.
    \end{itemize}
    \medskip

    \item \textbf{insertion}\\
    $q:=$
    \begin{verbatim}
        INSERT INTO R
        VALUES (a1, a2, ..., an)
    \end{verbatim}
    where $R$ is a relation in $\schRel$ and ${a_1, \ldots, a_n}$ are arbitrary values with $n = \ssize{\Attr{R}}$.\\

    Then,
    \begin{itemize}
        \item $\type{q}=\qins$;
        \item $\rel{q}=R$;
        \item $\predset{q}=\emptyset$;
        \item $\obsset{q}=\emptyset$;     
        \item $\modset{q}=\Attr{\rel{q}}$.
    \end{itemize}
    \medskip

    \item \textbf{key-based deletion}\\
    $q:=$
    \begin{verbatim}
        DELETE FROM R
         WHERE <key-condition(q)>
    \end{verbatim}
    Where $R$ is a relation in $\schRel$ and $\keycond{q}$ is a condition intended to find a tuple by its primary key attributes of $R$.\\

    Then,
    \begin{itemize}
        \item $\type{q}=\qkdel$;
        \item $\rel{q}=R$;
        \item $\predset{q}=\emptyset$ (as the selection is not predicate-based);
        \item $\obsset{q}=\emptyset$; and,     
        \item $\modset{q}=\Attr{\rel{q}}$.
    \end{itemize}
    \medskip

    \item \textbf{predicate-based deletion}\\
    $q:=$
    \begin{verbatim}
        DELETE FROM R
         WHERE <predicate-condition(q)>
    \end{verbatim}
    Where $R$ is a relation in $\schRel$, and $\predcond{q}$ is a condition over (a subset of) the attributes in $\Attr{R}$.\\

    Then,
    \begin{itemize}
        \item $\type{q}=\qpdel$;
        \item $\rel{q}=R$;
        \item $\predset{q}$ equals the attributes mentioned in $\predcond{q}$;
        \item $\obsset{q}=\emptyset$; and,     
        \item $\modset{q}=\Attr{\rel{q}}$.
    \end{itemize}
\end{itemize}

\medskip
The flow instructions loop and `|' in \btp correspond to
\begin{itemize}
    \item \textbf{loops}
    \begin{verbatim}
        REPEAT
            <subprogram>
        END REPEAT
    \end{verbatim}
    where $\subprogram$ is itself a transaction program.

    \item \textbf{conditional execution}
    \begin{verbatim}
        IF
            <subprogram_1>
        ELSE
            <subprogram_2>
        ENDIF
    \end{verbatim}
    where $\subprogram_1$ and $\subprogram_2$ are two (possibly empty) transaction programs.
\end{itemize}


\section{Proofs of Section~\ref{sec:ser:graph}}

\bigskip

\noindent
{\bf Proof of Lemma~\ref{lem:counterflow}}

\begin{proof}
    
    The proof is straightforward: In a schedule $\schedule$, all other types of dependencies $\dependson[\schedule]{b_i}{a_j}$ imply a version order on the versions of tuples read or written by operations $b_i$ and $a_j$ that is consistent with the direction of the dependency. Therefore, 
    if $\schedule$ is an \mvrc schedule the read last committed property implies $C_i <_s C_j$, thus that $\dependson[\schedule]{b_i}{a_j}$ is indeed not counterflow.
\end{proof}
\bigskip

\noindent
{\bf Proof of Theorem~\ref{theo:cg:cycles}}
\begin{proof}
    Let $\cyclesym = (T_1, b_1, a_2, T_2), \ldots, (T_n, b_n, a_1, T_2)$ be an arbitrary cycle in $\cg{\schedule}$.
    That one of the dependencies of $\cyclesym$ is counterflow follows directly from its definition, as otherwise
    $\CT[1]\schords\CT[2]\schords\cdots\schords\CT[n]\schords\CT[1]$ thus stating that the commit of $T_1$ occurs before itself in $\schedule$.
    Similarly, at least one of the dependencies in $\cyclesym$ is not counterflow, as otherwise $\CT[1]\schords\CT[n]\schords\cdots\schords\CT[2]\schords\CT[1]$, again stating that the commit of $\trans[1]$ occurs before itself.

    Before giving the main argument of the proof, 
     we first state two general properties over pairs of
    adjacent dependencies $\dependson[\schedule]{b_{i-1}}{a_i}$ and
    $\dependson[\schedule]{b_i}{a_{i+1}}$ in $\cyclesym$ with
    $\dependson[\schedule]{b_i}{a_{i+1}}$ being counterflow,
    $\dependson[\schedule]{b_{i-1}}{a_i}$ being non-counterflow, and with $b_{i-1}$ a
    write operation.

    Firstly, notice that then $\dependson[\schedule]{b_{i-1}}{a_i}$ must be a ww-, wr- or predicate
    wr-dependency, for which the definition of \mvrc schedules implies 
    \begin{equation}
        b_{i-1} \schords \CT[i-1] \schords a_i.\label{th:prop:1}
    \end{equation}

    Secondly, since $\dependson[\schedule]{b_{i}}{a_{i+1}}$ is counterflow, it
    must be an rw- or predicate
    rw-antidependency (due to Lemma~\ref{lem:counterflow}), which implies (by definition of
    \mvrc schedules) that
    \begin{equation}
        b_i \schords \CT[i+1].\label{th:prop:2}
    \end{equation}
    Indeed, $a_{i+1} <_{\trans[i+1]} \CT[i+1] \schords b_i$
    would imply that $\schvf(a_{i+1})$ occurs before the version(s) that $b_i$
    reads (by read-last committed), which contradicts that $\dependson[\schedule]{b_{i}}{a_{i+1}}$ is an rw-
    or predicate rw-antidependency. 

    We are now ready for the main argument of the proof and proceed with the assumption that for every pair
    $\dependson[\schedule]{b_{i-1}}{a_i}$ and
    $\dependson[\schedule]{b_i}{a_{i+1}}$ of adjacent dependencies in
    $\cyclesym$ with $\dependson[\schedule]{b_i}{a_{i+1}}$ a counterflow
    dependency, $\dependson[\schedule]{b_{i-1}}{a_i}$ is non-counterflow and
    $b_{i-1}$ is a write operation (as otherwise there is nothing to show).
    It then remains to argue that for at least one of these adjacent dependencies we
    have $b_i <_{\trans[i]} a_i$. The proof is by contradiction: we assume that $a_i
    <_{\trans[i]} b_i$ is always true for such pairs and given an inductive
    argument leading to a contradiction.

    For this, first take an arbitrary non-counterflow dependency
    $\dependson[\schedule]{b_{i-1}}{a_i}$ of cycle $\cyclesym$. (This dependency
    exists by our earlier made assumption and the fact that $\cyclesym$ contains
    at least two dependencies.) We will fix the
    commit operation $\CT[i-1]$ of $b_{i-1}$ and from now on refer to it by $\CT[]$.
    Now, if the dependency $\dependson[\schedule]{b_i}{a_{i+1}}$ adjacent to
    $\dependson[\schedule]{b_{i-1}}{a_i}$ in $\cyclesym$ is not counterflow, it
    is immediate that $\CT[] \le_\schedule \CT[i-1] \schords \CT[i] \schords
    \CT[i+1]$.
    If
    $\dependson[\schedule]{b_i}{a_{i+1}}$ is counterflow, it follows from
    our assumption $a_i
    <_{\trans[i]} b_i$ and observations (\ref{th:prop:1}) and (\ref{th:prop:2}) that $\CT[] \le_\schedule \CT[i-1]\schords \CT[i+1]$. 

    If $\CT[i+1] =
    \CT[]$, we have now proven that $\CT[] \schords \CT[]$, which is the desired
    contradiction. If $\CT[i+1] \ne \CT[]$, we repeat the procedure taking
    $\trans[i+1]$ as $\trans[i-1]$. Since
    $\CT[i]$ can never equal $\CT[]$ (since we started from a non-counterflow
    dependency) the procedure will eventually terminate with the desired
    contradiction, which concludes proof.
\end{proof}

\section{Proofs of Section~\ref{sec:robustness}}
\subsection{Proof of Proposition~\ref{prop:robust}}
\begin{proof}
    If $\progset'$ is robust against \mvrc, then every schedule in $\schedules{\progset'}{\mvrc}$ must be conflict serializable. It immediately follows that every schedule in $\schedules{\progset}{\mvrc}$ is conflict serializable as well.
\end{proof}
\section{Proofs of Section~\ref{sec:detecting:robustness}}

In this section, we say that to statements $q_i$ and $q_j$ \emph{allow a non-counterflow dependency} if either Table~(\ref{table:dep}) mentions true on the intersection of row $\type{q_i}$ and column $\type{q_j}$ \emph{or} it mentions $\bot$ and Algorithm~\ref{alg:graph}, function
{\sc ncDepConds}$(q_i, q_j)$ gives true.

Finally, we say that $q_i$ and $q_j$ \emph{allow a counterflow dependency} if either Table~(\ref{table:cdep}) mentions true on the
    intersection of row $\type{q_i}$ and column $\type{q_j}$; \emph{or} Algorithm~\ref{alg:graph}, function
{\sc cDepConds}$(q_i, q_j)$ gives true.

\subsection*{Proof of Proposition~\ref{pro:equi}}

\begin{proof}
        Direction (1) $\Rightarrow$ (2) is straightforward, hence we focus on
        (2) $\Rightarrow$ (1). The proof is by contraposition. That is, we show
        that if $\progset$ is not robust against \mvrc then also
        $\unfold{\progset}$ is not robust against \mvrc. 

        If $\progset$ is not robust against \mvrc there is a non conflict
        serializable schedule $\schedule$ in $\schedules{\progset}{\mvrc}$
        (cf., Definition~\ref{def:robustness}), which implies (using
        Theorem~\ref{theo:not-conflict-serializable}) that $\cg{\schedule}$
        contains a cycle $\cyclesym$. We remark that $\cyclesym$ must have a
        finite-length because $\schedule$ involves a finite number of
        transactions (by definition of $\schedules{\progset}{\mvrc}$). Let
        $\mathcal{T}$ be the set of transactions that $\schedule$ is defined
        over. Without loss of generality, we can assume that $\mathcal{T}$
        contains only transactions mentioned in $\cyclesym$. Indeed, all other
        transactions can be safely removed from the schedule while leaving the
        schedule valid under \mvrc.

        If every transaction on the $\cyclesym$ is an instantiation for an LTP in
        $\unfold{\progset}$ the proposition is immediate, therefore we continue
        with the assumption that there is at least one counterexample transaction 
        $\trans[i]$ with $(\trans[i-1], b_{i-1},
        a_i, \trans[i])$, $(\trans[i], b_i, a_{i+1}, \trans[i+1])$ its incoming
        and outgoing edge in $\cyclesym$. 

        Since $\trans[i]$ is an instantiation of a BTP P, we can assume existence of
        a mapping $\alpha$ that reveals the choices of the unfolding of $P$, and
        a mapping $\beta$ that maps $P$ and its fragments onto its corresponding
        instantiation.  In other words,
        $\alpha(loop(P)) = P_1P_2\ldots P_k$ for some integer $k$, 
        $\alpha(P_1 \mid P_2)$ equals $P_1$ or $P_2$, and so on. 
        On the other hand, $\beta(loop(P)) = \beta(P_1)\beta(P_2)\ldots \beta(P_k)$, $\beta(P_1 \mid P_2) =
        \beta(\alpha(P_1 \mid P_2))$, and so on.

        Based on $\alpha$ and $\beta$ we can construct an alternative mapping
        $\alpha'$ that defines an unfolding for the same $P$ but with different
        choices, so that operations $a_i$ and $b_i$ are still preserved in $\beta(\alpha'(P))$ but
        with $\alpha'(P)$ now representing an LTP from $\unfold{\progset}$.
        Indeed, for a BTP $P$ including operation $a_i$ or
        $b_i$ we construct $\alpha'$ in the following (inductive) way:
        if $P = loop(P')$ let $\alpha'(P)$ be the result of removing from 
        $\alpha(P) = P_1P_2\ldots P_k$ all $P_i$'s containing 
        neither $a_i$ nor $b_i$ (thus leaving a sequence of zero, one or two
        BTPs). For all other cases we let $\alpha'(P) = \alpha(P)$.
        Then let's call $\trans[i]' = \beta(\alpha'(P))$.

        Now, by removing from $\schedule$ all operations from $\trans[i]$ that
        are not in $\trans[i]'$, we obtain a schedule $\schedule'$ over
        $(\mathcal{T} \cup \{\trans[i]'\}) \setminus \{\trans[i]\}$ that is
        still valid under \mvrc and has a
        cycle $\cyclesym'$ equal to $\cyclesym$ except that $\trans[i]$ is replaced
        by $\trans[i]'$. 

        Since the construction does not influence the length of $\cyclesym$ and
        only changes a problematic transaction $\trans[i]$, we can repeat
        this procedure until all
        problematic transactions are removed, then eventually resulting in the
        desired non conflict
        serializable schedule from $\unfold{\progset}$, which concludes the
        proof.
\end{proof}

\subsection*{Proof of Proposition~\ref{prop:algocorrectness}}

\begin{proof}
    Let $\dependson[\schedule]{b_i}{a_j}$ be a dependency as defined in Condition~\ref{cond:sg:prop}. The proof is by case distinction.
    More precisely, we show for each dependency $\dependson[\schedule]{b_i}{a_j}$ that, if it is not counterflow, $q_i$ and $q_j$ allow a non-counterflow dependency
    (it then follows from the definition of $\sg{\progset}$ that $(P_i, q_i, \textit{not counterflow}, q_j, P_j)$ is an edge), 
    and if it is counterflow, that $q_i$ and $q_j$ allow a counterflow dependency (again implying by definition of $\sg{\progset}$ that then $(P_i, q_i, \textit{counterflow}, q_j, P_j)$ is an edge).

    \smallskip
    \noindent If \emph{$\dependson[\schedule]{b_i}{a_j}$ is a non-counterflow ww-dependency}, then, $b_i$ and $a_j$ are write operations, $\WGSet{b_i} \cap \WGSet{a_j} \neq \emptyset$, and $\schwvf(b_i) \schvord \schwvf(a_j)$. The latter implies, by definition of schedules, that $b_i$ is not a $\myDE$-operation and that $a_j$ is not an $\myIN$-operation. We can thus conclude by the definition of statement and instantiation of statement that 
    \begin{align*}
        \type{q_i} &\in \{\qkmod, \qpmod, \qins\}, \\
        \type{q_j} &\in \{\qkmod, \qpmod, \qkdel, \qpdel\},
    \end{align*}
    and that $\WriteSet{q_i} \cap \WriteSet{q_j} \neq \emptyset$ since $\WriteSet{q_i} = \WGSet{b_i}$ and $\WriteSet{q_j} = \WGSet{a_j}$.
    The fact that $q_i$ and $q_j$ allow a non-counterflow dependency is now straightforward.

    \smallskip
    \noindent If \emph{$\dependson[\schedule]{b_i}{a_j}$ is a non-counterflow wr-dependency}, then $b_i$ is a write operation and $a_j$ is a read operation with $\WGSet{b_i} \cap \RGSet{a_j} \neq \emptyset$, and $\schwvf(b_i) = \schrvf(a_j)$ or $\schwvf(b_i) \schvord \schrvf(a_j)$. Now, the definition of schedules implies that $b_i$ is not a $\myDE$-operation. From the definition of statement, and instantiation of statement, it follows that 
    \begin{align*}
        \type{q_i} &\in \{\qkmod, \qpmod, \qins\}, \\ 
        \type{q_j} &\in \{\qkmod, \qpmod, \qksel, \qpsel\},
    \end{align*}
    and that $\WriteSet{q_i} \cap \ReadSet{q_j} \neq \emptyset$ due to $\WriteSet{q_i} = \WGSet{b_i}$ and $\ReadSet{q_j} = \RGSet{a_j}$.
    That $q_i$ and $q_j$ allow a non-counterflow dependency is again straightforward.

    \smallskip
    \noindent If \emph{$\dependson[\schedule]{b_i}{a_j}$ is a non-counterflow rw-antidependency}, then $b_i$ is a read operation and $a_j$ is a write operations with $\RGSet{b_i} \cap \WGSet{a_j} \neq \emptyset$, and $\schrvf(b_i) \schvord \schwvf(a_j)$. This time, the definition of schedules implies that $a_j$ is not a $\myIN$-operation. From the definition of statement, and instantiation of statement, it follows that 
    \begin{align*}
        \type{q_i} &\in \{\qkmod, \qpmod, \qksel, \qpsel\}, \\
        \type{q_j} &\in \{\qkmod, \qpmod, \qkdel, \qpdel\},
    \end{align*}
    and $\ReadSet{q_i} \cap \WriteSet{q_j} \neq \emptyset$ due to $\ReadSet{q_i} = \RGSet{b_i}$ and $\WriteSet{q_j} = \WGSet{a_j}$.
    That $q_i$ and $q_j$ allow a non-counterflow dependency is again immediate from its definition.

    \smallskip
    \noindent If \emph{$\dependson[\schedule]{b_i}{a_j}$ is a non-counterflow predicate wr-dependency}, then 
        $b_i$ is a write operation on a tuple of type $R$, $a_j$ is a predicate read on relation $R$, $b_i$ is
        over a tuple $\x$ and $\schwvf(b_i) = \x_i$ or $\schwvf(b_i) \schvord
        \x_i$ with $\x_i$ the version of $\x$ in $\schpvf(a_j)$, and either $b_i$ is an $\myIN$ or $\myDE$ operation, or
        $\WGSet{b_i} \cap \PGSet{a_j} \neq \emptyset$. By definition of statement and instantiation of statement it follows that
    \begin{align*}
        \type{q_i} &\in \{\qins, \qkmod, \qpmod, \qkdel, \qpdel\}, \\
        \type{q_j} &\in \{\qpsel, \qpmod, \qpdel\},
    \end{align*}
    and that either $\type{q_i} \in \{\qins, \qkdel, \qpdel\}$ or $\WriteSet{q_i} \cap \PReadSet{q_j} \neq \emptyset$, since $\WriteSet{q_i} = \WGSet{b_i}$ and $\PReadSet{q_j} = \RGSet{a_j}$. As before, it now follows straightforwardly from the definition that $q_i$ and $q_j$ indeed allow a non-counterflow dependency.

    \smallskip
    \noindent If \emph{$\dependson[\schedule]{b_i}{a_j}$ is a non-counterflow predicate rw-antidependency}, then, $b_i$ is a predicate read on a relation $R$, $a_j$ is a write operation on a tuple of type $R$, $a_j$ is over a tuple $\x$ and $\x_i \schvord \schwvf(a_j)$ with $\x_i$ the version of $\x$ in $\schpvf(b_i)$, and either
        $a_j$ is an $\myIN$ or $\myDE$ operation or $\PGSet{b_i} \cap \WGSet{a_j} \neq \emptyset$.
    From the definition of statement, and instantiation of statement, it thus follows that 
    \begin{align*}
        \type{q_i} &\in \{\qpsel, \qpmod, \qpdel\}, \\
        \type{q_j} &\in \{\qins, \qkmod, \qpmod, \qkdel, \qpdel\},
    \end{align*}
    and either $\type{q_j} \in \{\qins, \qkdel, \qpdel\}$ or $\PReadSet{q_i} \cap \WriteSet{q_j} \neq \emptyset$, since $\PReadSet{q_i} = \WGSet{b_i}$ and $\WriteSet{q_j} = \RGSet{a_j}$. That $q_i$ and $q_j$ allow a non-counterflow dependency follows again by its definition.

    \smallskip\noindent
    At this point, we remark that we have considered all possible non-counterflow dependencies. For the counterflow dependencies, it follows from Lemma~\ref{lem:counterflow} that only two cases need to be considered:
    
    \smallskip
    \noindent If \emph{$\dependson[\schedule]{b_i}{a_j}$ is a counterflow rw-antidependency}, then $b_i$ is a read operation and $a_j$ is a write operations with $\RGSet{b_i} \cap \WGSet{a_j} \neq \emptyset$, and $\schrvf(b_i) \schvord \schwvf(a_j)$. As before, the definition of schedules implies that $a_j$ is not a $\myIN$-operation. From the definition of statement, and instantiation of statement, it follows that 
    \begin{align*}
        \type{q_i} &\in \{\qksel, \qpsel\}, \\
        \type{q_j} &\in \{\qkmod, \qpmod, \qkdel, \qpdel\},
    \end{align*}
    and $\ReadSet{q_i} \cap \WriteSet{q_j} \neq \emptyset$ due to $\ReadSet{q_i} = \RGSet{b_i}$ and $\WriteSet{q_j} = \WGSet{a_j}$.
    {We notice that $\type{q_i}$ can indeed not equal $\qkmod$ or $\qpmod$ because then by definition of instantiation of statement $q_i$, there must be a write operation $b'_i$ instantiated from $q_i$ over the same tuple as $b_i$ and $a_j$. Furthermore, since $b'_i$ and $b_i$ must be in the same atomic chunk and $\dependson{b_i}{a_j}$ is a counterflow rw-antidependency, we have either $b_i' \schords a_j \schords \CT[j] \schords \CT[i]$ or $a_j \schords b_i' \schords \CT[j] \schords \CT[i]$, where both cases imply a dirty write.
    }
    To see that $q_i$ and $q_j$ indeed allow a counterflow dependency it remains to verify that there is no foreign key
    $f \in \schFK$ and a pair of statements $q_k \in \prog[i]$ and $q_\ell \in
    \prog[j]$ with $\type{q_k}, \type{q_\ell} \in \{\qkupd, \qkdel, \qins\}$, $q_k <_{\prog[i]} q_i$ and $q_\ell <_{\prog[j]} q_j$ such
    that $q_k = f(q_i)$ and $q_\ell = f(q_j)$ are foreign key constraints for
    respectively $\prog[i]$ and $\prog[j]$. The argument is by contradiction: Assume
    that there is such a foreign key and pair of statements $q_k$ and $q_\ell$.
    Then the instantiations of $q_k$ and $q_\ell$ must involve write operations $b_i'$ and $a_j'$ over a common tuple $f(t)$, with $t$ the tuple that 
    $b_i$ and $a_j$ are over. But then the fact that $\dependson[\schedule]{b_i}{a_j}$ is counterflow means $b_i \schords a_j \schords \CT[i]$ or $a_j \schords b_i \schords \CT[j]$, implying 
    $b_i' \schords a_j' \schords\CT[i]$ or $a_j' \schords b_i' \schords\CT[j]$, thus the presence of a dirty write in $\schedule$, which contradicts with $\schedule$ being allowed under $\mvrc$.

    \smallskip
    \noindent If \emph{$\dependson[\schedule]{b_i}{a_j}$ is a counterflow predicate rw-antidependency}, then, $b_i$ is a predicate read on a relation $R$, $a_j$ is a write operation on a tuple of type $R$, $a_j$ is over a tuple $\x$ and $\x_i \schvord \schwvf(a_j)$ with $\x_i$ the version of $\x$ in $\schpvf(b_i)$, and either
        $a_j$ is an $\myIN$ or $\myDE$ operation or $\PGSet{b_i} \cap \WGSet{a_j} \neq \emptyset$.
    From the definition of statement, and instantiation of statement, it thus follows that 
    \begin{align*}
        \type{q_i} &\in \{\qpsel, \qpmod, \qpdel\}, \\
        \type{q_j} &\in \{\qins, \qkmod, \qpmod, \qkdel, \qpdel\},
    \end{align*}
    and either $\type{q_j} \in \{\qins, \qkdel, \qpdel\}$ or $\PReadSet{q_i} \cap \WriteSet{q_j} \neq \emptyset$, since $\PReadSet{q_i} = \WGSet{b_i}$ and $\WriteSet{q_j} = \RGSet{a_j}$. 

    \smallskip
    \noindent
    We have now considered all cases, which concludes the proof.
\end{proof}

\subsection*{Proof of Theorem~\ref{theo:sufcond:robustness}}

\begin{proof}
    The proof is by contraposition: We show that if the set of \ltp{}s
    $\progset$ is not robust against \mvrc then there is a cycle with the conditions of the theorem.

    If $\progset$ is not robust against
    $\mvrc$, Definition~\ref{def:robustness} implies existence of a schedule
    $\schedule$ in $\schedules{\progset}{\mvrc}$ that is not conflict
    serializable, thus (due to Theorem~\ref{theo:not-conflict-serializable})
    with $\cg{\schedule}$ containing a cycle $\cyclesym'$.  Since $\schedule$
    is allowed under \mvrc{} (by definition of $\schedules{\progset}{\mvrc}$) 
    cycle $\cyclesym'$ has the properties listed in
    Theorem~\ref{theo:cg:cycles}. It therefore remains to show only how these
    properties about dependencies between operations can be lifted to
    properties over edges in $\sg{\progset}$.

    The link is made by Proposition~\ref{prop:algocorrectness}. Indeed, take an
    arbitrary dependency $\dependson[\schedule]{b_i}{a_j}$ in $\schedule$, say
    with $b_i$ from transaction $\trans[i]$ and $a_j$ from transaction
    $\trans[j]$, and with $P_i$ and $P_j$ the programs in $\progset$ from which
    $\trans[i]$ and $\trans[j]$ were instantiated, and $q_i$ and $q_j$ the
    statements in respectively $P_i$ and $P_j$ from which operations $b_i$ and
    $a_j$ were instantiated. Then it is implied by
    Proposition~\ref{prop:algocorrectness} that there is an edge $(P_i, q_i, c,
    q_j, P_j)$. Furthermore, if $\dependson[\schedule]{b_i}{a_j}$ is
    counterflow then we can assume that $c =$ counterflow, and if
    $\dependson[\schedule]{b_i}{a_j}$ is non-counterflow, that $c =$
    non-counterflow. Since statements are instantiated as atomic chunks, all
    properties of the theorem now indeed follow straightforwardly from
    Theorem~\ref{theo:cg:cycles}. 
\end{proof}

\subsection*{Proof of Proposition~\ref{prop:soundness}}
\begin{proof}
    The result follows from Theorem~\ref{theo:sufcond:robustness}, as Algorithm~\ref{alg:robust:check}
    checks quite literally its conditions. More precisely, Algorithm~\ref{alg:robust:check} first computes $\sg{\progset}$, using Algorithm~\ref{alg:graph},
    with properties defined in Condition~\ref{cond:sg:prop} (cf, Proposition~\ref{prop:algocorrectness}).
    Then it searches for cycles with the properties of Theorem~\ref{theo:sufcond:robustness} on $\sg{\progset}$.
    For cycles with the first condition, we notice that existence of two adjacent counterflow edges implies existence of two adjacent counterflow edges $(P_3, q_3, \textit{counterflow}, q_4, P_4), (P_4, q'_4, \textit{counterflow}, q_5, P_5)$ that are preceded by a non-counterflow edge $(P_1, q_1, \textit{non-counterflow}, q_2, P_3)$. (Notice that $P_3$ here is intentional). Hence such a cycle will get detected by the algorithm. 
    Towards cycles with the second condition, let $(P_{i-1}, q_{i-1}, \textit{non-counterflow}, q_i, P_i)$ and $(P_i, q'_i, \textit{counterflow}, q_{i+1}, P_{i+1})$ be the pair of edges as specified by Theorem~\ref{theo:sufcond:robustness}. To show that the algorithm will detect such a cycle, we can assign $(P_{i-1}, q_{i-1}, \textit{non-counterflow}, q_i, P_i)$ to edges $(P_1, q_1, \textit{non-counterflow}, q_2, P_2)$ and $(P_3, q_3, \textit{c}, q_4, P_4)$, and assign $(P_i, q'_i, \textit{counterflow}, q_{i+1}, P_{i+1})$ to $(P_4, q_4', \textit{counterflow}, q_5, P_5)$. Note in particular that $P_3 = P_{i-1}$ is reachable from $P_2 = P_{i}$, as $P_{i-1}$ and $P_i$ are part of a cycle. The result is now immediate.
\end{proof}


\section{Benchmarks}
\label{sec:appendix:benchmarks}

\subsection{SmallBank Benchmark}
\label{sec:app:smallbank}

The SmallBank benchmark~\cite{Alomari:2008:CSP:1546682.1547288} is defined over a database schema consisting of three relations (underlined attributes are primary keys):
\begin{itemize}
    \item Account(\underline{Name}, CustomerID);
    \item Savings(\underline{CustomerID}, Balance); and
    \item Checking(\underline{CustomerID}, Balance).
\end{itemize}
The \Account table associates customer names with IDs; CustomerID is a {\tt UNIQUE} attribute. The other tables contain the balance (numeric value) of the savings and checking accounts of customers identified by their ID. {\Account(CustomerID) is a foreign key referencing both the columns
\Savings(CustomerID) and \Checking(CustomerID).
The application code can interact with the database only through the following transaction programs:
\begin{itemize}
    \item  Balance($N$): returns the total balance (savings \& checking) for a customer with name $N$.

    \item  DepositChecking($N$,$V$): makes a deposit of amount $V$ 
    on the checking account of the customer with name $N$.

    \item TransactSavings($N$,$V$): makes a deposit or withdrawal $V$ on the savings account of the customer with name $N$.

    \item Amalgamate($N_1$,$N_2$): transfers all the funds from 
    $N_1$ to 
    $N_2$.

    \item WriteCheck($N$,$V$): writes a check $V$ against the account of the customer with name $N$, penalizing if overdrawing.
\end{itemize}
The SQL code for each transaction program is given in 
Figure~\ref{fig:smallbank:SQL}.
The corresponding BTPs are summarized in Figure~\ref{fig:smallbank:btp},
and the summary graph constructed for this benchmark is visualized in Figure~\ref{fig:smallbank:sg:fullscale}.

\begin{figure*}
\small
\begin{minipage}[t]{\textwidth/2-2ex}
\begin{verbatim}
Balance(N):
    SELECT CustomerId INTO :x
      FROM Account
     WHERE Name=:N;
        
    SELECT Balance INTO :a 
      FROM Savings
     WHERE CustomerId=:x;
     
    SELECT Balance + :a 
      FROM Checking
     WHERE CustomerId=:x;
    COMMIT;

Amalgamate(N1,N2):
    SELECT CustomerId INTO :x1
      FROM Account
     WHERE Name=:N1;
    
    SELECT CustomerId INTO :x2
      FROM Account
     WHERE Name=:N2;
    
    UPDATE Savings AS new
       SET Balance = 0
      FROM Savings AS old
     WHERE new.CustomerId=:x1
           AND old.CustomerId
           = new.CustomerId
    RETURNING old.Balance INTO :a;
    
    UPDATE Checking AS new
       SET Balance = 0
      FROM Checking AS old
     WHERE new.CustomerId=:x1
           AND old.CustomerId
           = new.CustomerId
    RETURNING old.Balance INTO :b;
    
    UPDATE Checking
       SET Balance = Balance + :a + :b
     WHERE CustomerId=:x2;
\end{verbatim}
\end{minipage}   
\begin{minipage}[t]{\textwidth/2}
\begin{verbatim}
DepositChecking(N,V):
    SELECT CustomerId INTO :x
      FROM Account
     WHERE Name=:N;
     
    UPDATE Checking
       SET Balance = Balance + :V 
     WHERE CustomerId=:x;
    COMMIT;

TransactSavings(N,V):
    SELECT CustomerId INTO :x
      FROM Account
     WHERE Name=:N;
        
    UPDATE Savings
       SET Balance = Balance + :V 
     WHERE CustomerId=:x;
    COMMIT;

WriteCheck(N,V):
    SELECT CustomerId INTO :x
      FROM Account
     WHERE Name=:N;
        
    SELECT Balance INTO :a 
      FROM Savings
     WHERE CustomerId=:x;
     
    SELECT Balance INTO :b 
      FROM Checking
     WHERE CustomerId=:x;
     
    IF (:a + :b) < :V THEN 
        :V = :V + 1
    END IF;
    
    UPDATE Checking
       SET Balance = Balance - :V
     WHERE CustomerId=:x;
    COMMIT;
\end{verbatim}
\end{minipage}
\caption{SmallBank SQL Transaction Templates.}
\label{fig:smallbank:SQL}
\end{figure*}

\begin{figure}
{\small
\begin{tabular}[t]{| l | l | l | l | l | l |}
    \hline
    $q$ & $\type{q}$ &  $\rel{q}$ & $\predset{q}$ & $\obsset{q}$ & $\modset{q}$\\
    \hline
    \hline
    \multicolumn{6}{|c|}{\bf Amalgamate $:= q_{1}; q_{2}; q_{3}; q_{4}; q_{5}$}\\
    \hline
    $q_{1}$ & $\qksel$ & Account & $\bot$ & \{CustomerId\} & $\bot$\\
    \hline
    $q_{2}$ & $\qksel$ & Account & $\bot$ & \{CustomerId\} & $\bot$\\
    \hline
    $q_{3}$ & $\qkupd$ & Savings & $\bot$ & \{Balance\} & \{Balance\}\\
    \hline
    $q_{4}$ & $\qkupd$ & Checking & $\bot$ & \{Balance\} & \{Balance\}\\
    \hline
    $q_{5}$ & $\qkupd$ & Checking & $\bot$ & \{Balance\} & \{Balance\}\\
    \hline
    \multicolumn{6}{|c|}{\bf Balance $:= q_{6}; q_{7}; q_{8}$}\\
    \hline
    $q_{6}$ & $\qksel$ & Account & $\bot$ & \{CustomerId\} & $\bot$\\
    \hline
    $q_{7}$ & $\qksel$ & Savings & $\bot$ & \{Balance\} & $\bot$\\
    \hline
    $q_{8}$ & $\qksel$ & Checking & $\bot$ & \{Balance\} & $\bot$\\
    \hline
    \multicolumn{6}{|c|}{\bf DepositChecking $:= q_{9}; q_{10}$}\\
    \hline
    $q_{9}$ & $\qksel$ & Account & $\bot$ & \{CustomerId\} & $\bot$\\
    \hline
    $q_{10}$ & $\qkupd$ & Checking & $\bot$ & \{Balance\} & \{Balance\}\\
    \hline
    \multicolumn{6}{|c|}{\bf TransactSavings $:= q_{11}; q_{12}$}\\
    \hline
    $q_{11}$ & $\qksel$ & Account & $\bot$ & \{CustomerId\} & $\bot$\\
    \hline
    $q_{12}$ & $\qkupd$ & Savings & $\bot$ & \{Balance\} & \{Balance\}\\
    \hline
    \multicolumn{6}{|c|}{\bf WriteCheck $:= q_{13}; q_{14}; q_{15}; q_{16}$}\\
    \hline
    $q_{13}$ & $\qksel$ & Account & $\bot$ & \{CustomerId\} & $\bot$\\
    \hline
    $q_{14}$ & $\qksel$ & Savings & $\bot$ & \{Balance\} & $\bot$\\
    \hline
    $q_{15}$ & $\qksel$ & Checking & $\bot$ & \{Balance\} & $\bot$\\
    \hline
    $q_{16}$ & $\qkupd$ & Checking & $\bot$ & \{Balance\} & \{Balance\}\\
    \hline
    \end{tabular}
}
\caption{BTPs and statement details for the SmallBank benchmark.
\label{fig:smallbank:btp}}
\end{figure}

\begin{figure}
    \resizebox{.5\textwidth}{!}{
    \begin{tikzpicture}[sgnode/.style={draw}, sgedge/.style={draw,->},sgcedge/.style={sgedge,dashed},qnode/.style={font=\footnotesize}]
    \node[sgnode] (Amalgamate) at (-3,-2) {Amalgamate};
    \node[sgnode] (DepositChecking) at (-3,2) {DepositChecking};
    \node[sgnode] (WriteCheck) at (3,2) {WriteCheck};
    \node[sgnode] (TransactSavings) at (3,-2) {TransactSavings};
    \node[sgnode] (Balance) at (0,0) {Balance};

    \path[sgedge] (TransactSavings) to[loop below]
     (TransactSavings);
    
    \path[sgedge] (TransactSavings) to[bend left=40]
     (WriteCheck);
    \path[sgcedge] (WriteCheck) to[bend left=20]
     (TransactSavings);
    \path[sgedge] (WriteCheck) to[bend left=40]
     (TransactSavings);
    
    \path[sgedge] (TransactSavings) to[bend left=40]
     (Balance);
    \path[sgcedge] (Balance) to[bend left=20]
     (TransactSavings);
    \path[sgedge] (Balance) to[bend left=40]
     (TransactSavings);
    
    \path[sgedge] (TransactSavings) to[bend left=40]
     (Amalgamate);
    \path[sgedge] (Amalgamate) to[bend left=40]
     (TransactSavings);
    
    
    \path[sgedge] (WriteCheck) to[loop above,in=75,out=105,looseness=8]
     (WriteCheck);
    \path[sgedge] (WriteCheck) to[loop above,in=73,out=107,looseness=10]
     (WriteCheck);
    \path[sgedge] (WriteCheck) to[loop above,in=71,out=109,looseness=12]
     (WriteCheck);
    \path[sgcedge] (WriteCheck) to[loop right]
     (WriteCheck);
    
    \path[sgedge] (WriteCheck) to[bend left=40]
     (Balance);
    \path[sgcedge] (Balance) to[bend left=20]
     (WriteCheck);
    \path[sgedge] (Balance) to[bend left=40]
     (WriteCheck);
    
    \path[sgedge] (WriteCheck) to[bend left=32]
     (Amalgamate);
    \path[sgedge] (WriteCheck) to[bend left=34]
     (Amalgamate);
     \path[sgedge] (WriteCheck) to[bend left=36]
     (Amalgamate);
     \path[sgcedge] (WriteCheck) to[bend left=16]
     (Amalgamate);
    \path[sgcedge] (WriteCheck) to[bend left=18]
     (Amalgamate);
    \path[sgcedge] (WriteCheck) to[bend left=20]
     (Amalgamate);
    \path[sgedge] (WriteCheck) to[bend left=38]
     (Amalgamate);
    \path[sgedge] (WriteCheck) to[bend left=40]
     (Amalgamate);
    \path[sgedge] (Amalgamate) to[bend left=40]
     (WriteCheck);
    \path[sgedge] (Amalgamate) to[bend left=38]
     (WriteCheck);
    \path[sgedge] (Amalgamate) to[bend left=36]
     (WriteCheck);
    \path[sgedge] (Amalgamate) to[bend left=34]
     (WriteCheck);
    \path[sgedge] (Amalgamate) to[bend left=32]
     (WriteCheck);
    
    \path[sgedge] (WriteCheck) to[bend left=40]
     (DepositChecking);
    \path[sgedge] (WriteCheck) to[bend left=38]
     (DepositChecking);
    \path[sgcedge] (WriteCheck) to[bend left=20]
     (DepositChecking);
    \path[sgedge] (DepositChecking) to[bend left=40]
     (WriteCheck);
    \path[sgedge] (DepositChecking) to[bend left=38]
     (WriteCheck);
    
    
    \path[sgcedge] (Balance) to[bend left=20]
     (Amalgamate);
    \path[sgcedge] (Balance) to[bend left=18]
     (Amalgamate);
    \path[sgcedge] (Balance) to[bend left=16]
     (Amalgamate);
    \path[sgedge] (Balance) to[bend left=40]
     (Amalgamate);
    \path[sgedge] (Balance) to[bend left=38]
     (Amalgamate);
    \path[sgedge] (Balance) to[bend left=36]
     (Amalgamate);
    \path[sgedge] (Amalgamate) to[bend left=40]
     (Balance);
    \path[sgedge] (Amalgamate) to[bend left=38]
     (Balance);
    \path[sgedge] (Amalgamate) to[bend left=36]
     (Balance);
    
    \path[sgcedge] (Balance) to[bend left=20]
     (DepositChecking);
    \path[sgedge] (Balance) to[bend left=40]
     (DepositChecking);
    \path[sgedge] (DepositChecking) to[bend left=40]
     (Balance);
    
    \path[sgedge] (Amalgamate) to[loop below,in=255,out=285,looseness=8]
     (Amalgamate);
    \path[sgedge] (Amalgamate) to[loop below,in=253,out=287,looseness=10]
     (Amalgamate);
    \path[sgedge] (Amalgamate) to[loop below,in=251,out=289,looseness=12]
     (Amalgamate);
    \path[sgedge] (Amalgamate) to[loop below,in=249,out=291,looseness=14]
     (Amalgamate);
    \path[sgedge] (Amalgamate) to[loop below,in=247,out=293,looseness=16]
     (Amalgamate);
    
    \path[sgedge] (Amalgamate) to[bend left=40]
     (DepositChecking);
    \path[sgedge] (Amalgamate) to[bend left=38]
     (DepositChecking);
    \path[sgedge] (DepositChecking) to[bend left=40]
     (Amalgamate);
    \path[sgedge] (DepositChecking) to[bend left=38]
     (Amalgamate);
    
    \path[sgedge] (DepositChecking) to[loop above]
     (DepositChecking);
\end{tikzpicture}
    } 
    
    \caption{Summary graph for the SmallBank benchmark. Counterflow edges are represented by dashed edges. To facilitate the presentation, edge labels are not visualized.}
  \label{fig:smallbank:sg:fullscale}
  \end{figure}
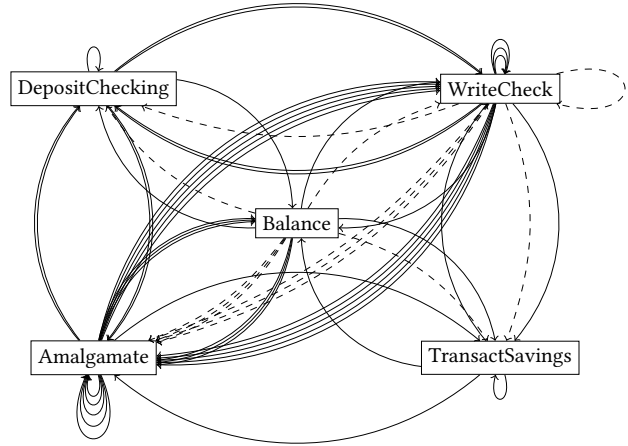

\subsection{TPC-C Benchmark}

The database schema of the TPC-C benchmark~\cite{TPCC} consists of nine relations (underlined attributes are primary keys):
\begin{itemize}
    \item Warehouse(\underline{w\_id}, w\_name, w\_street\_1, w\_street\_2, w\_city, w\_state, w\_zip, w\_tax, w\_ytd),
    \item District(\underline{d\_id, d\_w\_id}, d\_name, d\_street\_1, d\_street\_2, d\_city, d\_state, d\_zip, d\_tax, d\_ytd, d\_next\_o\_id),
    \item Customer(\underline{c\_id, c\_d\_id, c\_w\_id}, c\_first, c\_middle, c\_last, c\_street\_1, c\_street\_2, c\_city, c\_state, c\_zip, c\_phone, c\_since, c\_credit, c\_credit\_lim, c\_discount, c\_balance, c\_ytd\_payment, c\_payment\_cnt, c\_delivery\_cnt, c\_data),
    \item History(\underline{h\_c\_id, h\_c\_d\_id, h\_c\_w\_id, h\_d\_id, h\_w\_id}, h\_date, h\_amount, h\_data),
    \item New\_Order(\underline{no\_o\_id, no\_d\_id, no\_w\_id}),
    \item Orders(\underline{o\_id, o\_d\_id, o\_w\_id}, o\_c\_id, o\_entry\_id, o\_carrier\_id, o\_ol\_cnt, o\_all\_local),
    \item Order\_Line(\underline{ol\_o\_id, ol\_d\_id, ol\_w\_id, ol\_number}, ol\_i\_id, ol\_supply\_w\_id, ol\_delivery\_d, ol\_quantity, ol\_amount, ol\_dist\_info),
    \item Item(\underline{i\_id}, i\_im\_id, i\_name, i\_price, i\_data),
    \item Stock(\underline{s\_i\_id, s\_w\_id}, s\_quantity, s\_dist\_01, s\_dist\_02, s\_dist\_03, s\_dist\_04, s\_dist\_05, s\_dist\_06, s\_dist\_07, s\_dist\_08, s\_dist\_09, s\_dist\_10, s\_ytd, s\_order\_cnt, s\_remote\_cnt, s\_data).
\end{itemize}

The foreign keys are as follows:
\begin{itemize}
  \item $f_1$: District(d\_w\_id) $\rightarrow$ Warehouse(w\_id),
  \item $f_2$: Customer(c\_d\_id, c\_w\_id) $\rightarrow$ District(d\_id, d\_w\_id),
  \item $f_3$: History(h\_c\_id, h\_c\_d\_id, h\_c\_w\_id) $\rightarrow$ Customer(c\_id, c\_d\_id, c\_w\_id),
  \item $f_4$: History(h\_d\_id, h\_w\_id) $\rightarrow$ District(d\_id, d\_w\_id),
  \item $f_5$: New\_Order(no\_o\_id, no\_d\_id, no\_w\_id) $\rightarrow$ Orders(o\_id, o\_d\_id, o\_w\_id),
  \item $f_6$: Orders(o\_d\_id, o\_w\_id) $\rightarrow$ District(d\_id, d\_w\_id),
  \item $f_7$: Orders(o\_c\_id,o\_d\_id, o\_w\_id) $\rightarrow$ Customer(c\_id, c\_d\_id, c\_w\_id),
  \item $f_8$: Order\_Line(ol\_o\_id, ol\_d\_id, ol\_w\_id) $\rightarrow$ Orders(o\_id, o\_d\_id, o\_w\_id),
  \item $f_9$: Order\_Line(ol\_i\_id) $\rightarrow$ Item(i\_id),
  \item $f_{10}$: Order\_Line(ol\_suplpy\_w\_id) $\rightarrow$ Warehouse(w\_id),
  \item $f_{11}$: Stock(s\_i\_id) $\rightarrow$ Item(i\_id),
  \item $f_{12}$: Stock(s\_w\_id) $\rightarrow$ Warehouse(w\_id).
\end{itemize}

The TPC-C benchmark~\cite{TPCC} defines five different transaction programs that can be executed.
Below, we give an informal description of each program, and refer to~\cite{TPCC} for a more formal description:
\begin{enumerate}
    \item NewOrder (SQL code in Figure~\ref{fig:tpcc:SQL:neworder}): creates a new order for a given customer. The id for this order is obtained by increasing the d\_next\_o\_id attribute of the corresponding District tuple by one. Each order consists of a number of items with respective quantities. For each of these items, a new Order\_Line tuple is created and the related stock quantity is decreased.
    \item Payment (SQL code in Figure~\ref{fig:tpcc:SQL:payment}): represents a customer identified paying an amount. This payment is reflected in the database by increasing the balance of this customer. This amount is furthermore added to the YearToDate income of both the related warehouse and district.
    \item OrderStatus (SQL code in Figure~\ref{fig:tpcc:SQL:orderstatus}): collects information of the most recent order placed by a given customer.
    \item Delivery (SQL code in Figure~\ref{fig:tpcc:SQL:delivery}): delivers 10 open orders. The status of each order is updated. The total price of each order is deduced from the balance of the customer who placed this order.
    \item StockLevel (SQL code in Figure~\ref{fig:tpcc:SQL:stocklevel}): determines which recently sold items have a stock level below a specified threshold.
\end{enumerate}

The derived set of BTPs is given in Figure~\ref{fig:tpcc:btp}, and the constructed summary graph for this benchmark is visualized in Figure~\ref{fig:tpc-c:sg}.

\begin{figure*}
\small
\begin{minipage}[t]{\textwidth-2ex}
\begin{verbatim}
  NewOrder:

    SELECT c_discount, c_last, c_credit INTO :c_discount, :c_last, :c_credit
    FROM customer
    WHERE c_w_id = :w_id AND c_d_id = :d_id AND c_id = :c_id;
    
    SELECT w_tax INTO :w_tax
    FROM warehouse
    WHERE w_id = :w_id;
    
    UPDATE district
    SET d_next_o_id = d_next_o_id + 1
    WHERE d_id = :d_id AND d_w_id = :w_id
    RETURNING d_next_o_id, d_tax INTO :o_id, :d_tax
    
    INSERT INTO ORDERS (o_id, o_d_id, o_w_id, o_c_id, o_entry_d, o_ol_cnt, o_all_local)
    VALUES (:o_id , :d _id , :w _id , :c_id , :datetime, :o_ol_cnt, :o_all_local);
    
    INSERT INTO NEW_ORDER (no_o_id, no_d_id, no_w_id)
    VALUES (:o_id , :d _id , :w _id );
    
    FOR each item in the order:

      SELECT i_price, i_name , i_data INTO :i_price, :i_name, :i_data
      FROM item
      WHERE i_id = :ol_i_id;
      
      UPDATE stock
      SET s_quantity = :ol_quantity, s_ytd = :s_ytd, s_order_cnt = :s_order_cnt, s_remote_cnt = :s_remote_cnt
      WHERE s_i_id = :ol_i_id AND s_w_id = :ol_supply_w_id
      RETURNING s_quantity, s_ytd, s_order_cnt, s_remote_cnt, s_data, s_dist_01, s_dist_02, s_dist_03, s_dist_04, s_dist_05,
      s_dist_06, s_dist_07, s_dist_08, s_dist_09, s_dist_10
      INTO :s_quantity, :s_ytd, :s_order_cnt, :s_remote_cnt, :s_data, :s_dist_01, :s_dist_02, :s_dist_03, :s_dist_04,
      :s_dist_05, :s_dist_06, :s_dist_07, :s_dist_08, :s_dist_09, :s_dist_10;
    
      INSERT INTO order_line (ol_o_id, ol_d_id, ol_w_id, ol_number, ol_i_id, ol_supply_w_id, ol_quantity, ol_amount,
                              ol_dist_info)
      VALUES (:o_id, :d_id, :w_id, :ol_number, :ol_i_id, :ol_supply_w_id, :ol_quantity, :ol_amount, :ol_dist_info);
    
    ENDFOR
    
    COMMIT;
\end{verbatim}
\end{minipage}   
\caption{SQL code for NewOrder program in TPC-C benchmark.}
\label{fig:tpcc:SQL:neworder}
\end{figure*}

\begin{figure*}
\small
\begin{minipage}[t]{\textwidth-2ex}
\begin{verbatim}
  Payment:

    UPDATE warehouse
    SET w_ytd = w_ytd + :h_amount
    WHERE w_id=:w_id
    RETURNING w_street_1, w_street_2, w_city, w_state, w_zip, w_name
    INTO :w_street_1, :w_street_2, :w_city, :w_state, :w_zip, :w_name;

    UPDATE district SET d_ytd = d_ytd + :h_amount
    WHERE d_w_id=:w_id AND d_id=:d_id
    RETURNING d_street_1, d_street_2, d_city, d_state, d_zip, d_name
    INTO :d_street_1, :d_street_2, :d_city, :d_state, :d_zip, :d_name;

    IF <selection of customer by name instead of ID>:

      SELECT c_id
      INTO :c_id
      FROM customer
      WHERE c_w_id=:c_w_id AND c_d_id=:c_d_id AND c_last=:c_last;

    ENDIF

    UPDATE customer
    SET c_balance = c_balance - :h_amount,
        c_ytd_payment = c_ytd_payment + :h_amount,
        c_payment_cnt = c_payment_cnt + 1
    WHERE c_w_id = :c_w_id AND c_d_id = :c_d_id AND c_id = :c_id
    RETURNING c_first, c_middle, c_last, c_street_1, c_street_2, c_city, c_state, c_zip, c_phone, c_credit, c_credit_lim,
              c_discount, c_balance, c_since
    INTO :c_first, :c_middle, :c_last, :c_street_1, :c_street_2, :c_city, :c_state, :c_zip, :c_phone, :c_credit,
         :c_credit_lim, :c_discount, :c_balance, :c_since;

    IF <c_credit == "BC">:
      
      SELECT c_data
      INTO :c_data
      FROM customer
      WHERE c_w_id=:c_w_id AND c_d_id=:c_d_id AND c_id=:c_id;

      UPDATE customer
      SET c_data = :c_new_data
      WHERE c_w_id = :c_w_id AND c_d_id = :c_d_id AND c_id = :c_id;

    ENDIF

    INSERT INTO history (h_c_d_id, h_c_w_id, h_c_id, h_d_id, h_w_id, h_date, h_amount, h_data)
    VALUES (:c_d_id, :c_w_id, :c_id, :d_id, :w_id, :datetime, :h_amount, :h_data);

    COMMIT;
\end{verbatim}
\end{minipage}   
\caption{SQL code for Payment program in TPC-C benchmark.}
\label{fig:tpcc:SQL:payment}
\end{figure*}

\begin{figure*}
\small
\begin{minipage}[t]{\textwidth-2ex}
\begin{verbatim}
  OrderStatus:

    IF <selection of customer by name instead of ID>:

      SELECT c_balance, c_first, c_middle, c_id INTO :c_balance, :c_first, :c_middle, :c_id
      FROM customer
      WHERE c_last=:c_last AND c_d_id=:d_id AND c_w_id=:w_id;

    ELSE:

      SELECT c_balance, c_first, c_middle, c_last
      INTO :c_balance, :c_first, :c_middle, :c_last
      FROM customer
      WHERE c_id=:c_id AND c_d_id=:d_id AND c_w_id=:w_id;

    ENDIF

    SELECT o_id, o_carrier_id, o_entry_id
    INTO :o_id, :o_carrier_id, :entdate
    FROM orders
    WHERE o_w_id=:w_id AND o_d_id=:d_id AND o_c_id=:c_id;

    SELECT ol_i_id, ol_supply_w_id, ol_quantity,
    ol_amount, ol_delivery_d
    FROM order_line
    WHERE ol_o_id=:o_id AND ol_d_id=:d_id AND ol_w_id=:w_id;

    COMMIT;
\end{verbatim}
\end{minipage}   
\caption{SQL code for OrderStatus program in TPC-C benchmark.}
\label{fig:tpcc:SQL:orderstatus}
\end{figure*}

\begin{figure*}
\small
\begin{minipage}[t]{\textwidth-2ex}
\begin{verbatim}
  Delivery:

    FOR each district:

      SELECT no_o_id INTO :no_o_id
      FROM new_order
      WHERE no_d_id = :d_id AND no_w_id = :w_id;

      DELETE FROM new_order
      WHERE no_o_id = :no_o_id AND no_d_id = :d_id AND no_w_id = :w_id;

      SELECT o_c_id INTO :c_id
      FROM orders
      WHERE o_id = :no_o_id AND o_d_id = :d_id AND o_w_id = :w_id;

      UPDATE orders
      SET o_carrier_id = :o_carrier_id
      WHERE o_id = :no_o_id AND o_d_id = :d_id AND o_w_id = :w_id;

      UPDATE order_line
      SET ol_delivery_d = :datetime
      WHERE ol_o_id = :no_o_id AND ol_d_id = :d_id AND ol_w_id = :w_id;

      SELECT ol_amount
      FROM order_line
      WHERE ol_o_id = :no_o_id AND ol_d_id = :d_id AND ol_w_id = :w_id;

      UPDATE customer
      SET c_balance = c_balance + :ol_total, c_delivery_cnt += 1
      WHERE c_id = :c_id AND c_d_id = :d_id AND c_w_id = :w_id;

    ENDFOR

    COMMIT;
\end{verbatim}
\end{minipage}   
\caption{SQL code for Delivery program in TPC-C benchmark.}
\label{fig:tpcc:SQL:delivery}
\end{figure*}

\begin{figure*}
\small
\begin{minipage}[t]{\textwidth-2ex}
\begin{verbatim}
  StockLevel:

    SELECT d_next_o_id INTO :o_id
    FROM district
    WHERE d_w_id=:w_id AND d_id=:d_id;

    SELECT ol_i_id
    FROM order_line
    WHERE ol_w_id=:w_id AND ol_d_id=:d_id AND
    ol_o_id<:o_id AND ol_o_id>=:o_id-20

    SELECT s_i_id
    FROM stock
    WHERE s_w_id=:w_id AND
    s_quantity < :threshold;

    COMMIT;
\end{verbatim}
\end{minipage}   
\caption{SQL code for StockLevel program in TPC-C benchmark.}
\label{fig:tpcc:SQL:stocklevel}
\end{figure*}

\begin{figure}
{\footnotesize
\begin{tabular}[t]{| l | l | l | l | l | l |}
    \hline
    $q$ & $\type{q}$ &  $\rel{q}$ & $\predset{q}$ & $\obsset{q}$ & $\modset{q}$\\
    \hline
    \hline
    \multicolumn{6}{|c|}{\textbf{Delivery} $:= \text{loop}(q_{1}; q_{2}; q_{3}; q_{4}; q_{5}; q_{6}; q_{7})$}\\
    \hline
    $q_{1}$ & $\qpsel$ & New\_Order & \{no\_d\_id, no\_w\_id\} & \{no\_o\_id\} & $\bot$\\
    \hline
    $q_{2}$ & $\qkdel$ & New\_Order & $\bot$ & $\bot$ & \{no\_d\_id, no\_o\_id, no\_w\_id\}\\
    \hline
    $q_{3}$ & $\qksel$ & Orders & $\bot$ & \{o\_c\_id\} & $\bot$\\
    \hline
    $q_{4}$ & $\qkupd$ & Orders & $\bot$ & \{\} & \{o\_carrier\_id\}\\
    \hline
    $q_{5}$ & $\qpupd$ & Order\_Line & \{ol\_d\_id, ol\_o\_id, & \{\} & \{ol\_delivery\_d\}\\
     & & & ol\_w\_id\} & & \\
    \hline
    $q_{6}$ & $\qpsel$ & Order\_Line & \{ol\_d\_id, ol\_o\_id, & \{ol\_amount\} & $\bot$\\
     & & & ol\_w\_id\} & & \\
    \hline
    $q_{7}$ & $\qkupd$ & Customer & $\bot$ & \{c\_balance, c\_delivery\_cnt\} & \{c\_balance, c\_delivery\_cnt\}\\
    \hline
    \multicolumn{6}{|c|}{\textbf{NewOrder} $:= q_{8}; q_{9}; q_{10}; q_{11}; q_{12}; \text{loop}(q_{13}; q_{14}; q_{15})$}\\
    \hline
    $q_{8}$ & $\qksel$ & Customer & $\bot$ & \{c\_credit, c\_discount, c\_last\} & $\bot$\\
    \hline
    $q_{9}$ & $\qksel$ & warehouse & $\bot$ & \{w\_tax\} & $\bot$\\
    \hline
    $q_{10}$ & $\qkupd$ & District & $\bot$ & \{d\_next\_o\_id, d\_tax\} & \{d\_next\_o\_id\}\\
    \hline
    $q_{11}$ & $\qins$ & Orders & $\bot$ & $\bot$ & \{o\_all\_local, o\_c\_id, o\_d\_id, o\_entry\_id, o\_id, \\
     & & & & & o\_ol\_cnt, o\_w\_id\}\\
    \hline
    $q_{12}$ & $\qins$ & New\_Order & $\bot$ & $\bot$ & \{no\_d\_id, no\_o\_id, no\_w\_id\}\\
    \hline
    $q_{13}$ & $\qksel$ & Item & $\bot$ & \{i\_data, i\_name, i\_price\} & $\bot$\\
    \hline
    $q_{14}$ & $\qkupd$ & Stock & $\bot$ & \{s\_data, s\_dist\_01, s\_dist\_02, s\_dist\_03, s\_dist\_04, & \{s\_order\_cnt, s\_quantity, s\_remote\_cnt, s\_ytd\}\\
     & & & & s\_dist\_05, s\_dist\_06, s\_dist\_07, s\_dist\_08, s\_dist\_09, & \\
     & & & & s\_dist\_10, s\_order\_cnt, s\_quantity, s\_remote\_cnt, s\_ytd\} & \\
    \hline
    $q_{15}$ & $\qins$ & Order\_Line & $\bot$ & $\bot$ & \{ol\_amount, ol\_d\_id, ol\_dist\_info, ol\_i\_id, ol\_number, \\
     & & & & & ol\_o\_id, ol\_quantity, ol\_supply\_w\_id, ol\_w\_id\}\\
    \hline
    \multicolumn{6}{|c|}{\textbf{OrderStatus} $:= (q_{16} \mid q_{17}); q_{18}; q_{19}$}\\
    \hline
    $q_{16}$ & $\qpsel$ & Customer & \{c\_d\_id, c\_last, & \{c\_balance, c\_first, c\_id, c\_middle\} & $\bot$\\
     & & & c\_w\_id\} & & \\
    \hline
    $q_{17}$ & $\qksel$ & Customer & $\bot$ & \{c\_balance, c\_first, c\_last, c\_middle\} & $\bot$\\
    \hline
    $q_{18}$ & $\qpsel$ & Orders & \{o\_c\_id, o\_d\_id, & \{o\_carrier\_id, o\_entry\_id, o\_id\} & $\bot$\\
     & & & o\_w\_id\} & & \\
    \hline
    $q_{19}$ & $\qpsel$ & Order\_Line & \{ol\_d\_id, ol\_o\_id, & \{ol\_amount, ol\_delivery\_d, ol\_i\_id, ol\_quantity, & $\bot$\\
     & & & ol\_w\_id\} & ol\_supply\_w\_id\} & \\
    \hline
    \multicolumn{6}{|c|}{\textbf{Payment} $:= q_{20}; q_{21}; (q_{22} \mid \varepsilon); q_{23}; (q_{24}; q_{25} \mid \varepsilon); q_{26}$}\\
    \hline
    $q_{20}$ & $\qkupd$ & warehouse & $\bot$ & \{w\_city, w\_name, w\_state, w\_street\_1, w\_street\_2, & \{w\_ytd\}\\
     & & & & w\_ytd, w\_zip\} & \\
    \hline
    $q_{21}$ & $\qkupd$ & District & $\bot$ & \{d\_city, d\_name, d\_state, d\_street\_1, d\_street\_2, & \{d\_ytd\}\\
     & & & & d\_ytd, d\_zip\} & \\
    \hline
    $q_{22}$ & $\qpsel$ & Customer & \{c\_d\_id, c\_last, & \{c\_id\} & $\bot$\\
     & & & c\_w\_id\} & & \\
    \hline
    $q_{23}$ & $\qkupd$ & Customer & $\bot$ & \{c\_balance, c\_city, c\_credit, c\_credit\_lim, c\_discount, & \{c\_balance, c\_payment\_cnt, c\_ytd\_payment\}\\
     & & & & c\_first, c\_last, c\_middle, c\_phone, c\_since, c\_state, & \\
     & & & & c\_street\_1, c\_street\_2, c\_ytd\_payment, c\_zip\} & \\
    \hline
    $q_{24}$ & $\qksel$ & Customer & $\bot$ & \{c\_data\} & $\bot$\\
    \hline
    $q_{25}$ & $\qkupd$ & Customer & $\bot$ & \{\} & \{c\_data\}\\
    \hline
    $q_{26}$ & $\qins$ & History & $\bot$ & $\bot$ & \{h\_amount, h\_c\_d\_id, h\_c\_id, h\_c\_w\_id, h\_d\_id, \\
     & & & & & h\_data, h\_date, h\_w\_id\}\\
    \hline
    \multicolumn{6}{|c|}{\textbf{StockLevel} $:= q_{27}; q_{28}; q_{29}$}\\
    \hline
    $q_{27}$ & $\qksel$ & District & $\bot$ & \{d\_next\_o\_id\} & $\bot$\\
    \hline
    $q_{28}$ & $\qpsel$ & Order\_Line & \{ol\_d\_id, ol\_o\_id, & \{ol\_i\_id\} & $\bot$\\
     & & & ol\_w\_id\} & & \\
    \hline
    $q_{29}$ & $\qpsel$ & Stock & \{s\_quantity, s\_w\_id\} & \{s\_i\_id\} & $\bot$\\
    \hline
\end{tabular}
}
\caption{BTPs and statement details for the TPC-C benchmark.
\label{fig:tpcc:btp}}
\end{figure}

\begin{figure}
\begin{tikzpicture}[sgnode/.style={draw}, sgedge/.style={draw,->},sgcedge/.style={sgedge,dashed},qnode/.style={font=\footnotesize}]

  \node[sgnode] (OrderStatus1) at (-5,5) {$\text{OrderStatus}_1$};
  \node[sgnode] (OrderStatus2) at (-2,5) {$\text{OrderStatus}_2$};
  \node[sgnode] (Payment11) at (1,5) {$\text{Payment}_{1,1}$};
  \node[sgnode] (Payment12) at (3,5) {$\text{Payment}_{1,2}$};
  \node[sgnode] (Payment21) at (5,5) {$\text{Payment}_{2,1}$};
  \node[sgnode] (Payment22) at (7,5) {$\text{Payment}_{2,2}$};
  \node[sgnode] (StockLevel) at (-8,-4) {StockLevel};
  \node[sgnode] (NewOrder1) at (-5,-4) {$\text{NewOrder}_{1}$};
  \node[sgnode] (NewOrder2) at (-3,-4) {$\text{NewOrder}_{2}$};
  \node[sgnode] (NewOrder3) at (-1,-4) {$\text{NewOrder}_{3}$};
  \node[sgnode] (Delivery1) at (2,-4) {$\text{Delivery}_{1}$};
  \node[sgnode] (Delivery2) at (4,-4) {$\text{Delivery}_{2}$};
  \node[sgnode] (Delivery3) at (6,-4) {$\text{Delivery}_{3}$};
\path[sgedge] (Delivery2) to[loop below]
(Delivery2);
\path[sgedge] (Delivery2) to[loop below]
(Delivery2);
\path[sgedge] (Delivery2) to[loop below]
(Delivery2);
\path[sgedge] (Delivery2) to[loop below]
(Delivery2);
\path[sgedge] (Delivery2) to[loop below]
(Delivery2);
\path[sgcedge] (Delivery2) to[loop right,out=5,in=355,looseness=10]
(Delivery2);
\path[sgedge] (Delivery2) to[bend left=40]
(Delivery3);
\path[sgedge] (Delivery2) to[bend left=42]
(Delivery3);
\path[sgedge] (Delivery2) to[bend left=44]
(Delivery3);
\path[sgedge] (Delivery2) to[bend left=46]
(Delivery3);
\path[sgedge] (Delivery2) to[bend left=48]
(Delivery3);
\path[sgedge] (Delivery2) to[bend left=50]
(Delivery3);
\path[sgedge] (Delivery2) to[bend left=52]
(Delivery3);
\path[sgedge] (Delivery2) to[bend left=54]
(Delivery3);
\path[sgedge] (Delivery2) to[bend left=56]
(Delivery3);
\path[sgedge] (Delivery2) to[bend left=58]
(Delivery3);
\path[sgcedge] (Delivery2) to[bend left=20]
(Delivery3);
\path[sgcedge] (Delivery2) to[bend left=22]
(Delivery3);
\path[sgedge] (Delivery3) to[bend left=40]
(Delivery2);
\path[sgedge] (Delivery3) to[bend left=42]
(Delivery2);
\path[sgedge] (Delivery3) to[bend left=44]
(Delivery2);
\path[sgedge] (Delivery3) to[bend left=46]
(Delivery2);
\path[sgedge] (Delivery3) to[bend left=48]
(Delivery2);
\path[sgedge] (Delivery3) to[bend left=50]
(Delivery2);
\path[sgedge] (Delivery3) to[bend left=52]
(Delivery2);
\path[sgedge] (Delivery3) to[bend left=54]
(Delivery2);
\path[sgedge] (Delivery3) to[bend left=56]
(Delivery2);
\path[sgedge] (Delivery3) to[bend left=58]
(Delivery2);
\path[sgcedge] (Delivery3) to[bend left=20]
(Delivery2);
\path[sgcedge] (Delivery3) to[bend left=22]
(Delivery2);
\path[sgedge] (Delivery2) to[bend left=40]
(NewOrder1);
\path[sgcedge] (Delivery2) to[bend left=20]
(NewOrder1);
\path[sgedge] (NewOrder1) to[bend left=40]
(Delivery2);
\path[sgedge] (NewOrder1) to[bend left=42]
(Delivery2);
\path[sgedge] (NewOrder1) to[bend left=44]
(Delivery2);
\path[sgedge] (Delivery2) to[bend left=40]
(NewOrder2);
\path[sgedge] (Delivery2) to[bend left=42]
(NewOrder2);
\path[sgedge] (Delivery2) to[bend left=44]
(NewOrder2);
\path[sgcedge] (Delivery2) to[bend left=20]
(NewOrder2);
\path[sgcedge] (Delivery2) to[bend left=22]
(NewOrder2);
\path[sgcedge] (Delivery2) to[bend left=24]
(NewOrder2);
\path[sgedge] (NewOrder2) to[bend left=40]
(Delivery2);
\path[sgedge] (NewOrder2) to[bend left=42]
(Delivery2);
\path[sgedge] (NewOrder2) to[bend left=44]
(Delivery2);
\path[sgedge] (NewOrder2) to[bend left=46]
(Delivery2);
\path[sgedge] (NewOrder2) to[bend left=48]
(Delivery2);
\path[sgedge] (Delivery2) to[bend left=40]
(NewOrder3);
\path[sgedge] (Delivery2) to[bend left=42]
(NewOrder3);
\path[sgedge] (Delivery2) to[bend left=44]
(NewOrder3);
\path[sgedge] (Delivery2) to[bend left=46]
(NewOrder3);
\path[sgedge] (Delivery2) to[bend left=48]
(NewOrder3);
\path[sgcedge] (Delivery2) to[bend left=20]
(NewOrder3);
\path[sgcedge] (Delivery2) to[bend left=22]
(NewOrder3);
\path[sgcedge] (Delivery2) to[bend left=24]
(NewOrder3);
\path[sgcedge] (Delivery2) to[bend left=26]
(NewOrder3);
\path[sgcedge] (Delivery2) to[bend left=28]
(NewOrder3);
\path[sgedge] (NewOrder3) to[bend left=40]
(Delivery2);
\path[sgedge] (NewOrder3) to[bend left=42]
(Delivery2);
\path[sgedge] (NewOrder3) to[bend left=44]
(Delivery2);
\path[sgedge] (NewOrder3) to[bend left=46]
(Delivery2);
\path[sgedge] (NewOrder3) to[bend left=48]
(Delivery2);
\path[sgedge] (NewOrder3) to[bend left=50]
(Delivery2);
\path[sgedge] (NewOrder3) to[bend left=52]
(Delivery2);
\path[sgedge] (Delivery2) to[bend left=15]
(OrderStatus1);
\path[sgedge] (Delivery2) to[bend left=16]
(OrderStatus1);
\path[sgedge] (Delivery2) to[bend left=17]
(OrderStatus1);
\path[sgedge] (OrderStatus1) to[bend left=15]
(Delivery2);
\path[sgedge] (OrderStatus1) to[bend left=16]
(Delivery2);
\path[sgedge] (OrderStatus1) to[bend left=17]
(Delivery2);
\path[sgcedge] (OrderStatus1) to[bend left=5]
(Delivery2);
\path[sgcedge] (OrderStatus1) to[bend left=6]
(Delivery2);
\path[sgcedge] (OrderStatus1) to[bend left=7]
(Delivery2);
\path[sgedge] (Delivery2) to[bend left=15]
(OrderStatus2);
\path[sgedge] (Delivery2) to[bend left=16]
(OrderStatus2);
\path[sgedge] (Delivery2) to[bend left=17]
(OrderStatus2);
\path[sgedge] (OrderStatus2) to[bend left=15]
(Delivery2);
\path[sgedge] (OrderStatus2) to[bend left=16]
(Delivery2);
\path[sgedge] (OrderStatus2) to[bend left=17]
(Delivery2);
\path[sgcedge] (OrderStatus2) to[bend left=5]
(Delivery2);
\path[sgcedge] (OrderStatus2) to[bend left=6]
(Delivery2);
\path[sgcedge] (OrderStatus2) to[bend left=7]
(Delivery2);
\path[sgedge] (Delivery2) to[bend left=15]
(Payment11);
\path[sgedge] (Payment11) to[bend left=15]
(Delivery2);
\path[sgedge] (Delivery2) to[bend left=15]
(Payment12);
\path[sgedge] (Payment12) to[bend left=15]
(Delivery2);
\path[sgedge] (Delivery2) to[bend left=15]
(Payment21);
\path[sgedge] (Payment21) to[bend left=15]
(Delivery2);
\path[sgedge] (Delivery2) to[bend left=15]
(Payment22);
\path[sgedge] (Payment22) to[bend left=15]
(Delivery2);
\path[sgedge] (Delivery3) to[loop below]
(Delivery3);
\path[sgedge] (Delivery3) to[loop below]
(Delivery3);
\path[sgedge] (Delivery3) to[loop below]
(Delivery3);
\path[sgedge] (Delivery3) to[loop below]
(Delivery3);
\path[sgedge] (Delivery3) to[loop below]
(Delivery3);
\path[sgedge] (Delivery3) to[loop below]
(Delivery3);
\path[sgedge] (Delivery3) to[loop below]
(Delivery3);
\path[sgedge] (Delivery3) to[loop below]
(Delivery3);
\path[sgedge] (Delivery3) to[loop below]
(Delivery3);
\path[sgedge] (Delivery3) to[loop below]
(Delivery3);
\path[sgedge] (Delivery3) to[loop below]
(Delivery3);
\path[sgedge] (Delivery3) to[loop below]
(Delivery3);
\path[sgedge] (Delivery3) to[loop below]
(Delivery3);
\path[sgedge] (Delivery3) to[loop below]
(Delivery3);
\path[sgedge] (Delivery3) to[loop below]
(Delivery3);
\path[sgedge] (Delivery3) to[loop below]
(Delivery3);
\path[sgedge] (Delivery3) to[loop below]
(Delivery3);
\path[sgedge] (Delivery3) to[loop below]
(Delivery3);
\path[sgedge] (Delivery3) to[loop below]
(Delivery3);
\path[sgedge] (Delivery3) to[loop below]
(Delivery3);
\path[sgcedge] (Delivery3) to[loop below,out=5,in=355,looseness=10]
(Delivery3);
\path[sgcedge] (Delivery3) to[loop below,out=5,in=355,looseness=10]
(Delivery3);
\path[sgcedge] (Delivery3) to[loop below,out=5,in=355,looseness=10]
(Delivery3);
\path[sgcedge] (Delivery3) to[loop below,out=5,in=355,looseness=10]
(Delivery3);
\path[sgedge] (Delivery3) to[bend left=40]
(NewOrder1);
\path[sgedge] (Delivery3) to[bend left=42]
(NewOrder1);
\path[sgcedge] (Delivery3) to[bend left=20]
(NewOrder1);
\path[sgcedge] (Delivery3) to[bend left=22]
(NewOrder1);
\path[sgedge] (NewOrder1) to[bend left=40]
(Delivery3);
\path[sgedge] (NewOrder1) to[bend left=42]
(Delivery3);
\path[sgedge] (NewOrder1) to[bend left=44]
(Delivery3);
\path[sgedge] (NewOrder1) to[bend left=46]
(Delivery3);
\path[sgedge] (NewOrder1) to[bend left=48]
(Delivery3);
\path[sgedge] (NewOrder1) to[bend left=50]
(Delivery3);
\path[sgedge] (Delivery3) to[bend left=40]
(NewOrder2);
\path[sgedge] (Delivery3) to[bend left=42]
(NewOrder2);
\path[sgedge] (Delivery3) to[bend left=44]
(NewOrder2);
\path[sgedge] (Delivery3) to[bend left=46]
(NewOrder2);
\path[sgedge] (Delivery3) to[bend left=48]
(NewOrder2);
\path[sgedge] (Delivery3) to[bend left=50]
(NewOrder2);
\path[sgcedge] (Delivery3) to[bend left=20]
(NewOrder2);
\path[sgcedge] (Delivery3) to[bend left=22]
(NewOrder2);
\path[sgcedge] (Delivery3) to[bend left=24]
(NewOrder2);
\path[sgcedge] (Delivery3) to[bend left=26]
(NewOrder2);
\path[sgcedge] (Delivery3) to[bend left=28]
(NewOrder2);
\path[sgcedge] (Delivery3) to[bend left=30]
(NewOrder2);
\path[sgedge] (NewOrder2) to[bend left=40]
(Delivery3);
\path[sgedge] (NewOrder2) to[bend left=42]
(Delivery3);
\path[sgedge] (NewOrder2) to[bend left=44]
(Delivery3);
\path[sgedge] (NewOrder2) to[bend left=46]
(Delivery3);
\path[sgedge] (NewOrder2) to[bend left=48]
(Delivery3);
\path[sgedge] (NewOrder2) to[bend left=50]
(Delivery3);
\path[sgedge] (NewOrder2) to[bend left=52]
(Delivery3);
\path[sgedge] (NewOrder2) to[bend left=54]
(Delivery3);
\path[sgedge] (NewOrder2) to[bend left=56]
(Delivery3);
\path[sgedge] (NewOrder2) to[bend left=58]
(Delivery3);
\path[sgedge] (Delivery3) to[bend left=40]
(NewOrder3);
\path[sgedge] (Delivery3) to[bend left=42]
(NewOrder3);
\path[sgedge] (Delivery3) to[bend left=44]
(NewOrder3);
\path[sgedge] (Delivery3) to[bend left=46]
(NewOrder3);
\path[sgedge] (Delivery3) to[bend left=48]
(NewOrder3);
\path[sgedge] (Delivery3) to[bend left=50]
(NewOrder3);
\path[sgedge] (Delivery3) to[bend left=52]
(NewOrder3);
\path[sgedge] (Delivery3) to[bend left=54]
(NewOrder3);
\path[sgedge] (Delivery3) to[bend left=56]
(NewOrder3);
\path[sgedge] (Delivery3) to[bend left=58]
(NewOrder3);
\path[sgcedge] (Delivery3) to[bend left=20]
(NewOrder3);
\path[sgcedge] (Delivery3) to[bend left=22]
(NewOrder3);
\path[sgcedge] (Delivery3) to[bend left=24]
(NewOrder3);
\path[sgcedge] (Delivery3) to[bend left=26]
(NewOrder3);
\path[sgcedge] (Delivery3) to[bend left=28]
(NewOrder3);
\path[sgcedge] (Delivery3) to[bend left=30]
(NewOrder3);
\path[sgcedge] (Delivery3) to[bend left=32]
(NewOrder3);
\path[sgcedge] (Delivery3) to[bend left=34]
(NewOrder3);
\path[sgcedge] (Delivery3) to[bend left=36]
(NewOrder3);
\path[sgcedge] (Delivery3) to[bend left=38]
(NewOrder3);
\path[sgedge] (NewOrder3) to[bend left=40]
(Delivery3);
\path[sgedge] (NewOrder3) to[bend left=42]
(Delivery3);
\path[sgedge] (NewOrder3) to[bend left=44]
(Delivery3);
\path[sgedge] (NewOrder3) to[bend left=46]
(Delivery3);
\path[sgedge] (NewOrder3) to[bend left=48]
(Delivery3);
\path[sgedge] (NewOrder3) to[bend left=50]
(Delivery3);
\path[sgedge] (NewOrder3) to[bend left=52]
(Delivery3);
\path[sgedge] (NewOrder3) to[bend left=54]
(Delivery3);
\path[sgedge] (NewOrder3) to[bend left=56]
(Delivery3);
\path[sgedge] (NewOrder3) to[bend left=58]
(Delivery3);
\path[sgedge] (NewOrder3) to[bend left=60]
(Delivery3);
\path[sgedge] (NewOrder3) to[bend left=62]
(Delivery3);
\path[sgedge] (NewOrder3) to[bend left=64]
(Delivery3);
\path[sgedge] (NewOrder3) to[bend left=66]
(Delivery3);
\path[sgedge] (Delivery3) to[bend left=15]
(OrderStatus1);
\path[sgedge] (Delivery3) to[bend left=16]
(OrderStatus1);
\path[sgedge] (Delivery3) to[bend left=17]
(OrderStatus1);
\path[sgedge] (Delivery3) to[bend left=18]
(OrderStatus1);
\path[sgedge] (Delivery3) to[bend left=19]
(OrderStatus1);
\path[sgedge] (Delivery3) to[bend left=20]
(OrderStatus1);
\path[sgedge] (OrderStatus1) to[bend left=15]
(Delivery3);
\path[sgedge] (OrderStatus1) to[bend left=16]
(Delivery3);
\path[sgedge] (OrderStatus1) to[bend left=17]
(Delivery3);
\path[sgedge] (OrderStatus1) to[bend left=18]
(Delivery3);
\path[sgedge] (OrderStatus1) to[bend left=19]
(Delivery3);
\path[sgedge] (OrderStatus1) to[bend left=20]
(Delivery3);
\path[sgcedge] (OrderStatus1) to[bend left=5]
(Delivery3);
\path[sgcedge] (OrderStatus1) to[bend left=6]
(Delivery3);
\path[sgcedge] (OrderStatus1) to[bend left=7]
(Delivery3);
\path[sgcedge] (OrderStatus1) to[bend left=8]
(Delivery3);
\path[sgcedge] (OrderStatus1) to[bend left=9]
(Delivery3);
\path[sgcedge] (OrderStatus1) to[bend left=10]
(Delivery3);
\path[sgedge] (Delivery3) to[bend left=15]
(OrderStatus2);
\path[sgedge] (Delivery3) to[bend left=16]
(OrderStatus2);
\path[sgedge] (Delivery3) to[bend left=17]
(OrderStatus2);
\path[sgedge] (Delivery3) to[bend left=18]
(OrderStatus2);
\path[sgedge] (Delivery3) to[bend left=19]
(OrderStatus2);
\path[sgedge] (Delivery3) to[bend left=20]
(OrderStatus2);
\path[sgedge] (OrderStatus2) to[bend left=15]
(Delivery3);
\path[sgedge] (OrderStatus2) to[bend left=16]
(Delivery3);
\path[sgedge] (OrderStatus2) to[bend left=17]
(Delivery3);
\path[sgedge] (OrderStatus2) to[bend left=18]
(Delivery3);
\path[sgedge] (OrderStatus2) to[bend left=19]
(Delivery3);
\path[sgedge] (OrderStatus2) to[bend left=20]
(Delivery3);
\path[sgcedge] (OrderStatus2) to[bend left=5]
(Delivery3);
\path[sgcedge] (OrderStatus2) to[bend left=6]
(Delivery3);
\path[sgcedge] (OrderStatus2) to[bend left=7]
(Delivery3);
\path[sgcedge] (OrderStatus2) to[bend left=8]
(Delivery3);
\path[sgcedge] (OrderStatus2) to[bend left=9]
(Delivery3);
\path[sgcedge] (OrderStatus2) to[bend left=10]
(Delivery3);
\path[sgedge] (Delivery3) to[bend left=15]
(Payment11);
\path[sgedge] (Delivery3) to[bend left=16]
(Payment11);
\path[sgedge] (Payment11) to[bend left=15]
(Delivery3);
\path[sgedge] (Payment11) to[bend left=16]
(Delivery3);
\path[sgedge] (Delivery3) to[bend left=15]
(Payment12);
\path[sgedge] (Delivery3) to[bend left=16]
(Payment12);
\path[sgedge] (Payment12) to[bend left=15]
(Delivery3);
\path[sgedge] (Payment12) to[bend left=16]
(Delivery3);
\path[sgedge] (Delivery3) to[bend left=15]
(Payment21);
\path[sgedge] (Delivery3) to[bend left=16]
(Payment21);
\path[sgedge] (Payment21) to[bend left=15]
(Delivery3);
\path[sgedge] (Payment21) to[bend left=16]
(Delivery3);
\path[sgedge] (Delivery3) to[bend left=15]
(Payment22);
\path[sgedge] (Delivery3) to[bend left=16]
(Payment22);
\path[sgedge] (Payment22) to[bend left=15]
(Delivery3);
\path[sgedge] (Payment22) to[bend left=16]
(Delivery3);
\path[sgedge] (NewOrder1) to[loop below]
(NewOrder1);
\path[sgedge] (NewOrder1) to[bend left=40]
(NewOrder2);
\path[sgedge] (NewOrder2) to[bend left=40]
(NewOrder1);
\path[sgedge] (NewOrder1) to[bend left=40]
(NewOrder3);
\path[sgedge] (NewOrder3) to[bend left=40]
(NewOrder1);
\path[sgedge] (NewOrder1) to[bend left=15]
(OrderStatus1);
\path[sgedge] (OrderStatus1) to[bend left=15]
(NewOrder1);
\path[sgcedge] (OrderStatus1) to[bend left=5]
(NewOrder1);
\path[sgedge] (NewOrder1) to[bend left=15]
(OrderStatus2);
\path[sgedge] (OrderStatus2) to[bend left=15]
(NewOrder1);
\path[sgcedge] (OrderStatus2) to[bend left=5]
(NewOrder1);
\path[sgedge] (NewOrder1) to[bend left=40]
(StockLevel);
\path[sgedge] (StockLevel) to[bend left=40]
(NewOrder1);
\path[sgcedge] (StockLevel) to[bend left=20]
(NewOrder1);
\path[sgedge] (NewOrder2) to[loop below]
(NewOrder2);
\path[sgedge] (NewOrder2) to[loop below]
(NewOrder2);
\path[sgedge] (NewOrder2) to[bend left=40]
(NewOrder3);
\path[sgedge] (NewOrder2) to[bend left=42]
(NewOrder3);
\path[sgedge] (NewOrder2) to[bend left=44]
(NewOrder3);
\path[sgedge] (NewOrder3) to[bend left=40]
(NewOrder2);
\path[sgedge] (NewOrder3) to[bend left=42]
(NewOrder2);
\path[sgedge] (NewOrder3) to[bend left=44]
(NewOrder2);
\path[sgedge] (NewOrder2) to[bend left=15]
(OrderStatus1);
\path[sgedge] (NewOrder2) to[bend left=16]
(OrderStatus1);
\path[sgedge] (OrderStatus1) to[bend left=15]
(NewOrder2);
\path[sgedge] (OrderStatus1) to[bend left=16]
(NewOrder2);
\path[sgcedge] (OrderStatus1) to[bend left=5]
(NewOrder2);
\path[sgcedge] (OrderStatus1) to[bend left=6]
(NewOrder2);
\path[sgedge] (NewOrder2) to[bend left=15]
(OrderStatus2);
\path[sgedge] (NewOrder2) to[bend left=16]
(OrderStatus2);
\path[sgedge] (OrderStatus2) to[bend left=15]
(NewOrder2);
\path[sgedge] (OrderStatus2) to[bend left=16]
(NewOrder2);
\path[sgcedge] (OrderStatus2) to[bend left=5]
(NewOrder2);
\path[sgcedge] (OrderStatus2) to[bend left=6]
(NewOrder2);
\path[sgedge] (NewOrder2) to[bend left=40]
(StockLevel);
\path[sgedge] (NewOrder2) to[bend left=42]
(StockLevel);
\path[sgedge] (NewOrder2) to[bend left=44]
(StockLevel);
\path[sgedge] (StockLevel) to[bend left=40]
(NewOrder2);
\path[sgedge] (StockLevel) to[bend left=42]
(NewOrder2);
\path[sgedge] (StockLevel) to[bend left=44]
(NewOrder2);
\path[sgcedge] (StockLevel) to[bend left=20]
(NewOrder2);
\path[sgcedge] (StockLevel) to[bend left=22]
(NewOrder2);
\path[sgcedge] (StockLevel) to[bend left=24]
(NewOrder2);
\path[sgedge] (NewOrder3) to[loop below]
(NewOrder3);
\path[sgedge] (NewOrder3) to[loop below]
(NewOrder3);
\path[sgedge] (NewOrder3) to[loop below]
(NewOrder3);
\path[sgedge] (NewOrder3) to[loop below]
(NewOrder3);
\path[sgedge] (NewOrder3) to[loop below]
(NewOrder3);
\path[sgedge] (NewOrder3) to[bend left=15]
(OrderStatus1);
\path[sgedge] (NewOrder3) to[bend left=16]
(OrderStatus1);
\path[sgedge] (NewOrder3) to[bend left=17]
(OrderStatus1);
\path[sgedge] (OrderStatus1) to[bend left=15]
(NewOrder3);
\path[sgedge] (OrderStatus1) to[bend left=16]
(NewOrder3);
\path[sgedge] (OrderStatus1) to[bend left=17]
(NewOrder3);
\path[sgcedge] (OrderStatus1) to[bend left=5]
(NewOrder3);
\path[sgcedge] (OrderStatus1) to[bend left=6]
(NewOrder3);
\path[sgcedge] (OrderStatus1) to[bend left=7]
(NewOrder3);
\path[sgedge] (NewOrder3) to[bend left=15]
(OrderStatus2);
\path[sgedge] (NewOrder3) to[bend left=16]
(OrderStatus2);
\path[sgedge] (NewOrder3) to[bend left=17]
(OrderStatus2);
\path[sgedge] (OrderStatus2) to[bend left=15]
(NewOrder3);
\path[sgedge] (OrderStatus2) to[bend left=16]
(NewOrder3);
\path[sgedge] (OrderStatus2) to[bend left=17]
(NewOrder3);
\path[sgcedge] (OrderStatus2) to[bend left=5]
(NewOrder3);
\path[sgcedge] (OrderStatus2) to[bend left=6]
(NewOrder3);
\path[sgcedge] (OrderStatus2) to[bend left=7]
(NewOrder3);
\path[sgedge] (NewOrder3) to[bend left=40]
(StockLevel);
\path[sgedge] (NewOrder3) to[bend left=42]
(StockLevel);
\path[sgedge] (NewOrder3) to[bend left=44]
(StockLevel);
\path[sgedge] (NewOrder3) to[bend left=46]
(StockLevel);
\path[sgedge] (NewOrder3) to[bend left=48]
(StockLevel);
\path[sgedge] (StockLevel) to[bend left=40]
(NewOrder3);
\path[sgedge] (StockLevel) to[bend left=42]
(NewOrder3);
\path[sgedge] (StockLevel) to[bend left=44]
(NewOrder3);
\path[sgedge] (StockLevel) to[bend left=46]
(NewOrder3);
\path[sgedge] (StockLevel) to[bend left=48]
(NewOrder3);
\path[sgcedge] (StockLevel) to[bend left=20]
(NewOrder3);
\path[sgcedge] (StockLevel) to[bend left=22]
(NewOrder3);
\path[sgcedge] (StockLevel) to[bend left=24]
(NewOrder3);
\path[sgcedge] (StockLevel) to[bend left=26]
(NewOrder3);
\path[sgcedge] (StockLevel) to[bend left=28]
(NewOrder3);
\path[sgedge] (OrderStatus1) to[bend left=40]
(Payment11);
\path[sgcedge] (OrderStatus1) to[bend left=20]
(Payment11);
\path[sgedge] (Payment11) to[bend left=40]
(OrderStatus1);
\path[sgedge] (OrderStatus1) to[bend left=40]
(Payment12);
\path[sgcedge] (OrderStatus1) to[bend left=20]
(Payment12);
\path[sgedge] (Payment12) to[bend left=40]
(OrderStatus1);
\path[sgedge] (OrderStatus1) to[bend left=40]
(Payment21);
\path[sgcedge] (OrderStatus1) to[bend left=20]
(Payment21);
\path[sgedge] (Payment21) to[bend left=40]
(OrderStatus1);
\path[sgedge] (OrderStatus1) to[bend left=40]
(Payment22);
\path[sgcedge] (OrderStatus1) to[bend left=20]
(Payment22);
\path[sgedge] (Payment22) to[bend left=40]
(OrderStatus1);
\path[sgedge] (OrderStatus2) to[bend left=40]
(Payment11);
\path[sgcedge] (OrderStatus2) to[bend left=20]
(Payment11);
\path[sgedge] (Payment11) to[bend left=40]
(OrderStatus2);
\path[sgedge] (OrderStatus2) to[bend left=40]
(Payment12);
\path[sgcedge] (OrderStatus2) to[bend left=20]
(Payment12);
\path[sgedge] (Payment12) to[bend left=40]
(OrderStatus2);
\path[sgedge] (OrderStatus2) to[bend left=40]
(Payment21);
\path[sgcedge] (OrderStatus2) to[bend left=20]
(Payment21);
\path[sgedge] (Payment21) to[bend left=40]
(OrderStatus2);
\path[sgedge] (OrderStatus2) to[bend left=40]
(Payment22);
\path[sgcedge] (OrderStatus2) to[bend left=20]
(Payment22);
\path[sgedge] (Payment22) to[bend left=40]
(OrderStatus2);
\path[sgedge] (Payment11) to[loop above]
(Payment11);
\path[sgedge] (Payment11) to[loop above]
(Payment11);
\path[sgedge] (Payment11) to[loop above]
(Payment11);
\path[sgedge] (Payment11) to[loop above]
(Payment11);
\path[sgedge] (Payment11) to[loop above]
(Payment11);
\path[sgedge] (Payment11) to[loop above]
(Payment11);
\path[sgedge] (Payment11) to[bend left=40]
(Payment12);
\path[sgedge] (Payment11) to[bend left=42]
(Payment12);
\path[sgedge] (Payment11) to[bend left=44]
(Payment12);
\path[sgedge] (Payment12) to[bend left=40]
(Payment11);
\path[sgedge] (Payment12) to[bend left=42]
(Payment11);
\path[sgedge] (Payment12) to[bend left=44]
(Payment11);
\path[sgedge] (Payment11) to[bend left=40]
(Payment21);
\path[sgedge] (Payment11) to[bend left=42]
(Payment21);
\path[sgedge] (Payment11) to[bend left=44]
(Payment21);
\path[sgedge] (Payment11) to[bend left=46]
(Payment21);
\path[sgedge] (Payment11) to[bend left=48]
(Payment21);
\path[sgedge] (Payment11) to[bend left=50]
(Payment21);
\path[sgedge] (Payment21) to[bend left=40]
(Payment11);
\path[sgedge] (Payment21) to[bend left=42]
(Payment11);
\path[sgedge] (Payment21) to[bend left=44]
(Payment11);
\path[sgedge] (Payment21) to[bend left=46]
(Payment11);
\path[sgedge] (Payment21) to[bend left=48]
(Payment11);
\path[sgedge] (Payment21) to[bend left=50]
(Payment11);
\path[sgedge] (Payment11) to[bend left=40]
(Payment22);
\path[sgedge] (Payment11) to[bend left=42]
(Payment22);
\path[sgedge] (Payment11) to[bend left=44]
(Payment22);
\path[sgedge] (Payment22) to[bend left=40]
(Payment11);
\path[sgedge] (Payment22) to[bend left=42]
(Payment11);
\path[sgedge] (Payment22) to[bend left=44]
(Payment11);
\path[sgedge] (Payment12) to[loop above]
(Payment12);
\path[sgedge] (Payment12) to[loop above]
(Payment12);
\path[sgedge] (Payment12) to[loop above]
(Payment12);
\path[sgedge] (Payment12) to[bend left=40]
(Payment21);
\path[sgedge] (Payment12) to[bend left=42]
(Payment21);
\path[sgedge] (Payment12) to[bend left=44]
(Payment21);
\path[sgedge] (Payment21) to[bend left=40]
(Payment12);
\path[sgedge] (Payment21) to[bend left=42]
(Payment12);
\path[sgedge] (Payment21) to[bend left=44]
(Payment12);
\path[sgedge] (Payment12) to[bend left=40]
(Payment22);
\path[sgedge] (Payment12) to[bend left=42]
(Payment22);
\path[sgedge] (Payment12) to[bend left=44]
(Payment22);
\path[sgedge] (Payment22) to[bend left=40]
(Payment12);
\path[sgedge] (Payment22) to[bend left=42]
(Payment12);
\path[sgedge] (Payment22) to[bend left=44]
(Payment12);
\path[sgedge] (Payment21) to[loop above]
(Payment21);
\path[sgedge] (Payment21) to[loop above]
(Payment21);
\path[sgedge] (Payment21) to[loop above]
(Payment21);
\path[sgedge] (Payment21) to[loop above]
(Payment21);
\path[sgedge] (Payment21) to[loop above]
(Payment21);
\path[sgedge] (Payment21) to[loop above]
(Payment21);
\path[sgedge] (Payment21) to[bend left=40]
(Payment22);
\path[sgedge] (Payment21) to[bend left=42]
(Payment22);
\path[sgedge] (Payment21) to[bend left=44]
(Payment22);
\path[sgedge] (Payment22) to[bend left=40]
(Payment21);
\path[sgedge] (Payment22) to[bend left=42]
(Payment21);
\path[sgedge] (Payment22) to[bend left=44]
(Payment21);
\path[sgedge] (Payment22) to[loop above]
(Payment22);
\path[sgedge] (Payment22) to[loop above]
(Payment22);
\path[sgedge] (Payment22) to[loop above]
(Payment22);
\end{tikzpicture}
\caption{Summary graph for the TPC-C benchmark. Counterflow edges are represented by dashed edges. To facilitate the presentation, edge labels are not visualized.}
\label{fig:tpc-c:sg}
\end{figure}

\subsection{Auction($n$) Benchmark}
Figure~\ref{fig:auction:n:sg} illustrates the general structure of summary graphs for Auction($n$) for arbitrary values of $n$.

\begin{figure}
\resizebox{0.99\textwidth}{!}{
\begin{tikzpicture}[sgnode/.style={draw}, sgedge/.style={draw,->},sgcedge/.style={sgedge,dashed},qnode/.style={font=\footnotesize}]
  \node[sgnode] (FindBids1) at (-7.5,0) {$\text{FB}^1$};
  \node[sgnode] (PlaceBid11) at (-6,0) {$\text{PB}^1_1$};
  \node[sgnode] (PlaceBid12) at (-4.5,0) {$\text{PB}^1_2$};
  \node[sgnode] (FindBids2) at (-2.5,0) {$\text{FB}^2$};
  \node[sgnode] (PlaceBid21) at (-1,0) {$\text{PB}^2_1$};
  \node[sgnode] (PlaceBid22) at (0.5,0) {$\text{PB}^2_2$};
  \node[sgnode] (FindBids3) at (2.5,0) {$\text{FB}^3$};
  \node[sgnode] (PlaceBid31) at (4,0) {$\text{PB}^3_1$};
  \node[sgnode] (PlaceBid32) at (5.5,0) {$\text{PB}^3_2$};
  \node[sgnode] (FindBids4) at (10,0) {$\text{FB}^n$};
  \node[sgnode] (PlaceBid41) at (11.5,0) {$\text{PB}^n_1$};
  \node[sgnode] (PlaceBid42) at (13,0) {$\text{PB}^n_2$};
  \path (PlaceBid32) to node {$\ldots$} (FindBids4);
\path[sgedge] (FindBids1) to[loop above]
(FindBids1);
\path[sgedge] (FindBids1) to[bend left=40]
(FindBids2);
\path[sgedge] (FindBids2) to[bend left=40]
(FindBids1);
\path[sgedge] (FindBids1) to[bend left=40]
(FindBids3);
\path[sgedge] (FindBids3) to[bend left=40]
(FindBids1);
\path[sgedge] (FindBids1) to[bend left=40]
(FindBids4);
\path[sgedge] (FindBids4) to[bend left=40]
(FindBids1);
\path[sgedge] (FindBids1) to[bend left=40]
(PlaceBid11);
\path[sgedge] (FindBids1) to[bend left=48]
(PlaceBid11);
\path[sgcedge] (FindBids1) to[bend left=20]
(PlaceBid11);
\path[sgedge] (PlaceBid11) to[bend left=40]
(FindBids1);
\path[sgedge] (PlaceBid11) to[bend left=48]
(FindBids1);
\path[sgedge] (FindBids1) to[bend left=40]
(PlaceBid12);
\path[sgedge] (PlaceBid12) to[bend left=40]
(FindBids1);
\path[sgedge] (FindBids1) to[bend left=40]
(PlaceBid21);
\path[sgedge] (PlaceBid21) to[bend left=40]
(FindBids1);
\path[sgedge] (FindBids1) to[bend left=40]
(PlaceBid22);
\path[sgedge] (PlaceBid22) to[bend left=40]
(FindBids1);
\path[sgedge] (FindBids1) to[bend left=40]
(PlaceBid31);
\path[sgedge] (PlaceBid31) to[bend left=40]
(FindBids1);
\path[sgedge] (FindBids1) to[bend left=40]
(PlaceBid32);
\path[sgedge] (PlaceBid32) to[bend left=40]
(FindBids1);
\path[sgedge] (FindBids1) to[bend left=40]
(PlaceBid41);
\path[sgedge] (PlaceBid41) to[bend left=40]
(FindBids1);
\path[sgedge] (FindBids1) to[bend left=40]
(PlaceBid42);
\path[sgedge] (PlaceBid42) to[bend left=40]
(FindBids1);
\path[sgedge] (FindBids2) to[loop above]
(FindBids2);
\path[sgedge] (FindBids2) to[bend left=40]
(FindBids3);
\path[sgedge] (FindBids3) to[bend left=40]
(FindBids2);
\path[sgedge] (FindBids2) to[bend left=40]
(FindBids4);
\path[sgedge] (FindBids4) to[bend left=40]
(FindBids2);
\path[sgedge] (FindBids2) to[bend left=40]
(PlaceBid11);
\path[sgedge] (PlaceBid11) to[bend left=40]
(FindBids2);
\path[sgedge] (FindBids2) to[bend left=40]
(PlaceBid12);
\path[sgedge] (PlaceBid12) to[bend left=40]
(FindBids2);
\path[sgedge] (FindBids2) to[bend left=40]
(PlaceBid21);
\path[sgedge] (FindBids2) to[bend left=48]
(PlaceBid21);
\path[sgcedge] (FindBids2) to[bend left=20]
(PlaceBid21);
\path[sgedge] (PlaceBid21) to[bend left=40]
(FindBids2);
\path[sgedge] (PlaceBid21) to[bend left=48]
(FindBids2);
\path[sgedge] (FindBids2) to[bend left=40]
(PlaceBid22);
\path[sgedge] (PlaceBid22) to[bend left=40]
(FindBids2);
\path[sgedge] (FindBids2) to[bend left=40]
(PlaceBid31);
\path[sgedge] (PlaceBid31) to[bend left=40]
(FindBids2);
\path[sgedge] (FindBids2) to[bend left=40]
(PlaceBid32);
\path[sgedge] (PlaceBid32) to[bend left=40]
(FindBids2);
\path[sgedge] (FindBids2) to[bend left=40]
(PlaceBid41);
\path[sgedge] (PlaceBid41) to[bend left=40]
(FindBids2);
\path[sgedge] (FindBids2) to[bend left=40]
(PlaceBid42);
\path[sgedge] (PlaceBid42) to[bend left=40]
(FindBids2);
\path[sgedge] (FindBids3) to[loop above]
(FindBids3);
\path[sgedge] (FindBids3) to[bend left=40]
(FindBids4);
\path[sgedge] (FindBids4) to[bend left=40]
(FindBids3);
\path[sgedge] (FindBids3) to[bend left=40]
(PlaceBid11);
\path[sgedge] (PlaceBid11) to[bend left=40]
(FindBids3);
\path[sgedge] (FindBids3) to[bend left=40]
(PlaceBid12);
\path[sgedge] (PlaceBid12) to[bend left=40]
(FindBids3);
\path[sgedge] (FindBids3) to[bend left=40]
(PlaceBid21);
\path[sgedge] (PlaceBid21) to[bend left=40]
(FindBids3);
\path[sgedge] (FindBids3) to[bend left=40]
(PlaceBid22);
\path[sgedge] (PlaceBid22) to[bend left=40]
(FindBids3);
\path[sgedge] (FindBids3) to[bend left=40]
(PlaceBid31);
\path[sgedge] (FindBids3) to[bend left=48]
(PlaceBid31);
\path[sgcedge] (FindBids3) to[bend left=20]
(PlaceBid31);
\path[sgedge] (PlaceBid31) to[bend left=40]
(FindBids3);
\path[sgedge] (PlaceBid31) to[bend left=48]
(FindBids3);
\path[sgedge] (FindBids3) to[bend left=40]
(PlaceBid32);
\path[sgedge] (PlaceBid32) to[bend left=40]
(FindBids3);
\path[sgedge] (FindBids3) to[bend left=40]
(PlaceBid41);
\path[sgedge] (PlaceBid41) to[bend left=40]
(FindBids3);
\path[sgedge] (FindBids3) to[bend left=40]
(PlaceBid42);
\path[sgedge] (PlaceBid42) to[bend left=40]
(FindBids3);
\path[sgedge] (FindBids4) to[loop above]
(FindBids4);
\path[sgedge] (FindBids4) to[bend left=40]
(PlaceBid11);
\path[sgedge] (PlaceBid11) to[bend left=40]
(FindBids4);
\path[sgedge] (FindBids4) to[bend left=40]
(PlaceBid12);
\path[sgedge] (PlaceBid12) to[bend left=40]
(FindBids4);
\path[sgedge] (FindBids4) to[bend left=40]
(PlaceBid21);
\path[sgedge] (PlaceBid21) to[bend left=40]
(FindBids4);
\path[sgedge] (FindBids4) to[bend left=40]
(PlaceBid22);
\path[sgedge] (PlaceBid22) to[bend left=40]
(FindBids4);
\path[sgedge] (FindBids4) to[bend left=40]
(PlaceBid31);
\path[sgedge] (PlaceBid31) to[bend left=40]
(FindBids4);
\path[sgedge] (FindBids4) to[bend left=40]
(PlaceBid32);
\path[sgedge] (PlaceBid32) to[bend left=40]
(FindBids4);
\path[sgedge] (FindBids4) to[bend left=40]
(PlaceBid41);
\path[sgedge] (FindBids4) to[bend left=48]
(PlaceBid41);
\path[sgcedge] (FindBids4) to[bend left=20]
(PlaceBid41);
\path[sgedge] (PlaceBid41) to[bend left=40]
(FindBids4);
\path[sgedge] (PlaceBid41) to[bend left=48]
(FindBids4);
\path[sgedge] (FindBids4) to[bend left=40]
(PlaceBid42);
\path[sgedge] (PlaceBid42) to[bend left=40]
(FindBids4);
\path[sgedge] (PlaceBid11) to[loop above,looseness=7]
(PlaceBid11);
\path[sgedge] (PlaceBid11) to[loop above,looseness=9]
(PlaceBid11);
\path[sgedge] (PlaceBid11) to[loop above,looseness=11]
(PlaceBid11);
\path[sgedge] (PlaceBid11) to[loop above,looseness=13]
(PlaceBid11);
\path[sgedge] (PlaceBid11) to[bend left=40]
(PlaceBid12);
\path[sgedge] (PlaceBid11) to[bend left=48]
(PlaceBid12);
\path[sgedge] (PlaceBid12) to[bend left=40]
(PlaceBid11);
\path[sgedge] (PlaceBid12) to[bend left=48]
(PlaceBid11);
\path[sgedge] (PlaceBid11) to[bend left=40]
(PlaceBid21);
\path[sgedge] (PlaceBid21) to[bend left=40]
(PlaceBid11);
\path[sgedge] (PlaceBid11) to[bend left=40]
(PlaceBid22);
\path[sgedge] (PlaceBid22) to[bend left=40]
(PlaceBid11);
\path[sgedge] (PlaceBid11) to[bend left=40]
(PlaceBid31);
\path[sgedge] (PlaceBid31) to[bend left=40]
(PlaceBid11);
\path[sgedge] (PlaceBid11) to[bend left=40]
(PlaceBid32);
\path[sgedge] (PlaceBid32) to[bend left=40]
(PlaceBid11);
\path[sgedge] (PlaceBid11) to[bend left=40]
(PlaceBid41);
\path[sgedge] (PlaceBid41) to[bend left=40]
(PlaceBid11);
\path[sgedge] (PlaceBid11) to[bend left=40]
(PlaceBid42);
\path[sgedge] (PlaceBid42) to[bend left=40]
(PlaceBid11);
\path[sgedge] (PlaceBid12) to[loop above]
(PlaceBid12);
\path[sgedge] (PlaceBid12) to[bend left=40]
(PlaceBid21);
\path[sgedge] (PlaceBid21) to[bend left=40]
(PlaceBid12);
\path[sgedge] (PlaceBid12) to[bend left=40]
(PlaceBid22);
\path[sgedge] (PlaceBid22) to[bend left=40]
(PlaceBid12);
\path[sgedge] (PlaceBid12) to[bend left=40]
(PlaceBid31);
\path[sgedge] (PlaceBid31) to[bend left=40]
(PlaceBid12);
\path[sgedge] (PlaceBid12) to[bend left=40]
(PlaceBid32);
\path[sgedge] (PlaceBid32) to[bend left=40]
(PlaceBid12);
\path[sgedge] (PlaceBid12) to[bend left=40]
(PlaceBid41);
\path[sgedge] (PlaceBid41) to[bend left=40]
(PlaceBid12);
\path[sgedge] (PlaceBid12) to[bend left=40]
(PlaceBid42);
\path[sgedge] (PlaceBid42) to[bend left=40]
(PlaceBid12);
\path[sgedge] (PlaceBid21) to[loop above,looseness=7]
(PlaceBid21);
\path[sgedge] (PlaceBid21) to[loop above,looseness=9]
(PlaceBid21);
\path[sgedge] (PlaceBid21) to[loop above,looseness=11]
(PlaceBid21);
\path[sgedge] (PlaceBid21) to[loop above,looseness=13]
(PlaceBid21);
\path[sgedge] (PlaceBid21) to[bend left=40]
(PlaceBid22);
\path[sgedge] (PlaceBid21) to[bend left=48]
(PlaceBid22);
\path[sgedge] (PlaceBid22) to[bend left=40]
(PlaceBid21);
\path[sgedge] (PlaceBid22) to[bend left=48]
(PlaceBid21);
\path[sgedge] (PlaceBid21) to[bend left=40]
(PlaceBid31);
\path[sgedge] (PlaceBid31) to[bend left=40]
(PlaceBid21);
\path[sgedge] (PlaceBid21) to[bend left=40]
(PlaceBid32);
\path[sgedge] (PlaceBid32) to[bend left=40]
(PlaceBid21);
\path[sgedge] (PlaceBid21) to[bend left=40]
(PlaceBid41);
\path[sgedge] (PlaceBid41) to[bend left=40]
(PlaceBid21);
\path[sgedge] (PlaceBid21) to[bend left=40]
(PlaceBid42);
\path[sgedge] (PlaceBid42) to[bend left=40]
(PlaceBid21);
\path[sgedge] (PlaceBid22) to[loop above]
(PlaceBid22);
\path[sgedge] (PlaceBid22) to[bend left=40]
(PlaceBid31);
\path[sgedge] (PlaceBid31) to[bend left=40]
(PlaceBid22);
\path[sgedge] (PlaceBid22) to[bend left=40]
(PlaceBid32);
\path[sgedge] (PlaceBid32) to[bend left=40]
(PlaceBid22);
\path[sgedge] (PlaceBid22) to[bend left=40]
(PlaceBid41);
\path[sgedge] (PlaceBid41) to[bend left=40]
(PlaceBid22);
\path[sgedge] (PlaceBid22) to[bend left=40]
(PlaceBid42);
\path[sgedge] (PlaceBid42) to[bend left=40]
(PlaceBid22);
\path[sgedge] (PlaceBid31) to[loop above,looseness=7]
(PlaceBid31);
\path[sgedge] (PlaceBid31) to[loop above,looseness=9]
(PlaceBid31);
\path[sgedge] (PlaceBid31) to[loop above,looseness=11]
(PlaceBid31);
\path[sgedge] (PlaceBid31) to[loop above,looseness=13]
(PlaceBid31);
\path[sgedge] (PlaceBid31) to[bend left=40]
(PlaceBid32);
\path[sgedge] (PlaceBid31) to[bend left=48]
(PlaceBid32);
\path[sgedge] (PlaceBid32) to[bend left=40]
(PlaceBid31);
\path[sgedge] (PlaceBid32) to[bend left=48]
(PlaceBid31);
\path[sgedge] (PlaceBid31) to[bend left=40]
(PlaceBid41);
\path[sgedge] (PlaceBid41) to[bend left=40]
(PlaceBid31);
\path[sgedge] (PlaceBid31) to[bend left=40]
(PlaceBid42);
\path[sgedge] (PlaceBid42) to[bend left=40]
(PlaceBid31);
\path[sgedge] (PlaceBid32) to[loop above]
(PlaceBid32);
\path[sgedge] (PlaceBid32) to[bend left=40]
(PlaceBid41);
\path[sgedge] (PlaceBid41) to[bend left=40]
(PlaceBid32);
\path[sgedge] (PlaceBid32) to[bend left=40]
(PlaceBid42);
\path[sgedge] (PlaceBid42) to[bend left=40]
(PlaceBid32);
\path[sgedge] (PlaceBid41) to[loop above,looseness=7]
(PlaceBid41);
\path[sgedge] (PlaceBid41) to[loop above,looseness=9]
(PlaceBid41);
\path[sgedge] (PlaceBid41) to[loop above,looseness=11]
(PlaceBid41);
\path[sgedge] (PlaceBid41) to[loop above,looseness=13]
(PlaceBid41);
\path[sgedge] (PlaceBid41) to[bend left=40]
(PlaceBid42);
\path[sgedge] (PlaceBid41) to[bend left=48]
(PlaceBid42);
\path[sgedge] (PlaceBid42) to[bend left=40]
(PlaceBid41);
\path[sgedge] (PlaceBid42) to[bend left=48]
(PlaceBid41);
\path[sgedge] (PlaceBid42) to[loop above]
(PlaceBid42);
\end{tikzpicture}
} 
\caption{Summary graph for the Auction($n$) benchmark. Counterflow edges are represented by dashed edges. To facilitate the presentation, edge labels are not visualized.}
\label{fig:auction:n:sg}
\end{figure}

}{}

\end{document}